\documentclass[reqno,12pt]{article}
\usepackage{amsmath,amsfonts,amssymb,amsthm,amstext,amscd}
\usepackage[all]{xy}
\usepackage{hyperref}
\usepackage{epsfig}
\usepackage{graphicx}
\usepackage{tikz}
\usetikzlibrary{decorations.pathmorphing}
\usepackage{caption}
\usepackage{subcaption}
\usepackage{lipsum}
\usepackage{psfrag}
\usepackage{comment}
\usepackage{cite}

\makeatletter \@addtoreset{equation}{section}

\makeatletter\renewcommand\section{\@startsection {section}{1}{\z@}%
                                                                                                                                                                                                                                                                                                                                                                                                                                                                                                                                                                                                                                                                                                                                                                                                                                                                                                                                                                                                                                                                                                                                                                                                                                                                                                                                                                                                                                                                                                                                                                                                                                                                                                                                                                                                                                                                                                   {-3.5ex \@plus -1ex \@minus -.2ex}
                                                                                                                                                                                                                                                                                                                                                                                                                                                                                                                                                                                                                                                                                                                                                                                                                                                                                                                                                                                                                                                                                                                                                                                                                                                                                                                                                                                                                                                                                                                                                                                                                                                                                                                                                                                                                                                                                                   {2.3ex \@plus.2ex}%
                                                                                                                                                                                                                                                                                                                                                                                                                                                                                                                                                                                                                                                                                                                                                                                                                                                                                                                                                                                                                                                                                                                                                                                                                                                                                                                                                                                                                                                                                                                                                                                                                                                                                                                                                                                                                                                                                                   {\normalfont\large\bfseries}}
\renewcommand\subsection{\@startsection{subsection}{2}{\z@}%
                                                                                                                                                                                                                                                                                                                                                                                                                                                                                                                                                                                                                                                                                                                                                                                                                                                                                                                                                                                                                                                                                                                                                                                                                                                                                                                                                                                                                                                                                                                                                                                                                                                                                                                                                                                                                                                                                                                                                                                                                                                     {-3.25ex\@plus -1ex \@minus -.2ex}%
                                                                                                                                                                                                                                                                                                                                                                                                                                                                                                                                                                                                                                                                                                                                                                                                                                                                                                                                                                                                                                                                                                                                                                                                                                                                                                                                                                                                                                                                                                                                                                                                                                                                                                                                                                                                                                                                                                                                                                                                                                                     {1.5ex \@plus .2ex}%
                                                                                                                                                                                                                                                                                                                                                                                                                                                                                                                                                                                                                                                                                                                                                                                                                                                                                                                                                                                                                                                                                                                                                                                                                                                                                                                                                                                                                                                                                                                                                                                                                                                                                                                                                                                                                                                                                                                                                                                                                                                     {\normalfont\bfseries}}

\newcommand{\be}{\begin{equation}}
\newcommand{\ee}{\end{equation}}
\newcommand{\bea}{\begin{eqnarray}}
\newcommand{\eea}{\end{eqnarray}}
\newcommand{\bse}{\begin{subequations}}
\newcommand{\ese}{\end{subequations}}
\newcommand{\beqa}{\begin{eqnarray}}
\newcommand{\eeqa}{\end{eqnarray}}
\newcommand{\beqar}{\begin{eqnarray*}}
\newcommand{\eeqar}{\end{eqnarray*}}
\newcommand{\bi}{\begin{itemize}}
\newcommand{\ei}{\end{itemize}}
\newcommand{\bn}{\begin{enumerate}}
\newcommand{\en}{\end{enumerate}}

\newcommand{\ba}{\begin{array}}
\newcommand{\ea}{\end{array}}
\newcommand{\bc}{\begin{center}}
\newcommand{\ec}{\end{center}}

\newcommand{\cf}{{\em cf.}\ }
\newcommand{\ie}{{\em i.e.}\ }

\def\pd{\partial}

\newcommand{\p}{\partial}

\newcommand{\mn}{{\mu\nu}}

\newcommand{\de}{\delta}

\newcommand{\vn}{{\vec{n}}}
\newcommand{\eps}{\epsilon}
\newcommand{\bomega}{{\boldsymbol \omega}}
\newcommand{\bTheta}{{\boldsymbol \Theta}}

\newcommand{\cL}{{\mathcal L}}

\newcommand{\bm}{\boldsymbol}
\def\d{\delta}

\def\om{\omega}
\def\nhegalgebra{\widehat{\mathcal{V}_{\vec{k},S}}}
\def\nhegps{\mathcal{G}_{\{ p \}}[F]}

\newtheorem{theorem}{Theorem}

\begin{document}

\begin{titlepage}

\begin{flushright}\vspace{-3cm}
{\small
IPM/P-2015/nnn \\
\today }\end{flushright}

\begin{center}

\centerline{{\Large{\bf{Wiggling Throat of Extremal Black Holes}}}} \vspace{10mm}

\centerline{\large{\bf{G. Comp\`{e}re\footnote{e-mail:gcompere@ulb.ac.be}$\, ^{\dag}$, K. Hajian\footnote{e-mail:kamalhajian@physics.sharif.edu}$\, ^{\ddag, \S}$, A. Seraj\footnote{e-mail:ali\_seraj@ipm.ir}$\,^{\dag}$ $^{\!\!\!\!,\ddag}$, M.M. Sheikh-Jabbari
\footnote{e-mail:
jabbari@theory.ipm.ac.ir}$\,^\ddag$}}}

\vspace{5mm}
\normalsize
\bigskip\medskip
{$^\dag$ \it \textit{Universit\'{e} Libre de Bruxelles and International Solvay Institutes
CP 231 B-1050 Brussels, Belgium
}\\
\smallskip
$^\ddag$ \textit{School of Physics, Institute for Research in Fundamental
Sciences (IPM), \\P.O.Box 19395-5531, Tehran, Iran} \\ \smallskip
$^\S$ \textit{Department of Physics, Sharif University of Technology,\\
 P.O. Box 11365-8639, Tehran, Iran}}

\vspace{5mm}

\begin{abstract}
\noindent 
{ We construct the classical phase space of geometries in the near-horizon region of vacuum extremal black holes as announced in [arXiv:1503.07861]. Motivated by the uniqueness theorems for such solutions and for perturbations around them, we build a family of metrics depending upon a single periodic function defined on the torus spanned by the $U(1)$ isometry directions. We show that this set of metrics is equipped with a consistent symplectic structure and hence defines a phase space. The phase space forms a representation of an infinite dimensional algebra of so-called symplectic symmetries. The symmetry algebra is an extension of the Virasoro algebra whose central extension is the black hole entropy. We motivate the choice of diffeomorphisms leading to the phase space and explicitly derive the symplectic structure, the algebra of symplectic symmetries and the corresponding conserved charges. We also discuss a formulation of these charges with a Liouville type stress-tensor on the torus defined by the $U(1)$ isometries and outline possible future directions.}

\end{abstract}


\end{center}

\end{titlepage}
\setcounter{footnote}{0}
\renewcommand{\baselinestretch}{1.05}  

\addtocontents{toc}{\protect\setcounter{tocdepth}{2}}
\tableofcontents


\section{Introduction and Summary of Results }

The only known microscopic models of black holes in string theory describe  supersymmetric black holes  \cite{Strominger:1996sh,Dabholkar:2014ema}, which are necessarily also extremal.  Remarkably, the extremal Kerr black hole is a close model to some of observed astrophysical black holes \cite{McClintock:2006xd,2011ApJ...742...85G,Brenneman:2006hw}. However, the extremal Kerr black hole cannot be supersymmetric and hence known string theory descriptions do not apply to these realistic black holes. It is then natural to ask which methods, independent of supersymmetry, can provide us with relevant microscopic information about non-supersymmetric extremal black holes.

On the other hand, given the fact that  black holes, in general, admit a thermodynamical description at the semiclassical level \cite{Bardeen:1973gs}, and in particular have entropy \cite{Bekenstein:1973ur},  one is motivated to explore how much extra information about the microscopic description of black holes one may be able to extract from a low energy description as a solution of classical gravity.   The main aim of this paper, which was announced in \cite{Compere:2015mza}, is to make steps in this direction. In particular, we present  a consistent proposal for the classical phase space and symmetries of the gravitational field around extremal spinning black holes in four and higher dimensions using covariant phase space methods. Given the phase space, the symplectic structure and its symmetries, one can apply usual quantization procedures. The latter may then provide a setup to explore the microscopic description of extremal black holes.

After the seminal work of Wald \cite{Wald:1993nt}, we have learned that the entropy of a black hole with a Killing horizon  may be viewed as a conserved Noether-Wald charge \cite{Iyer:1994ys} associated with the Killing vector field generating the horizon. We also know that the temperature attributed to the thermodynamic description of a black hole is a quantity which can be read only, { up to a conventional normalization usually imposed at spatial infinity,} from the form of the metric near the horizon. Moreover, other quantities which appear as chemical potentials in the  thermodynamical description of black holes like horizon angular velocity and horizon electric potential, are quantities attributed to the horizon. These and other facts about black holes have led to the idea that the information about  black hole microstates is completely encoded in the (quantum and classical) near horizon data. { If this idea is correct, no information is needed in the surroundings of the black hole nor in its interior to describe the black hole microstates. The near horizon geometry for a generic black hole is not a decoupled region of the black hole in the sense of geodesic completeness.} Nonetheless, for the class of extremal black holes, i.e. black holes with degenerate (non-bifurcate Killing) horizon or, equivalently, black holes at zero Hawking temperature, a near horizon limit exists and yields a new class of solutions { decoupled from the asymptotic region,} the Near Horizon Extremal Geometries (NHEG). Therefore, within the mindset alluded above, it is natural to explore the NHEG family in search for a formulation of (extremal) black hole microstate problem. This is the setup we will analyze here.

At the classical level, there are uniqueness theorems for extremal black holes and their near horizon geometries.  In particular, the extremal Kerr black hole is the unique asymptotically flat, stationary vacuum solution to four dimensional Einstein's equations \cite{Amsel:2009et}. It admits a near-horizon limit with enhanced  $SL(2,\mathbb R) \times U(1)$ isometry \cite{Bardeen:1999px}, which is again an Einstein vacuum solution. This new geometry is  the unique solution with this set of isometries \cite{Kunduri:2007vf}. Similar statements extend to $d$ dimensional Einstein vacuum solutions with  $SL(2,\mathbb R) \times U(1)^{d-3}$ isometry \cite{Kunduri:2007vf, Kunduri:2013gce}. This is the class of NHEG's we will be focusing on in this work.

Killing horizons (the codimension one null surface generated by a Killing vector) and bifurcation horizons (codimension two intersections of future and past branches of Killing horizons) play a crucial role in the thermodynamic analysis of black holes and in defining the conserved charges. Although not black holes (in the absence of an event horizon), the NHEG have an infinite set of bifurcation surfaces with unit surface gravity \cite{Hajian:2013lna},  as we will review and detail in section \ref{sec-NHEG-review}. Moreover, one can define the entropy as a conserved Noether-Wald charge on any of these bifurcation horizons upon using a specific linear combination of   $SL(2,\mathbb R) \times U(1)^{d-3}$ isometries as generator \cite{Hajian:2013lna, Hajian:2014twa}. Invariance under  
$SL(2,\mathbb{R})$  then ensures that the conserved charge is independent of the choice of bifurcation surface. The near-horizon geometry and its enhanced symmetries allow to find the precise symmetry canonically associated with entropy in the strict extremal limit, thereby completing Wald's program \cite{Wald:1993nt}.  

Appearance of an AdS$_2$ factor in the geometry (associated with the $SL(2,\mathbb R)$ isometry), may prompt the idea of using  an AdS/CFT correspondence \cite{Aharony:1999ti} in exploring the black hole microstates. This idea seems  to be full of obstacles given the issues with defining quantum (gravity) theories on AdS$_2$; e.g. see \cite{Sen:2008yk, Sen:2008vm,Strominger:1998yg,Maldacena:1998uz,Hartman:2008dq,Castro:2014ima}. Another related proposal put forward in \cite{Guica:2008mu}, is considering perturbations with prescribed falloff behavior on the near-horizon limit of extremal Kerr and studying their asymptotic symmetry group, with the idea to promote the asymptotic symmetry group to the symmetry of the quantum Hilbert space of microstates.  
{Nonetheless, it was realized that this proposal which is usually dubbed as Kerr/CFT cannot be a full-fledged correspondence because of the following conceptual problems:} the near horizon limit does not admit consistent back-reacted local bulk dynamics \cite{Amsel:2009ev,Dias:2009ex}, it does not admit axisymmetric and stationary configurations other than the background itself \cite{Amsel:2009ev}, and it does not admit perturbations which asymptotically respect the background isometries \cite{Hajian:2014twa}. Given these background and perturbation uniqueness theorems, one is hence led to considering perturbations only generated through diffeomorphisms.

Symmetries and their associated conserved charges have been an important guiding principle, especially in modern physics. Within the set of diffeomorphisms relevant to describe generally covariant gravitational theories and in the context of near-horizon extremal geometries,  two classes of symmetries, namely isometries and asymptotic symmetries, have been largely studied in the literature  \cite{Bardeen:1999px,Kunduri:2007vf, Kunduri:2013gce,Guica:2008mu}. A third class of symmetries, dubbed symplectic symmetries, was introduced in \cite{Compere:2015mza} and is our main focus in this paper. Our main result is the construction of the NHEG phase space, including the symplectic structure and its conserved charges, using these symmetries.

Before stating the summary of our results, we pause for explaining the difference between symplectic  symmetries (appearing in this work) and asymptotic symmetries (e.g. appearing in the Kerr/CFT setup \cite{Guica:2008mu,Bredberg:2011hp,Compere:2012jk}). In general, gauge systems such as gravity, admit a conserved symplectic structure, which allows one to define a (not necessarily conserved) surface charge associated with any gauge parameter, such as a diffeomorphism generator. A symplectic symmetry is defined as a gauge parameter such that the symplectic structure, when contracted with the corresponding gauge transformation, vanishes on-shell but not off-shell. Such symplectic symmetries are large gauge transformations, similar to asymptotic symmetries, but they are defined everywhere in spacetime, not only in an asymptotic region. They are associated with nontrivial conserved surface charges. The existence of symplectic symmetries implies the existence of boundary conditions where the asymptotic symmetries are the symplectic symmetries, but not necessarily the other way around. AdS$_3$ Einstein gravity provides an example of symplectic symmetries: the two Virasoro algebras found as asymptotic symmetries by Brown and Henneaux \cite{Brown:1986nw} can be promoted to symplectic symmetries \cite{Compere:2014cna}. We expect that symplectic symmetries might arise when bulk propagating degrees of freedom are absent. Motivated by the lack of consistent dynamical degrees of freedom in the near-horizon limit of extremal black holes \cite{Amsel:2009ev,Dias:2009ex,Hajian:2014twa}, it is then natural to search for symplectic symmetries in such near-horizon geometries, too. That is exactly what we will do in this work.

\subsection{Summary of results}

In this work we focus on the class of $d$ dimensional Near Horizon Extremal Geometries, which are solutions to vacuum Einstein gravity and have $SL(2,\mathbb R) \times U(1)^{d-3}$ isometry. These geometries are 
specified by $k^i,\ i=1,2,\cdots, d-3$, which will be collectively denoted as $\vec{k}$  and a set of functions of the coordinate $\theta$.\footnote{The dimensionless vector $\vec{k} $ physically represents the linear change of angular velocity close to extremality, normalized using the Hawking temperature,  $\vec{\Omega} = \vec{\Omega}_{ext} + \frac{2 \pi}{\hbar} \vec{k} \;T_H + O(T_H^2)$, see e.g. \cite{Compere:2012jk,Johnstone:2013ioa}. For the extremal Kerr black hole, $k=1$.} There are then $d-3$ conserved charges $\vec{J}$, associated with $U(1)^{d-3}$.
The NHEG has an entropy $S$ which is related to the other parameters as $\frac{\hbar}{2\pi} S=  k^i J_i= \vec{k}\cdot\vec{J}$ \cite{Astefanesei:2006dd,Hajian:2013lna}.

Our main result are: 
\begin{enumerate}
\item The existence of the NHEG phase space $\mathcal{G}[F]$, i.e. a set of   diffeomorphic metrics with $SL(2,\mathbb R) \times U(1)^{d-3}$ isometry which depend upon an arbitrary periodic function  $F=F(\vec{\varphi})$ on the $d-3$ torus spanned by the $U(1)$ isometries dubbed the \emph{wiggle function}. Symplectic symmetries can be defined as the set of diffeomorphisms which can arbitrarily change the wiggle function.
\item The phase space is equipped with a consistent symplectic structure through which we define conserved surface charges associated to any each
symplectic symmetry. 
\item We work out the algebra of these conserved charges, the NHEG algebra $\widehat{\mathcal{V}_{\vec{k},S}}$ whose generators $L_{\vec{n}}$,  
$\vec{n} \in \mathbb Z^{d-3}$, satisfy 
\bea\label{NHEG-algebra_i}
[ L_{\vec{m}}, L_{\vec{n}} ] = \vec{k} \cdot (\vec{m}- \vec{n}) L_{\vec{m}+\vec{n}} +  \frac{S}{2\pi}(\vec{k}\cdot \vec{m})^3 \delta_{\vec{m}+\vec{n},0}\,.
\eea
The NHEG algebra generators commute with the isometries leading to the ``full NHEG symmetry algebra''
\bea\label{FULL-NHEG-algebra}
\text{Full NHEG Symmetry Algebra}= SL(2,\mathbb R) \times U(1)^{d-3} \times \widehat{\mathcal{V}_{\vec{k},S}}.
\eea
\item We give an explicit construction of the charges over the phase space from a one-dimensional ``Liouville stress-tensor'' for a fundamental boson field ${\Psi}$, which is constructed from the wiggle function $F(\vec{\varphi})$. 

\end{enumerate}
It is instructive to make a few short comments here: 
\begin{itemize}
\item As it is seen, the algebra $\widehat{\mathcal{V}_{\vec{k},S}}$ is the familiar Virasoro algebra in four dimensions while in higher dimensions $\widehat{\mathcal{V}_{\vec{k},S}}$ is a generalization of Virasoro algebra, which to our knowledge has not appeared before in the literature of physics or mathematics. Although the ``higher rank Virasoro algebras'' have appeared in the mathematics literature  \cite{Patera91,Mazorchuk98,2006math......7614G}, none of them explicitly depend upon a vector $\vec{k} \in \mathbb R^{d-3}$. 
\item As is made explicit in \eqref{FULL-NHEG-algebra}, the symmetry algebra $\widehat{\mathcal{V}_{\vec{k},S}}$, even in four dimensions, is not an extension of the $U(1)$ symmetries of the
background. Explicitly, $L_{\vec{0}}$ is not the angular momenta $\vec{J}$, or a linear combination thereof.

\item The black hole entropy $S$ appears a central term, consistently with the entropy law $\frac{\hbar}{2\pi}  S = \vec{k} \cdot \vec{J}$ and the fact that the angular momenta commute with the Virasoro generators $[\vec{J},L_{\vec{n}}] = 0$.
\end{itemize}

\subsection{Outline}

Section \ref{sec-NHEG-review} is meant to provide the minimum needed information about the NHEG background and to fix the notations and conventions. In particular, we review some basic facts about the family of near-horizon geometries: their isometries, causal structure, and the laws of NHEG mechanics. 

In section \ref{sec-NHEG-phase-space} we discuss how we construct the family of geometries which will be promoted as the elements of the NHEG phase space. These geometries are built through a specific one-function family of diffeomorphisms. We first fix the form of infinitesimal coordinate transformations, generators of the phase space, by providing physically motivated requirements. Then, we work out the finite coordinate transformations through the ``exponentiation'' procedure that we explain.

A phase space is a configuration space equipped with a symplectic structure. We specify the symplectic structure on the set of geometries that we built in section \ref{sec-symplectic structure}. We first briefly review the covariant phase space method and then construct a conserved, consistent symplectic structure for our problem and discuss how the surface ``symplectic charges'' can be read from the symplectic structure. 
In the appendix \ref{appendix:charges} we give a more detailed  discussion on the general construction of the symplectic structure and its consistency relations, and how to compute the surface charges,  their algebra and central extension.

In section \ref{sec-charges-and-algebra} we apply the construction given in section \ref{sec-symplectic structure} and appendix \ref{appendix:charges} to the specific NHEG phase space, compute the charges, their algebra and the central element. Moreover, we give an explicit representation of the charges over the phase space in terms of the single periodic wiggle function $F(\vec{\varphi})$ specifying the geometries in the phase space. We also discuss the semi-classically quantized NHEG algebra.

In the last section \ref{sec-discussion}, we further discuss the results and the physical implications of the NHEG phase space and algebra and  discuss various ways in which our construction can be extended.

In  appendix \ref{Appendix-proof-details} we have gathered some technical details of the computations and the proofs. 
In appendix \ref{append-Kerr-CFT} we discuss the  alternative possible diffeomorphism in our motivated class which leads to a consistent phase space. For this case, similarly to the Kerr/CFT proposal, the symplectic symmetry is just a Virasoro algebra. The form of our generators is slightly different than the one in the original Kerr/CFT \cite{Guica:2008mu}, allowing us to construct a phase space consisting of smooth geometries specified by a single function of one periodic coordinate. Due to this similarity, we call this phase space ``the Kerr/CFT phase space''.

\section{Quick Review on NHEG}\label{sec-NHEG-review}

The near horizon extremal geometries (NHEG) are generic classes of geometries with at least $SL(2,\mathbb{R}) \times U(1)$ isometry. These geometries, as the name suggests, may appear in the near horizon limit of extremal black holes, while they may also be viewed as independent classes of geometries. Here we will mainly adopt the latter viewpoint. In this work, for concreteness and technical simplicity, we will focus on a special class of the NHEG which are Einstein vacuum solutions in generic $d$ dimensions with $SL(2,\mathbb{R})\times U(1)^{d-3}$ isometry. The general metric for this class of NHEG is 
\begin{align}\label{NHEG-metric}
	{ds}^2&=\Gamma(\theta)\left[-r^2dt^2+\frac{dr^2}{r^2}+d\theta^2+\sum_{i,j=1}^{d-3}\gamma_{ij}(\theta)(d\varphi^i+k^irdt)(d\varphi^j+k^jrdt)\right]
\end{align}
where
\begin{align}\label{ranges}
t\in (-\infty,+\infty),\qquad r\in \{r<0 \} \text{ or } \{r>0\},\qquad \theta\in[0,\theta_{Max} ], \qquad \varphi^i\sim \varphi^i+2\pi,
\end{align}
and $k^i$ are given constants. { We fix the orientation to be $\eps_{tr\theta \varphi^1 \dots \varphi^{d-3}}=+1$.} 
The geometry is a warped fibred product over an AdS$_2$ factor, spanned by $t,r$, with a Euclidean smooth and compact codimension two surface $\mathcal{H}$, covered by $\theta,\varphi^i$; i.e. $\mathcal{H}$ are constant $t,r$ surfaces. Notably, due to the $SL(2,\mathbb{R})$ isometry of the background, constant $t=t_{\mathcal H},r=r_{\mathcal H}$ surfaces for any value of $t_{\mathcal H}, r_{\mathcal H}$, all give isometric surfaces $\mathcal{H}$.

\begin{figure}[htb]	
	\captionsetup{width=.8\textwidth}
	\centering
		\centering
		\begin{tikzpicture}[scale=0.8]
		\path
		( -2,0)  coordinate (I)
		(2,-4) coordinate (II)
		(2, 4) coordinate (III)
		(-2,8)  coordinate (IV)
		(2,8) coordinate (V)
		(-2,-4) coordinate (VI)
		;
		\draw (IV) --node[midway, below, sloped] {$r=-\infty$} (I);
		\draw (V) -- (III);
		\draw [ultra thick, draw=black, fill=gray!20] (I) -- node[midway, above, sloped] {$r=0$} (II) -- node[midway, above, sloped] {$r=\infty$} (III) --  node[midway, below, sloped] {$r=0$} (I);
		\draw (III) -- (IV);
		\draw (I) -- (VI);
		\node at (.5,0) {${\mathbf{I}}$};
		\node at (-0.5,4) {${\mathbf{II}}$};
		\draw[>=stealth,->] (-2,5.1)--(-2,5);
		\draw[>=stealth,->] (-2,1.1)--(-2,1);
		\draw[>=stealth,->] (-2,-2.9)--(-2,-3);
		\draw[>=stealth,->] (2,4.9)--(2,5);
		\draw[>=stealth,->,ultra thick] (2,0.9)--(2,1);
		\draw[>=stealth,->, ultra thick] (2,-3.1)--(2,-3);		
		\end{tikzpicture}
	\caption{Penrose diagram for NHEG, suppressing the $\theta,\varphi^i$ directions. The positive and negative $r$ values of the coordinates used in \eqref{NHEG-metric} respectively cover ${\mathbf{I}}$ and ${\mathbf{II}}$ regions in the above figure. The two boundaries are mapped onto each other by an $r$--$\vec{\varphi}$ inversion symmetry \eqref{r-phi-inversion}. {The arrows on the boundaries shows the flow of time $t$. Note also that flow of time  is reversed between regions I and II.}}	\label{fig:penrose diagram}
\end{figure}
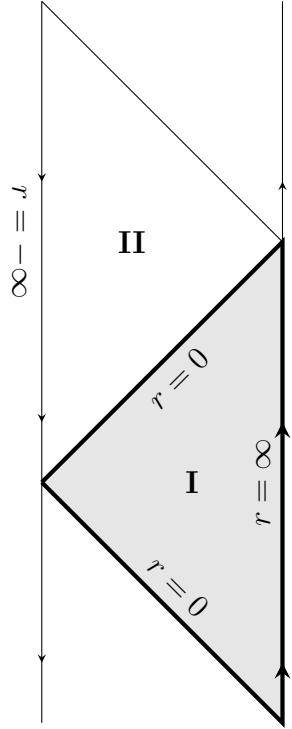

The first two terms of the above metric form an AdS$_2$ in the Poincar\'{e} patch; $r=0$ is the Poincar\'{e} horizon. The metric however extends beyond the horizon. 
The AdS$_2$ metric has two disjoint boundaries. Covering the global coordinate system with families of Poincar\'e patches, one can assign these boundaries at $r=\pm\infty$, as has been depicted in Fig. \ref{fig:penrose diagram} (see also \cite{Aharony:1999ti}). The range of the $\theta$ coordinate is fixed requiring that $\mathcal{H}$ is a smooth and compact manifold. Note that $\mathcal{H}$ can take various topologies \cite{Hollands:2009ng}. Requiring the geometry to be smooth and Lorentzian implies  $\Gamma(\theta)>0$ and  the eigenvalues of $\gamma_{ij}$ to be real and nonnegative. Moreover, smoothness and absence of conical singularity of $\mathcal{H}$  implies that: (1) At most one of the eigenvalues of $\gamma_{ij}(\theta)$ matrix can be vanishing around a given $\theta=\theta_0$ coordinate; (2) if at $\theta_0$ we have a vanishing eigenvalue, it should behaves as $(\theta-\theta_0)^2+\mathcal{O}(\theta-\theta_0)^3$. { Note that the coefficient of $(\theta-\theta_0)^2$ should be exactly one.}

The geometry is completely determined by the functions $\Gamma(\theta), \gamma_{ij}(\theta)$ and the $d-3$ constants $k^i$ which are determined through the Einstein field equations. There are many constraints and the number of independent parameters in any dimension is not easily determined. After detailed analysis, it was found in \cite{Hollands:2009ng} following \cite{Kunduri:2008rs} that there are $(d-2)(d-3)/2$ independent continuous parameters and two discrete parameters that specify a given NHEG. The discrete parameters specify the topology which can be either $S^2 \times T^{d-4}$, or $S^3 \times T^{d-5}$, or quotients thereof, $L(p,q) \times T^{d-5}$ where $L(p,q)$ is a Lens space. In four dimensions, there is only one continuous parameter which is the entropy or angular momentum (remember that $k=1$ in that case). In five dimensions, there are three possible topologies $S^2 \times S^1$, $S^3$ and $L(p,q)$ and three continuous parameters. 

\paragraph{NHEG isometries.} The NHEG background \eqref{NHEG-metric} enjoys $SL(2,\mathbb{R})\times U(1)^{d-3}$ isometry. 
The  $SL(2,\mathbb{R})$ isometries generated by Killing vectors $\xi_a$ with $ a\in \{-,0,+\}$, 
\be\label{xi1-xi2}
\begin{split}
\xi_- &=\partial_t\,,\qquad \xi_0=t\partial_t-r\partial_r,\qquad	\xi_+ =\dfrac{1}{2}(t^2+\frac{1}{r^2})\partial_t-tr\partial_r-\frac{1}{r}{k}^i{\p}_{\varphi^i},
\end{split}
\ee
and the  $U(1)^{d-3}\;$ isometries  by Killing vectors $\mathrm{m}_i$ with $i\in \{1,\cdots, d-3 \}$, 
\begin{align}\label{U(1)-generators}
\mathrm{m}_i=\p_{\varphi^i}.
\end{align}
The isometry algebra is then
\begin{align}\label{commutation relation}
[\xi_0,\xi_-]=-\xi_-,\qquad [\xi_0,\xi_+]=\xi_+, \qquad [\xi_-,\xi_+]=\xi_0\,,\qquad [\xi_a,\mathrm{m}_i]=0.
\end{align}
That is, if we view $\xi_0$ as the scaling operator, $\xi_-,\xi_+$ are respectively lowering and raising operators in $SL(2,\mathbb{R})$. We also note that $\xi_-,\xi_0$ form a two dimensional subalgebra of $SL(2,\mathbb{R})$. For further use we define the structure constants $f_{ab}^{\;\;\; c}$ from $[\xi_a,\xi_b] = f_{ab}^{\;\; \;c}\xi_c$. 

\textbf{Notations:} \emph{Hereafter, we will denote all the $d-3$ indices by vector sign; e.g. $k^i$ will be denoted by $\vec{k}$, $\varphi^i$ by $\vec{\varphi}$, ${\p}_{\varphi^i}$ by $\vec{\partial}_{\varphi}$  and when there is a summation over $i$-indices it will be denoted by dot-product; e.g. ${k}^i{\p}_{\varphi^i}=\vec{k}\cdot\vec{\partial}_{\varphi}=\vec{k}\cdot\vec{\mathrm{m}}$.}

The NHEG also enjoys various $Z_2$ { isometries}. The two which will be relevant for our later analysis are $r$--$\vec{\varphi}$ and $t$--$\vec{\varphi}$-inversions. The $t$--$\vec{\varphi}$-inversion, 
\be\label{t-phi-inversion}
(t,\varphi^i) \ \ \to\ \ (-t,-\varphi^i).
\ee
is reminiscent of similar symmetry in the (extremal) black hole (see \cite{Schiffrin:2015yua} for a recent discussion) whose near horizon limit leads to the NHEG. One may readily check that under the above $Z_2$, 
$\xi_0$ do not change while $\xi_-, \xi_+, \vec{\mathrm{m}}$ change sign.
Another $Z_2$ isometry is the $r$--$\vec{\varphi}$-inversion,
\be\label{r-phi-inversion}
(r,\varphi^i) \ \ \to\ \ (-r,-\varphi^i).
\ee
This $Z_2$ exchanges the two boundaries of AdS$_2$ (\emph{cf.} Fig. \ref{fig:penrose diagram}). Under the $r$--$\vec{\varphi}$-inversion \eqref{r-phi-inversion}, the $SL(2,\mathbb{R})$ Killing vectors \eqref{xi1-xi2} remain invariant.

The space-time inversion PT provides yet another $Z_2$ isometry.
 
\paragraph{NHEG examples in $4d$ and $5d$.} As some examples of NHEG, let us consider the near horizon geometry of extremal Kerr black hole (NHEK) in four dimensions \cite{Bardeen:1999px} and extremal Myers-Perry black hole in five dimensions \cite{Kunduri:2007vf,Kunduri:2008rs}. For NHEK we have
\begin{align}\label{NHEK-solution}
\Gamma = J\frac{1+\cos^2\theta}{2},\qquad \gamma_{11}=\left(\dfrac{2\sin\theta}{1+\cos^2\theta}\right)^2,\qquad k=1,
\end{align}
where $J$ is a constant equal to the angular momentum of the corresponding black hole. The range of polar coordinate is $\theta\in [0,\pi]$. Near the roots of $\gamma_{11}$ which occur at $\theta=0,\pi$, it clearly satisfies the smoothness condition and the compact surface $\mathcal{H}$, whose area is $4\pi J$, is topologically a two-sphere. 

For the $5d$ doubly spinning extremal Myers-Perry near-horizon geometry we have
\begin{equation}
\begin{split}
\Gamma &= \frac{1}{4} (a+b) (a\cos^2\frac{\theta}{2}+b\sin^2\frac{\theta}{2}),\qquad k^1=\frac{1}{2}\sqrt{\frac{b}{a}},\qquad k^2=\frac{1}{2}\sqrt{\frac{a}{b}},\\\\
\gamma_{ij}&=\dfrac{4}{(a\cos^2\frac{\theta}{2}+b\sin^2\frac{\theta}{2})^2}
\begin{pmatrix}
 a (a+b\sin^2\frac{\theta}{2}) \sin ^2\frac{\theta}{2} &  a b \cos ^2\frac{\theta}{2} \sin ^2\frac{\theta}{2} \\ \ \ & \ \ \\
 a b \cos ^2\frac{\theta}{2} \sin ^2\frac{\theta}{2} &  b \cos ^2\frac{\theta}{2} (b+a\cos^2\frac{\theta}{2}) \\
\end{pmatrix},
\end{split}
\end{equation}
where $a>0,b>0$ are constants related to the angular momenta, and $\theta\in [0,\pi]$. Note that $k^1k^2=\frac14$ and hence $k^1$ and $k^2$ are not independent. One can compute the eigenvalues $\lambda_{1,2}(\theta)$ of the matrix $\gamma_{ij}$. Then we observe that one of the eigenvalues is always positive, while the other eigenvalue (say $\lambda_2$) vanishes at $\theta=0,\pi$. Near these poles we find
\begin{align}
\lambda_2=\theta^2+\mathcal{O}(\theta^3),\qquad \lambda_2=(\pi-\theta)^2+\mathcal{O}((\pi-\theta)^3)
\end{align}
satisfying the regularity condition. The $3d$ surface $\mathcal{H}$ is hence topologically $S^3$ and it is area is $2\pi^2\cdot \sqrt{ab}(a+b)^2$.

\subsection{Killing horizons}\label{sec-Killing-horizon}

The Petrov classification has been extended to higher dimensions \cite{Coley:2004jv}. NHEG is a Petrov type D spacetime \cite{Godazgar:2009fi}. It has two real principal null directions which turn out to be congruences of torsion,  expansion and twist free geodesics \cite{Durkee:2010ea}. They are generated by
\begin{align}\label{ells}
\begin{split}
\ell_+&=\left(\dfrac{1}{r}\partial_t +r\partial_r-\vec{k}\cdot\vec{\partial}_{\varphi}\right)\,, \\
\ell_-&=\left(\dfrac{1}{r}\partial_t -r\partial_r-\vec{k}\cdot\vec{\partial}_{\varphi}\right)\,.
\end{split}
\end{align}
These vector fields are respectively normal to the hypersurfaces, 
\begin{equation}
\begin{split}
\mathcal{N}_+ \; : \quad v\equiv t+\dfrac{1}{r}= const\equiv t_{\mathcal H}+\dfrac{1}{r_{\mathcal H}}= v_{\mathcal H}\, ,\\
\mathcal{N}_- \; :\quad  u\equiv t-\dfrac{1}{r}= const \equiv t_{\mathcal H}-\dfrac{1}{r_{\mathcal H}}= u_{\mathcal H}\,.
\end{split}
\end{equation}
One may readily see that $\ell_+\cdot dv=\ell_-\cdot du=0$ and that $\mathcal{N}_\pm$ are therefore null hypersurfaces. Intersection of these two hypersurfaces is a $d-2$ dimensional compact surface $\mathcal{H}$, identified by $t=t_{\mathcal H}, r=r_{\mathcal H}$. Note that both $\ell_\pm$ are  normal to $\mathcal{H}$ and its binormal tensor is 
\begin{align}\label{binormal} 
\boldsymbol{\eps}_\perp=\Gamma dt \wedge dr =\frac{\Gamma}{2}  r^2d v \wedge du,
\end{align}
normalized such that $\eps^\perp_\mn \eps_\perp^\mn=-2$. We note that under the $t$--$\vec{\varphi}$-inversion or $r$--$\vec{\varphi}$-inversion symmetries \eqref{t-phi-inversion}-\eqref{r-phi-inversion}, $\ell_\pm\leftrightarrow -\ell_{\mp}$.

The surface $\mathcal{H}$ is similar to the bifurcation surface of a Killing horizon in black hole geometries, in the sense that it has two normal null vectors. In what follows we make this statement precise and prove the existence of bifurcate Killing horizon at each point $t_{\mathcal H}, r_{\mathcal H}$ \cite{Hajian:2013lna,Hajian:2014twa}.  (Similar arguments can be found in \cite{Bengtsson:2005zj} for warped $AdS_3$ geometries.) 

\paragraph{Killing Horizon Generator.} By definition, $\mathcal{N}=\{\mathcal{N}_+ \cup \mathcal{N}_-\}$ is the Killing horizon of the Killing vector field $\zeta$, provided that the vector $\zeta$ is normal to $\mathcal{N}$. Let us now consider the Killing vector $\zeta_{\mathcal H}$ \cite{Hajian:2014twa}
\begin{align}\label{zeta-H}
\zeta_{\mathcal H}&
=n_{\mathcal H}^a\xi_a - \vec{k}\cdot\vec{\mathrm{m}},
\end{align}
where $n_{\mathcal H}^a$ are given by the following functions computed at the constant value $t=t_{\mathcal H},\,r=r_{\mathcal H}$ 
\begin{align}\label{n-a-r-t}
n^-=-\frac{t^2r^2-1}{2r}\,, \qquad n^0=t\,r\,,\qquad n^+=-r.
\end{align}
It can be shown that these functions form the coadjoint representation of $SL(2,\mathbb{R})$ as follows.	
The space of functions of $t,r$ forms a vector space in $\mathbb R$. The $SL(2,\mathbb{R})$ action is defined by  $\xi_a f(t,r)=\xi_a^\mu \p_\mu f(t,r)$. Now consider the subspace spanned by the three functions $n_a$ (with lower indices) defined as
\begin{align}
n_+= \frac{t^2r^2-1}{2r}, \quad n_0= t\,r ,\quad n_-= r.
\end{align}
One can check that the action of $SL(2,\mathbb R)$ vectors $\xi_a$ on the functions $n_b$ is given by a matrix whose components are the $SL(2,\mathbb R)$ structure constants, 
\begin{align}
\xi_a n_b&= f_{ab}^{\;\;\; \;c} \;n_c.
\end{align}
Therefore, the subspace spanned by $\{n_+,n_0,n_-\}$ forms the adjoint representation space of the $SL(2,\mathbb R)$ algebra. The functions $n^a$ are then defined as $n^a = K^{ab}n_b$, using the Killing form of $SL(2,\mathbb{R})$ in $(-,0,+)$ basis
\begin{align}
K_{ab}=K^{ab}=\begin{pmatrix}
0&0&-1\\
0&1&0\\
-1&0&0
\end{pmatrix}.
\end{align}
Accordingly the functions $n^a$ form the coadjoint representation. Since the Killing vectors $\xi_a$ \eqref{xi1-xi2} also form an adjoint representation of $SL(2,\mathbb{R})$, one can consider the direct product $n_a \otimes \xi_b$ which can be decomposed into $\textbf{3}\otimes \textbf{3}= \textbf{5} \oplus \textbf{3}\oplus \textbf{1}$.  The singlet $\textbf{1}$  is given by the vector $n^a\, \xi_a^{\;\mu} = K^{ab}n_b \xi_a^{\;\mu}$. This is obviously a singlet representation, since it is constructed by the contraction of the Killing form with two vectors. Indeed it can be shown that $n^a\, \xi_a=\vec{k}\cdot\vec{\mathrm{m}}$ and therefore the Killing vector $\zeta_{\cal H}$ vanishes on the surface $\cal H$.

The three vector $n^a$ can also be interpreted as the position vector of an $AdS_2$ surface
embedded in a three dimensional flat space $\mathbb{R}^{2,1}$ with the metric given by $-K_{ab}$. Explicitly
\begin{align}
n^2\equiv -K_{ab}n^a n^b=2n^+n^--(n^0)^2=-1.
\end{align}
The vector $n^a_{\cal H}$ is a specific point on this surface, but any other point can be obtained by an $SL(2,\mathbb{R})$ group action on this vector.\\

Returning back to \eqref{zeta-H}, one can check that
\begin{align}
\zeta_{\mathcal H}\big{|}_{\mathcal{N}_\pm}=\frac{r-r_{\mathcal H}}{r} \ell_\pm.
\end{align}
Note also that $\zeta_{\mathcal H}$ vanishes at the bifurcation surface $\mathcal{H}$. Therefore, $\mathcal{N}$ is the ``Killing horizon" of $\zeta_{\mathcal H}$, and $\mathcal{H}$ is its bifurcation surface. The choice of $t_{\mathcal H},r_{\mathcal H}$ is arbitrary in the above argument, so there are \textit{infinitely many} Killing horizons, bifurcating at any compact surface determined by $t_{\mathcal H},r_{\mathcal H}$.

It is important to note that although the extremal black hole does not possess any bifurcate Killing horizon, the corresponding near horizon geometry has an infinite number of them. The reason why one can find bifurcate Killing horizons in NHEG but not in extremal black hole geometry traces back to the enhancement of symmetries in the near horizon geometry. We explicitly used this fact in construction of the vector $\zeta_{\mathcal H}$.

Another important feature about the vector $\zeta_{\mathcal H}$ is that on $\mathcal{H}$, 
\bea\label{propH}
\nabla_{[\mu}{\zeta_{\mathcal H}}_{\nu]}=\eps^\perp_{\mu\nu} 
\eea
where $\boldsymbol\eps_\perp$ is the binormal tensor \eqref{binormal}. We can use this fact to compute the \textit{surface gravity} on the bifurcation surface of the Killing horizon:
\begin{align}\label{kappa-NHEG}
\kappa^2&=-\dfrac{1}{2}|\nabla{\zeta_{\mathcal H}}|^2=1.
\end{align}
The above gives the value of $\kappa^2$. As in the usual black hole cases, $\zeta_{\mathcal H}$ is the generator of a bifurcate Killing horizon with future and past oriented branches.\footnote{In the black hole terminology, the future (past) oriented branch of horizons corresponds to the black (white) hole. However here there is no event horizon.} One can then show that the value of $\kappa$ is $+1$ for the future oriented branch and $-1$ for the past oriented branch.
As  a consequence of $SL(2,\mathbb{R})$ invariance the surface gravity is a constant and independent of $t_{\mathcal H}$ and $r_{\mathcal H}$. As in the Rindler space, one can associate an Unruh-type temperature \cite{Unruh:1976db} to the Killing horizons. This temperature is simply $\frac{\hbar}{2\pi}$ and constant over $\mathcal{H}$.

\paragraph{Gaussian null coordinates.} 
Another coordinate system of interest is the Gaussian null coordinate system (GNC) (also called ingoing Eddington-Finkelstein coordinates)\footnote{In our construction of the phase space we mainly use  Poincar\'e coordinates. However, we will make some remarks about the usage of other coordinate systems as a starting point for constructing the phase space in the discussion section.}. This coordinates are obtained by the following transformations
\begin{align}
v=t+\dfrac{1}{r},\qquad \phi^i=\varphi^i+k^i \ln r,\qquad r\rightarrow r,\qquad \theta\rightarrow \theta,
\end{align}
Therefore the metric takes the form
\begin{align}\label{NHEG-metric-EF}
{ds}^2=\Gamma(\theta)\left[-r^2dv^2-2\,dr\, dv+d\theta^2+\sum_{i,j=1}^{d-3}\gamma_{ij}(\theta)(d\phi^i+k^irdv)(d\phi^j+k^jrdv)\right]
\end{align}
with $v\in (-\infty,\infty)$ and the range of the other coordinates is the same as \eqref{ranges}. In the same way one can express the metric in outgoing Eddington-Finkelstein coordinates by replacing $t$ with $u=t-1/r$.
In GNC coordinates, the Killing vectors are { redefined as}
\be
\begin{split}
	\xi_- &=\partial_v\,,\qquad \xi_0=v\partial_v-r\partial_r,\qquad 	\xi_+ =\frac{v^2}{2}\partial_v-(vr+1)\partial_r+v\vec{k}\cdot \vec{\p}_\phi,\qquad {\mathrm m}_i=\p_{\phi^i}\,.
\end{split}
\ee
{Note that the above  are related to the Killing vectors \eqref{xi1-xi2}-\eqref{U(1)-generators} by an automorphism of the algebra of isometries.} As a check, the commutation relations \eqref{commutation relation} still hold. 

\begin{figure}[!h]	
	\captionsetup{width=.8\textwidth}
	\centering
\includegraphics[scale=0.36, angle=45]{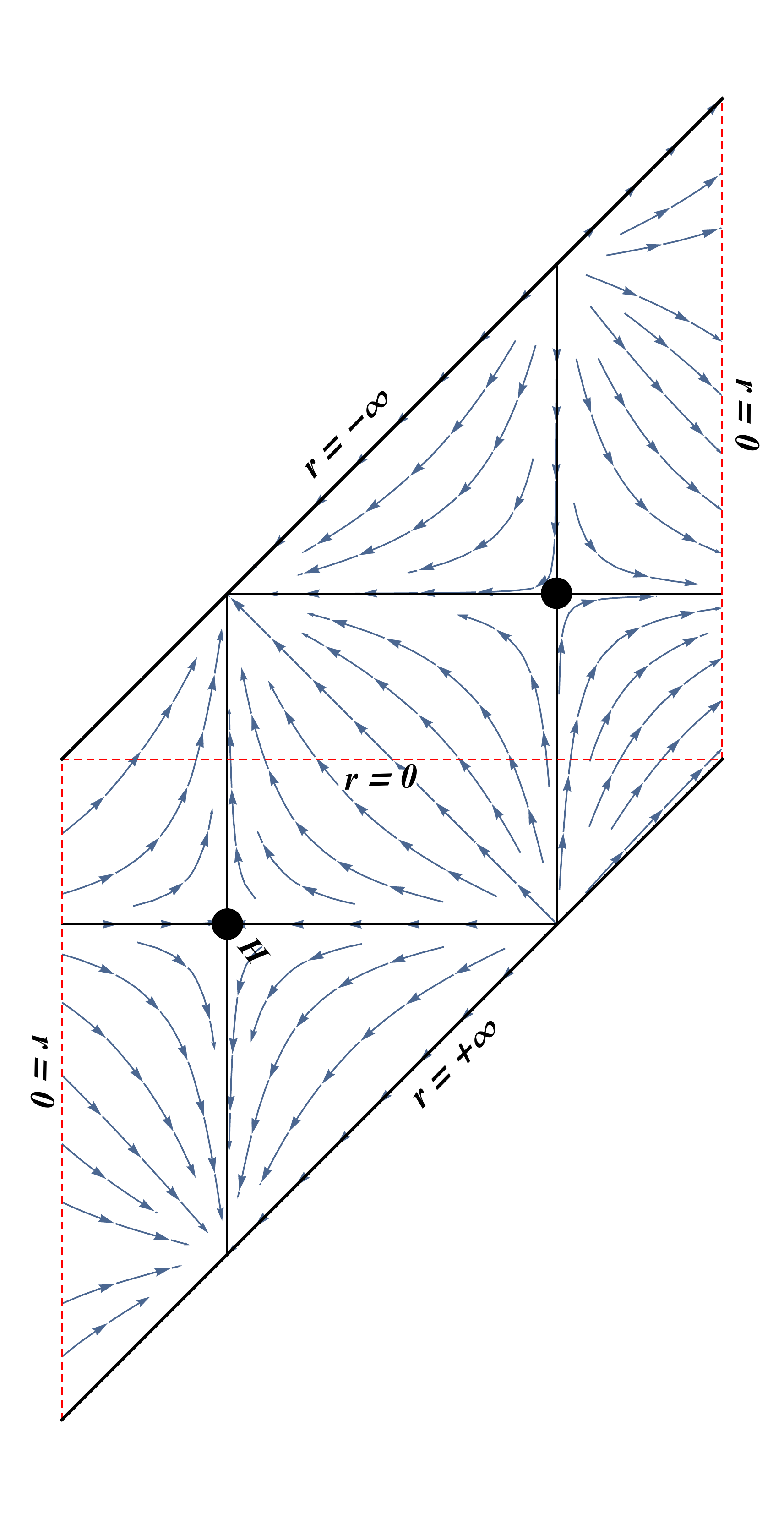}
	\caption{Flow of the Killing vector $\zeta_{\mathcal H}$. The two black dots denote the codimension two bifurcation surfaces and the $45^\circ$ lines intersecting at  them are the Killing horizons $\mathcal{N}$. Under  $r$--$\vec{\varphi}$ inversion \eqref{r-phi-inversion} the upper and lower triangles separated by $r=0$ line are mapped to each other. The Killing vector is mapped as $\zeta_{\mathcal H}\to -\zeta_{\mathcal H}$ under $r$--$\vec{\varphi}$ inversion.}\label{fig:Flow}
\end{figure}

\paragraph{Kruskal-type coordinates and causal structure.} To gain a better intuition about the Killing horizons of the NHEG it is useful to draw the flow of $\zeta_{\mathcal H}$ over the spacetime in a Kruskal-type coordinate $(u,v,\theta,\varphi^i)$. To this end, 
we note that for the $r\geq 0$ ($r\leq 0$) region, $v\geq u$ ($v\leq u$). Also, $v=u$ represents the asymptotic (large $r$ region) of spacetime. Also, $u=const, v=const$ represent null hypersurfaces. In these coordinates the Killing vectors are
\begin{align}
\xi_-&=\partial_u+\partial_v\,,\qquad \xi_0=u\partial_u+v\partial_v\,,\qquad \xi_+=\dfrac{1}{2}\left(u^2\partial_u+v^2\partial_v\right)+\frac12(u-v)\vec{k}\cdot\vec{\p}_{\varphi}
\end{align}
and
\begin{align}
n^-&=\dfrac{uv}{u-v}\,,\qquad n^0=-\dfrac{u+v}{u-v}\,,\qquad n^+=\dfrac{2}{u-v}\,,
\end{align}
therefore
\begin{align}
\zeta_{\mathcal H}&=\dfrac{1}{u_{\mathcal H}-v_{\mathcal H}}\bigg[(u-u_{\mathcal H})(u-v_{\mathcal H})\partial_u +(v-v_{\mathcal H})(v-u_{\mathcal H})\partial_v+\big((u-u_{\mathcal H})-(v-v_{\mathcal H})\big)\vec{k}\cdot\vec{\p}_{\varphi}\bigg].
\end{align}
It is clearly seen that this vector vanishes at $u=u_{\mathcal H},v=v_{\mathcal H}$. The flow of $\zeta_{\mathcal H}$ 
is depicted in Fig. \ref{fig:Flow}.

\subsection{NHEG entropy}

Given a generic diffeomorphic invariant Lagrangian, the entropy, which is the conserved Noether-Wald charge for a bifurcate Killing horizon,  is defined as \cite{Wald:1993nt,Iyer:1994ys} 
\begin{equation}
\begin{split}\label{eq Wald entropy}
\frac{S}{2 \pi}  &=-\frac{1}{\hbar}\oint_{\mathcal H} \boldsymbol{\eps}_\mathcal{H} \frac{\delta \mathcal{L}}{\delta R_{\mu \nu \alpha \beta}}\eps^\perp_{\mu \nu}\eps^\perp_{\alpha \beta},
\end{split}
\end{equation}
where $\boldsymbol{\eps}_\mathcal{H} $ is the volume form on $\mathcal H$ and $\eps^\perp_{\mu \nu}$ the binormal normalized as $\eps^\perp_{\mu \nu} \eps_\perp^{\mu \nu} = -2$. In Einstein theory, this definition reduces to the familiar Bekenstein entropy (the area law).  For extremal black holes, there is no bifurcation surface and the derivation of entropy as a Noether charge breaks down. However, physically one would expect that the entropy should be a continuous function for near-extremal black holes and hence the entropy for extremal black holes may be  obtained from a limiting procedure starting from near-extremal black holes.

Now for extremal black holes, the near horizon geometry possesses infinitely many bifurcate Killing horizons with Killing generator \eqref{zeta-H}. The fact that on the bifurcation surface $\mathcal{H}$ the Killing vector $\zeta_{\mathcal H}$ vanishes, and that $\nabla_{[\mu}{\zeta_{\mathcal H}}_{\nu]}=\eps^\perp_{\mu\nu} $ allows one to prove that the entropy is given by the Noether charge associated with $\zeta_{\mathcal H}$ and coincides with \eqref{eq Wald entropy} where $\boldsymbol{\eps}_\mathcal{H}=\Gamma^{\frac{d-2}{2}} \sqrt{\gamma}\,d\theta\, d\vec{\varphi}$ is the volume form of any surface $\mathcal{H}$  \cite{Hajian:2013lna}. This last result completes Wald's program for defining the entropy as a Noether charge in the case of extremal black holes by using the additional $SL(2,\mathbb R)$ symmetry in the near-horizon region.

\subsection{Laws of NHEG mechanics}
For a general theory of pure gravity determined by a diffeomorphism invariant Lagrangian $\mathcal{L}$, and admitting a solution of the form \eqref{NHEG-metric}, one can prove the following ``Laws of NHEG Mechanics'' \cite{Hajian:2013lna} (see also \cite{Astefanesei:2006dd}).
\paragraph{Zeroth Law:}
$k^i$ should necessarily be constant as a result of $SL(2,\mathbb{R})$ invariance of the background. Moreover, the surface gravity is constant over any  $\mathcal{H}$-surface (\emph{cf.} \eqref{kappa-NHEG} and discussions below it).
\paragraph{The Entropy Law:}
\begin{equation}\label{entropy law}
\frac{\hbar}{2\pi} S =\vec{k}\cdot\vec{J}-\oint_{\mathcal H} \sqrt{-g}\mathcal{L},
\end{equation}
where the angular momentum $J_i$ is the conserved charge corresponding to the $\mathrm{m}_i$ isometry. For NHEG's which are Einstein vacuum solutions, like the class we have focused on here, $\sqrt{-g}\mathcal{L}=0$ on-shell and hence the entropy law reduces to $\frac{\hbar}{2\pi}S=\vec{k}\cdot\vec{J}$.

\paragraph{Entropy Perturbation Law.} Consider a generic perturbation $\delta\Phi$ over NHEG solution satisfying the linearized field equations. One can associate charge perturbations $\delta{\vec{J}}$ and $\delta S$ to these perturbations.    Assuming that perturbations are invariant under $\xi_-,\xi_0$ Killing vectors \eqref{xi1-xi2}, $[\xi_-,\delta\Phi]=[\xi_0,\delta\Phi]=0$, one can prove the following relation \cite{Hajian:2013lna,Hajian:2014twa} \begin{align}\label{EPL}
\frac{\hbar}{2\pi} \delta S=\vec{k}\cdot\delta \vec{J}.
\end{align}

We make the intriguing comment that the factor of $\frac{\hbar}{2\pi}$ in  
\eqref{entropy law} and  \eqref{EPL} could be attributed to the ``Unruh-type'' temperature of the NHEG background (\emph{cf.} discussions below equation \eqref{kappa-NHEG}).

\section{NHEG Phase Space}\label{sec-NHEG-phase-space}
 
In this section we present the construction of the phase space geometries. 
We find it useful to the reader to first start with a qualitative presentation before deriving the details.

{
	
\subsection{Overview on the NHEG phase space}
	A phase space is a configuration space of fields equipped with a 
		 finite and conserved symplectic structure.} Due to the absence of finite energy propagating degrees of freedom in the NHEG background, we propose to build the NHEG phase space using the set of geometries obtained by specific coordinate transformations of the background \eqref{NHEG-metric}. It should be emphasized that in a diffeomorphic invariant theory not all coordinate transformations are necessarily pure gauge transformations. In a gauge theory, one can associate surface charges to local gauge transformations. Those with vanishing charge are defined to be pure gauge, while those with well-defined, finite and conserved nonvanishing charges describe physically distinct configurations in the phase space. Other gauge transformations are not allowed. In the following we will define such non-trivial diffeomorphisms associated with conserved surface charges.

In this context, the most common and better known setup is the asymptotic symmetry method. Here, we will rather follow a different approach which we could name the \emph{symplectic symmetry method}. In the asymptotic symmetry method, one defines the phase space through appropriately prescribed asymptotic boundary conditions. Diffeomorphisms which preserve the boundary conditions are said to be allowed. Allowed infinitesimal diffeomorphisms are either nontrivial, if associated with well-defined, finite and conserved charges defining the asymptotic symmetries, or they are trivial (or  equivalently pure gauge), if associated with vanishing charges over the phase space. In the symplectic symmetry method, we instead specify a class of infinitesimal diffeomorphisms everywhere in spacetime and exponentiate them to find finite coordinate transformations upon which we build the phase space. A requirement on the infinitesimal diffeomorphisms is that they are non-trivial; i.e. associated with well-defined conserved charges even though they are not isometries.\footnote{We refer to such vectors $\chi$ as symplectic symmetries since the presymplectic structure $\boldsymbol\omega [\delta\Phi,\mathcal L_\chi \Phi ,\Phi]$ (defined in appendix A.) is zero on-shell everywhere while the surface charges built from the symplectic structure are non-vanishing (see section \ref{secint} for further details).} Symplectic symmetries are therefore extensions into the bulk of asymptotic symmetries defined at infinity; any symplectic symmetry is necessarily also asymptotic but not vice-versa. In the symplectic symmetry method, one never defines the set of pure gauge transformations, which at any rate do not contain physical information. In a sense the phase space built from the symplectic symmetries defines physical perturbations  in a fixed gauge and all the physical information is contained in the symmetries.

While the construction of the family of diffeomorphisms and the associated symplectic structure on the phase space are intertwined,  for the clarity of the presentation, we first present a (mostly self-contained) derivation of the family of diffeomorphisms and the resulting family of geometries while we will discuss the construction of the symplectic structure in section \ref{sec-symplectic structure}. It is however important to keep in mind that to these  diffeomorphisms there should  be associated  finite, conserved, well-defined and non-vanishing surface charges derived from the symplectic structure, as we will discuss in section \ref{sec-charges-and-algebra}. The latter property justifies considering these diffeomorphisms as physically relevant.

	As already mentioned before, we restrict ourselves to solutions of the $d$ dimensional Einstein vacuum theory
	\be\label{EH-action}
	\mathcal{S}=\frac{1}{16\pi G}\oint d^dx \sqrt{-g}R,
	\ee
	with $SL(2,\mathbb{R})\times U(1)^{d-3}$ isometry. These solutions are uniquely identified, \emph{up to coordinate transformations}, by the topology of the $\mathcal H$ surfaces and by $(d-2)(d-3)/2$ continuous parameters collectively denoted as $\{ p \}$ including the angular momenta $J_i$, $i=1,\cdots d-3$ \cite{Kunduri:2008rs,Hollands:2009ng}.

At the infinitesimal level, a coordinate transformation is generated by a vector field  $\chi^\mu$ through $x^\mu\rightarrow x^\mu+ \chi^\mu$. We denote all dynamical fields as $\Phi$. In this paper $\Phi$ is only the metric, but we keep that notation to facilitate possible generalizations with additional fields. An active coordinate transformation generates a perturbation, denoted as $\delta_\chi \Phi$, which is the Lie derivative of the dynamical field $\delta_\chi \Phi=\mathcal{L}_\chi \Phi$. Such a perturbation automatically obeys the linearized field equations as a consequence of general covariance. 

In the following, we will first single out the infinitesimal diffeomorphisms around the background using a set of physical requirements. We use the background in the fixed coordinate system $(t,r,\theta,\varphi^i)$. 
Arbitrary field configurations of the phase space are then produced by finite coordinate transformations, obtained by the \textit{exponentiation} of these infinitesimal coordinate transformations. To this end we require the functional form of the vector field $\chi$ to be preserved along any element of the phase space.  We will finally comment on the isometries of the phase space and on the algebra of infinitesimal diffeomorphisms at the end of the section.

\textbf{Notations.} \emph{For the sake of clarity, we will use the following convention from now on: all quantities associated with the background metric \eqref{NHEG-metric} will be defined with an overline. In particular, the metric \eqref{NHEG-metric} will be denoted as $\bar\Phi  \equiv \bar g_{\mu\nu}$ and infinitesimal diffeomorphisms around the background will be generated by $ \overline \chi^\mu $. Instead, we denote a generic element of the phase space as $\Phi$ and an infinitesimal diffeomorphism tangent to the phase space as $\chi$.
}

\subsection{Generator of infinitesimal transformations}
\label{Construction of chi}

We start with the most general diffeomorphism generator around the background $\overline \chi$ and determine the generator of our infinitesimal transformations through  the six conditions listed below.

\paragraph{ (1) $[\overline \chi,\overline\xi_{0}]=[\overline \chi,\overline\xi_{-}]=0$.} These conditions are supported as follows:
\begin{description}
\item[1.1) $\mathcal H$-independent charges.] Any conserved charge is defined through integrating over a $d-2$ dimensional bifurcation surface $\mathcal{H}$. However, there are infinitely many of such surfaces at any given $t_{\mathcal H},r_{\mathcal H}$.  We require that all such conserved charges be equal. Since two points $t,r$ and $t_{\mathcal H},r_{\mathcal H}$ can be mapped through a diffeomorphism generated by $\overline \xi_-,\overline \xi_0$, we require that these vectors commute with $\overline \chi$. 

\item[1.2) Perturbations $\delta_{\overline \chi}\Phi$ in the $SL(2,\mathbb R)$ { lowest} and zero weight representation.] As mentioned in the introduction, we construct the phase space such that the field perturbations around the background  $\delta_{\overline \chi}\Phi$ have vanishing $SL(2,\mathbb R)$ charges. A sufficient condition for the latter is that $\delta_{\overline \chi}\Phi$ are invariant under $\overline \xi_{0,-}$; i.e. $\mathcal{L}_{\overline \xi_{-,0}}\delta_{\overline \chi}\Phi=0$. It then implies that $\mathcal L_{\overline \xi_{0,-}} \Phi = 0$ on the entire phase space generated by ${\chi}$ and the associated charges will be zero on the entire phase space.\footnote{ The charges associated with $\xi_+$ will then also turn out to be zero, as we will explain around \eqref{Komar-integral}. Note that imposing instead $[\overline\chi,\overline\xi_+]=0$ would imply $\overline\chi^t=\overline\chi^r=0$ which would be unnecessarily too restrictive.} In the appendix \ref{theorem xi12}, we have proved that this condition implies $[\overline\chi, \overline\xi_-]=0,\ [\overline{\chi},\overline\xi_0]=\beta^i {\overline{\mathrm m}}_i$, with constant $\beta_i$, after discarding vectors $\overline\chi$ which are linear combinations of the $SL(2,\mathbb R)$ algebra. We then fix the constants $\beta^i=0$ since exponentiating such generators would lead to logarithmic terms which would be very irregular at the Poincar\'e horizon. These perturbations are therefore { lowest} weight because annihilated by $\overline \xi_-$ and of weight zero because annihilated by $\overline \xi_0$. 

\item[1.3) Finiteness of energy of perturbations.] As argued in \cite{Hajian:2014twa} only perturbations with  $\mathcal{L}_{\overline\xi_{-,0}}\delta_{\overline\chi}\Phi=0$ can be related to finite energy perturbations around the original extremal black hole whose near horizon limit gives the near horizon extremal geometry in question.\footnote{It was shown in \cite{Hajian:2014twa} that the necessary and sufficient condition for the entropy perturbation law (EPL) \eqref{EPL} is $\xi_-,\xi_0$ invariance of the perturbations. Nonetheless, as we will argue, here  we are dealing with perturbations with vanishing entropy and angular momenta variations $\delta J_i=\delta S=0$ and the EPL is trivially satisfied.}
\end{description}
This condition fixes the $t$ and $r$ dependence of all components  of ${\overline \chi}$:
\be\label{chi-xi1,2}
{\overline \chi}=\frac{1}{r}\eps^t\p_t+r\eps^r\p_r+\eps^\theta\p_\theta+\vec{\epsilon}\cdot\vec{\p}_{\varphi},
\ee
where the  $\epsilon$-coefficients are only functions of $\theta,\vec{\varphi}$. Also, it implies that $\xi_- = \overline \xi_-$ and $\xi_0 = \overline \xi_0$ on any element of the phase space. Therefore, $\overline\xi_-$, $\overline\xi_0$ will be Killing isometries of each element of the phase space.

\paragraph{(2) $\nabla_\mu {\overline \chi}^\mu=0$.} We require the volume element $\boldsymbol{\epsilon}$,
\begin{equation}\label{volume-form-d}
\boldsymbol{\epsilon}=\frac{\sqrt{-g}}{d\,!}{\epsilon}_{\mu_1\mu_2\cdots\mu_d}dx^{\mu_1}\wedge dx^{\mu_2}\wedge\cdots \wedge dx^{\mu_d},
\end{equation}
to be the same for all elements in the phase space; i.e.  $\delta_{{\overline \chi}}\boldsymbol{\epsilon}=0$. Since $\boldsymbol{\epsilon}$ is covariant, $\delta_{{\overline \chi}}\boldsymbol{\epsilon}=\mathcal{L}_{\overline \chi} \boldsymbol{\epsilon}$. On the other hand,
\begin{equation}
\mathcal{L}_{\overline \chi} \boldsymbol{\epsilon}={\overline \chi}\cdot d\boldsymbol{\epsilon}+d({\overline \chi}\cdot \boldsymbol{\epsilon})=d({\overline \chi}\cdot \boldsymbol{\epsilon})=\star(\nabla_\mu {\overline \chi}^\mu),
\end{equation}
{where $\star$ is the standard $d$ dimensional Hodge dual.} Therefore, $\mathcal{L}_{\overline \chi} \boldsymbol{\epsilon}=0$ is equivalent to $\nabla_\mu {\overline \chi}^\mu=0$.

\paragraph{(3) $\delta_{\overline \chi}\mathbf{L}=0$,} where $\mathbf{L}=\frac{1}{16\pi G} R\boldsymbol{\epsilon}$ is the Einstein-Hilbert  Lagrangian $d$-form evaluated on the NHEG background \eqref{NHEG-metric} before imposing the equations of motion. (The functional form of $\Gamma(\theta)$ and $\gamma_{ij}(\theta)$ is therefore arbitrary except for the regularity conditions.)  Since $\mathbf{L}$ is a scalar density built from the metric, it is invariant under the background $SL(2,\mathbb{R})\times U(1)^{d-3}$ isometries and only admits $\theta$ dependence.

The above properties (2) and (3) lead to 
\begin{equation}
\eps^\theta=0, \qquad \quad \eps^{r}=-\vec{\partial}_{\varphi}\cdot \vec{\epsilon}\,.
\end{equation}

\paragraph{(4) $ \eps^t = - b \, \vec{\partial}_{\varphi}\cdot \vec{\epsilon},\, b=\pm 1$.}
This condition can be motivated from two different perspectives: 
\begin{description}

\item[4.1) Preservation of a null geodesic congruence.] As discussed in section \ref{sec-Killing-horizon}, the NHEG has two expansion, rotation and shear free null geodesic congruences generated by $\ell_+$ and $\ell_-$ which are respectively normal to constant $v = t+\frac{1}{r}$ and $u=t-\frac{1}{r}$ surfaces \cite{Durkee:2010ea}. We request that either $\mathcal{L}_{\overline \chi} v=0$ or $\mathcal{L}_{\overline \chi} u=0$, yielding the above condition with $b=\pm 1$ for the choice of $\ell_\pm$. It implies that each element in the phase space will admit one of the branches of their bifurcate horizon $\mathcal{N}_+$ or $\mathcal{N}_-$, respectively.

\item[4.2) Regularity of $\mathcal H$ surfaces.] As we will discuss in section \ref{sec finite trans}, this condition ensures that constant $t,r$ surfaces $\mathcal H$ are regular without singularities at poles on each element of the phase space. Fixing instead $b=0$ as done in \cite{Guica:2008mu} will lead to surfaces $\mathcal H$ with singularities.

\end{description}

The two possibilities $b = \pm 1$ are related to each other by either a $t$--$\vec{\varphi}$ or $r$--$\vec{\varphi}$ inversion symmetry of the background (\emph{cf.} discussions of the previous section). The two phase spaces built with either of these choices are mapped to each other by this $Z_2$ symmetry. 
Without loss of generality we choose $b=+1$.

\paragraph{(5) $\vec{\eps}$ are $\theta$-independent and periodic functions of $\varphi^i$.} We impose these conditions as they guarantee (i) smoothness of the $t,r$ constant surfaces ${\cal H}$ of each element of the phase space, as we will show below in section \ref{sec finite trans}, and (ii) constancy of the angular momenta $\vec{J}$ and the volume of ${\cal H}$ over the phase space, as we will also show in section \ref{sec:iso}.

\paragraph{(6)  Finiteness, conservation and regularity of the symplectic structure.} 
These final conditions crucially depend on the definition of the symplectic structure which is presented in section \ref{sec-symplectic structure}. Our analysis reveals that additional conditions are required in order to obtain a well-defined  symplectic structure. After fixing the ambiguities in the boundary terms of the symplectic structure, we found two classes of generators: 
\begin{itemize}
\item[6.1)] $\vec{\eps} \cdot \vec{\p}_{{\varphi}} = \eps(\phi)\p_\phi$ where $\phi$ is a specific $SL(d-3,\mathbb Z)$ choice of circle in the $(d\!-\!3)$-torus spanned by $\vec{\varphi}$ and $\eps(\phi)$ is a periodic function of $\phi$. 
\item[6.2)] $\vec{\eps} = \vec{k} \eps(\varphi^1,\dots \varphi^{d-3})$, where $\eps$ is a function periodic in all its $d-3$ variables.
\end{itemize}
In four dimensions, where $\vec{k}$ has one component and $k=1$, the above two classes are identical. In higher dimensions however, the two classes are distinct and mutually incompatible because the Lie bracket between one generator ${\overline \chi}$ with $\vec{\eps}$ defined from the first class 6.1) with another generator ${\overline \chi}$ with  $\vec{\eps}$ defined from the second class 6.2) does not belong to any of these classes.

The first choice leads to a Kerr/CFT type diffeomorphism, which may be used to construct \emph{the Kerr/CFT phase space}. We discuss this in the appendix \ref{append-Kerr-CFT}. The second choice leads to the NHEG phase space which is the main focus of our paper and will be described here and in the next two sections. We also show in appendix  \ref{append-Kerr-CFT} that no phase space exists which contains both the classes 6.1) and 6.2), assuming the same definition for the symplectic structure.

As a result, we end up with the following NHEG phase space generator 
\begin{equation}\label{ASK}
\boxed{\overline \chi[{\epsilon}(\vec{\varphi})]={\epsilon}\vec{k}\cdot\vec{\pd}_\varphi-\vec{k}\cdot\vec{\pd}_\varphi{\epsilon}\;(\dfrac{b}{r}\pd_{t}+r\pd_{r})}
\end{equation}
with $b= \pm 1$ which generates the infinitesimal perturbations tangent to the phase space around the background, $\delta\Phi[{\epsilon}(\vec{\varphi})]=\mathcal{L}_{\overline\chi} \bar{\Phi}$.

\subsection{Finite transformations and generic metric of the phase space}\label{sec finite trans}

We define the NHEG phase space from the \textit{exponentiation} of the vector field $\overline \chi$ with an arbitrary periodic function $\overline \epsilon(\vec{\varphi})$. At the infinitesimal level, one applies the coordinate transformation
\begin{align}
\overline x&\rightarrow x = \overline x-\overline \chi( \overline x).
\end{align}
To define the finite coordinate transformation $ \overline x \rightarrow x(\overline x)$ we need to specify the vector field $\chi$ for an arbitrary element of the phase space. For this purpose, we impose that the vector $\chi$ keeps its functional form identical to the one of $\overline \chi$, though with a possibly different function, which we denote by $\epsilon(\vec{\varphi})$. More precisely, we require that the coordinate transformation maps the vector $\chi[{\epsilon}(\varphi)]$ to the vector $\overline \chi[\overline {\epsilon}(\bar{\varphi})]$ defined on the background as
\begin{align}\label{chi-vs-chi'}
\chi^\mu[{\epsilon}(\varphi)]&=\dfrac{\pd x^\mu}{\pd \bar{x}^\alpha}\overline \chi^\alpha[\overline\epsilon(\bar{\varphi})].
\end{align}
In this section we keep the $b$ parameter in \eqref{ASK} unfixed (without setting it to $\pm 1$). This will allow us to derive the property \textbf{4.2)} claimed in the previous subsection.

The finite coordinate transformation ought to take the form
\begin{align}\label{finite ansatz}
\bar{\varphi}^i=\varphi^i + k^i F(\vec{\varphi}), \qquad \bar{r} =re^{-{\Psi(\vec{\varphi})}},\qquad \bar{t} =t-\frac{b}{r}(e^{\Psi(\vec{\varphi})}-1),
\end{align}
with functions $F(\varphi^i)$ and ${\Psi}(\varphi^i)$ periodic in all of their arguments in order to ensure smoothness. 
Indeed, the form of the finite coordinate transformation \eqref{finite ansatz} is constrained by the following facts: (1) $\vec{\eps}$ is proportional to $\vec{k}$ and hence $\varphi^i-\bar\varphi^i$ is also proportional to $k^i$; (2) $\overline \chi$ commutes with $\xi_-$ and therefore the time dependence is trivial; (3) there is no $\theta$ dependence; (4) $\overline \chi$ commutes with $\xi_0$ and therefore the radial dependence is uniquely fixed; (5) since $\overline  \chi$ commutes with the vector
\bea
\eta_b \equiv \frac{b}{r}\p_t+r\p_r, \label{defetab}
\eea
the coordinate
$$
v_b \equiv t+\frac{b}{r},
$$
is invariant.\footnote{In other words, in the coordinates $(v_b,r,\theta,\varphi^i)$ the generator $\overline \chi$ has $\vec{\pd}_\varphi$ and ${\pd}_r$ components. Therefore the coordinate $v_b$ is not affected by the exponentiation of $\overline \chi$.} Note that $v_b$ for $b=\pm 1$ reduces to $v$ and $u$. This finally fixes the form \eqref{finite ansatz} where we can check that $v_b = \bar t+\frac{b}{\bar r}$. 

The remaining question is how to relate the functions $F(\vec{\varphi})$ and  ${\Psi}(\vec{\varphi})$ such that \eqref{chi-vs-chi'} is satisfied. The answer is unique and given by
\begin{align}\label{Psi-def}
e^{\Psi}=1+\vec{k}\cdot\vec{\pd}_\varphi F.
\end{align}
We prove this equation in appendix \ref{proof1}. We also note that the arguments of $\chi$ and $\bar{\chi}$, respectively $\eps(\vec{\varphi})$ and $\bar\eps(\vec{\bar\varphi})$ are related as 
\be\label{eps-eps-bar}
\bar\eps(\vec{\bar\varphi})=e^{\Psi}\ \eps(\vec{\varphi}).
\ee
Therefore, from now on we will denote the phase space as $\mathcal{G}_{\{ p \}}[F]$ as a function of the initial parameters of the NHEG background and as a function of the function $F(\vec{\varphi})$ which we will dub the \emph{wiggle function}.

Using the finite coordinate transformations we can finally derive the one-function family of metrics which constitute the phase space in the $(t,r,\theta,\varphi^i)$ coordinate system:
\begin{align}\label{g-F}
ds^2=\Gamma(\theta)&\Big[-\left( \boldsymbol\sigma - b d \Psi \right)^2+\Big(\dfrac{dr}{r}-d{\Psi}\Big)^2
+d\theta^2+\gamma_{ij}(d\tilde{\varphi}^i+{k^i}\boldsymbol{\sigma})(d\tilde{\varphi}^j+{k^j}\boldsymbol{\sigma})\Big],
\end{align}
where $v_b=t+\frac{b}{r}$ and
\be\label{tilde-varphi}
\boldsymbol{\sigma}=e^{-{\Psi}}rdv_b+b\dfrac{dr}{r},\qquad \tilde{\varphi}^i=\varphi^i+k^i (F-b{\Psi})\,.
\ee
We note that, by virtue of periodicity of $F$ and ${\Psi}$, all angular variables $\bar{\varphi}^i$, $\varphi^i$ and $\tilde{\varphi}^i$ have $2\pi$ periodicity.
 
As a cross-check one can readily observe that $\xi_-=\p_t$ and $\xi_0=t\p_t-r\p_r$ are isometries of the metric \eqref{g-F}. Moreover, one can check that for $|b|=1$, constant $v_b$ are null surfaces at which $\p_r$ becomes null.

We will be defining the conserved charges through integration of $(d-2)$-forms on the constant $t,r$ surfaces ${\cal H}$ whose metric is
\be\label{H-surface}
ds^2_{\cal H}=\Gamma(\theta)\left[(1-b^2)d{\Psi}^2+d\theta^2+\gamma_{ij}(\theta)\,d\tilde{\varphi}^i\, d\tilde{\varphi}^j\right].
\ee
For a generic function $F(\vec{\varphi})$ (and hence ${\Psi}$), the above metric \eqref{H-surface} does have the same  metric and topology as the constant $t,r$ surfaces on the background \eqref{NHEG-metric} if and only if $|b|=1$. This provides the justification for the requirements 4.2) and 3). 

We also comment that even at $b=1$, \eqref{H-surface} comes with the coordinate $\tilde{\varphi}^i$ \eqref{tilde-varphi}. Therefore, the volume form of \eqref{H-surface} differs from that of constant $t,r$ surfaces of \eqref{NHEG-metric} by the Jacobian of transformation matrix $M_i^j$
\be\label{Jacobian}
M_i^j=\frac{\p\tilde{\varphi}^i}{\p\bar{\varphi}^j}=\delta^i_j-k^i Y_j,\qquad Y_j=\p_j{\Psi}+  \vec{k}\cdot\vec{\p}_\varphi (e^{-{\Psi}})\ \p_j F,
\ee
and hence 
\be\label{detM}
\det{M}=1-\vec{k}\cdot \vec{Y}=1+\vec{k}\cdot\vec{\p}_\varphi(e^{-{\Psi}}).
\ee
Since this is one of the main results of this paper, we write again the final metric over the final phase space (with $b=1$) as
\begin{equation}\label{finalphasespace}
\boxed{ds^2=\Gamma(\theta)\Big[-\left( \boldsymbol\sigma -  d \Psi \right)^2+\Big(\dfrac{dr}{r}-d{\Psi}\Big)^2+d\theta^2+\gamma_{ij}(d\tilde{\varphi}^i+{k^i}\boldsymbol{\sigma})(d\tilde{\varphi}^j+{k^j}\boldsymbol{\sigma})\Big], }
\end{equation}
\bea
\boxed{\boldsymbol{\sigma}=e^{-{\Psi}}rd(t+\frac{1}{r})+\dfrac{dr}{r},\qquad \tilde{\varphi}^i=\varphi^i+k^i (F-{\Psi})\,,\qquad e^{\Psi}=1+\vec{k}\cdot\vec{\pd}_\varphi F . }
\eea

\subsection{Algebra of generators}

One can expand the periodic function $\epsilon(\vec{\varphi})$ in its Fourier modes:
\begin{equation}
\epsilon(\vec{\varphi})=- \sum _{\vec{n}} c_{\vec{n}}\,e^{- i(\vec{n}\cdot \vec{\varphi})}\,
\end{equation}
for some constants $c_{\vec{n}}$ and $\vec{n}\equiv (n_1,n_2,\dots,n_n)$, $n_i\in \mathbb{Z}$.\footnote{ The sign conventions are fixed such that the algebra takes the form \eqref{chi-algebra} and such that the central charge takes the form \eqref{CC}.} 
 Therefore the generator $\chi$ decomposes as
\begin{equation}\label{Ln expansion}
\chi=\sum _{\vec{n}} c_{\vec{n}}\chi_{_{\vec{n}}}\,,
\end{equation}
where
\begin{equation}\label{Final Ln}
\chi_{_{\vec{n}}}=-e^{-i(\vec{n}\cdot\vec{\varphi})}\bigg(i(\vec{n}\cdot\vec{k})(\frac{1}{r}\partial_t+ r\partial_r)+\vec{k}\cdot\vec{\partial_{\varphi}} \bigg)\,.
\end{equation}
The Lie bracket between two such Fourier modes is given by
\begin{equation}\label{chi-algebra}
i \left[\chi_{_{\vec{m}}},\chi_{_{\vec{n}}}\right]_{L.B.}= \vec{k}\cdot(\vec{m}-\vec{n})\chi_{_{\vec{m}+\vec{n}}}\,.
\end{equation}
{ Since the generators do not explicitly depend upon the metric field, the total bracket defined in \eqref{totalbracket} coincides with the Lie bracket. We will discuss the representation of this algebra by conserved charges in section \ref{sec-charges-and-algebra}. }

\subsection{$SL(2,\mathbb{R})\times U(1)^{d-3}$ isometries of the phase space}
\label{sec:iso}

Since the whole phase space is constructed by coordinate transformations from the NHEG  background \eqref{NHEG-metric}, they will still have the same isometries.  
 The isometries in the phase space are defined by the pushforward of the background isometries under the coordinate transformations. Explicitly,
\begin{align}
\bar{\xi}&=\bar{\xi}^\nu \frac{\p}{\p \bar{x}^\nu}=\left(\bar{\xi}^\nu \dfrac{\p x^\mu}{\p \bar{x}^\nu}\right)\frac{\p}{\p {x}^\mu}\,.\nonumber
\end{align}
As a result, the Killing vectors  are defined as
\begin{align}
\xi^\mu&=\dfrac{\pd x^\mu }{\pd \bar{x}^\nu}\bar{\xi}^\nu
\end{align}
where $\bar{\xi}^\nu$ are defined in \eqref{xi1-xi2}. Note that the transformation matrix $\frac{\pd x^\mu }{\pd \bar{x}^\nu}$ is a function of $F(\vec{\varphi})$ and hence $\xi^\mu$ constitute  \textit{field dependent} isometries on (each point of) the phase space.

After a straightforward computation, the $SL(2,\mathbb{R})\times U(1)^{d-3}$ isometries are explicitly 
\begin{align}\label{m-barm}
\xi_- =\partial_t\,,\qquad \xi_0&=t\partial_t-r\partial_r,\qquad \xi_+ =\dfrac{1}{2}(t^2+\frac{1}{r^2})\partial_t-tr\partial_r-\frac{1}{r}{k}^i{\p}_{\varphi^i}+\dfrac{1}{r}\vec{k}\cdot\vec{\p}_\varphi (F-{\Psi}){\eta_+},\nonumber\\
\mathrm{m}_i&=\left(\delta^j_i-e^{-\Psi}k^j\p_i F\right)\p_{\varphi^j} +(\p_i {\Psi}-e^{-\Psi}\vec{k}\cdot\vec{\p}_\varphi  {\Psi}\p_i F) {\eta_+},
\end{align}
where $\eta_+ = \eta_{b=1}$ is defined in \eqref{defetab}, see also appendix \ref{Appendix-eta}. As a consequence of the construction,  $\xi_-,\xi_0$ are not field dependent; i.e. they are independent of the function $F$, but other isometries are field dependent. 

{The angular momenta $J_i$ are by definition the conserved charges associated with the Killing vectors $\mathrm{m}_i$ whereas the charge $H_{\vec{0}}$ is associated with $k^i \p_{\varphi^i}$ which is not a Killing vector. Despite the fact that the vectors $\mathrm{m}_i$ are field dependent \cf\eqref{m-barm}, their conserved charge $J_i$ is fixed on the whole phase space as we will demonstrate in section \ref{sec-NHEG-Algebra}. On the contrary, the vector $k^i \p_{\varphi^i}$ has fixed components over the phase space but its conserved charge $H_{\vec{0}}$ varies over the phase space, as a consequence of the symmetry algebra as discussed in section \ref{sec-NHEG-Algebra}. 

Moreover, using \eqref{H-surface} and \eqref{detM} one may readily show that the area of the bifurcation surface $\mathcal{H}$, and hence the entropy $S$, is independent of the function $F$ and therefore is the same over the phase space. Indeed, the area of  $\mathcal{H}$ (at $b=1$) is given by
\be\label{sameA}
A_{\mathcal{H}}=\int d\theta\Gamma^{\frac{d-2}{2}} \sqrt{\det\gamma}\cdot \int \prod_i d\tilde{\varphi}^i=\int d\theta\Gamma^{\frac{d-2}{2}} \sqrt{\det\gamma}\cdot \int \prod_i d\bar{\varphi}^i,
\ee
and therefore equal to the one of the background. In the last equality we used the fact that the Jacobian of the transformation from $\tilde{\varphi}$ to $\bar{\varphi}$ is one plus a total derivative, as given in \eqref{detM}. Therefore, the phase space consists of metrics with equal $S$ and $J_i$.  }

\definecolor{light gray}{RGB}{220,220,220}
\begin{figure}[!bht]
	\captionsetup{width=0.8\textwidth}	
	\centering
	\begin{tikzpicture}[scale=1.2]
	\fill[even odd rule,light gray]
	(0,0) to (5,1.5) to  (11,1.5) to
	(6,0) to (0,0);
	\draw[very thick] (5.55,0) -- (5.55,-1) node[align=center,anchor=north] { };
	\draw[very thin,white] (1,0) to (6,1.5);
	\draw[very thin,white] (2,0) to (7,1.5);
	\draw[very thin,white] (3,0) to (8,1.5);
	\draw[very thin,white] (4,0) to (9,1.5);
	\draw[very thin,white] (5,0) to (10,1.5);
	\draw[very thin,white] (0.8,0.25) to (6.8,0.25);
	\draw[very thin,white] (1.6,0.5) to (7.6,0.5);
	\draw[very thin,white] (2.5,0.75) to (8.5,0.75);
	\draw[very thin,white] (3.3,1) to (9.3,1);
	\draw[very thin,white] (4.1,1.25) to (10.1,1.25);
	\draw[very thick,->] (5.55,0.75) -- (5.55,3) node[anchor=north west] {$J_i$};
	\fill[black] (5.55,0.75) circle (0.05) node[anchor=north] {\footnotesize $g[F=0]=\bar{g}$};
	\fill (8,1.25) circle (0.05) node[anchor=north] {\footnotesize $g[F]$};
	\draw (2,0) node[anchor=north] {$\mathcal{G}_{\{ p \}}[F]$};
	\end{tikzpicture}
	\caption{\footnotesize  A schematic depiction of the NHEG phase space $\mathcal{G}_{\{ p \}}[F]$. The vertical axis shows different background NHEG solutions of the form \eqref{NHEG-metric} specified by different angular momenta $J_i$, and the horizontal plane shows the phase space constructed by  the action of the finite coordinate transformation \eqref{finite ansatz}. Each geometry in the phase space is  identified by a periodic function $F(\vec{\varphi})$ and admits the same angular momenta $J_i$ and entropy.}\label{fig phase space}
\end{figure}
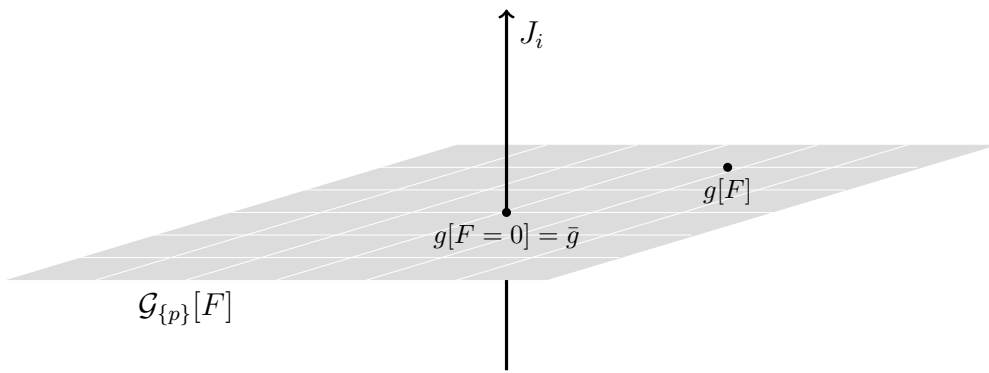

\paragraph{Summary of the section:}  The NHEG phase space $\mathcal{G}_{\{ p \}}[F]$ is a one-function  family of everywhere smooth metrics given in \eqref{finalphasespace}. These are obtained through finite coordinate transformations \eqref{finite ansatz} acting on the NHEG background \eqref{NHEG-metric}, which is the $F=0$ element in the phase space. All the metrics of the form \eqref{finalphasespace} have the same angular momentum and same parameters $\vec{k}$. By the entropy law \eqref{entropy law}, they have the same entropy. This  last observation is schematically depicted in Fig \ref{fig phase space}.

\section{Symplectic Structure}\label{sec-symplectic structure}

The set $\mathcal{G}_{\{ p \}}[F] $ consisting of field configurations (metrics) \eqref{g-F}, can be viewed as a manifold, where each point of this manifold represents a metric $g[F]$ over the spacetime, determined by the functional form of the wiggle function $F[\varphi^i]$. In order for $\nhegps$ to be a \textit{phase space}, it should be accompanied by a \textit{symplectic structure}. That is, a finite, closed and nondegenerate two-form which is the integral of a $d-1$ spacetime form and two-form in field variations, the \emph{presymplectic form}. The aim of this section is to define the presymplectic form on the set of metrics \eqref{g-F}.

The ADM formulation of gravity \cite{Arnowitt:1962hi} provides a way to construct the phase space and its symplectic structure, see also \cite{Regge:1974zd,Brown:1986ed}. Such Hamiltonian methods are not covariant by construction since they split space and time.  The covariant phase space method, developed in \cite{Lee:1990nz,Wald:1993nt} and refined in \cite{Barnich:2001jy,Barnich:2007bf,Compere:2008us}, is a prescription to construct the phase space in a covariant fashion. A self-contained brief review on this topic is given in appendix \ref{appendix:charges}.

In the particular case at hand, a complete basis of one-forms at any point of $\mathcal{G}_{\{ p \}}[F] $, is given by the Lie derivative of fields with respect to generators $\chi_{\vec{n}}$ \eqref{Final Ln}. In other words, we can expand any variation $\de\Phi$ as
\begin{align}\label{tangent space basis}
	\de\Phi&=\sum_{\vec{n}} c_{\vec{n}}\, \mathcal L_{\chi_{{\vec{n}}}} \Phi .
\end{align}
From the fundamental theorem of the covariant phase space formalism, the Theorem 1 in appendix \ref{appendix:charges}, for a given Lagrangian the symplectic structure is equal on-shell to a sum of boundary terms. Such boundary terms are surface charges which are the integral of surface charge $d-2$ forms. All dynamical information about the phase space is therefore encoded in these surface charges.

More precisely, since the geometries in $\mathcal{G}_{\{ p \}}[F] $ have two spatial boundaries due to the $AdS_2$ factor, the symplectic structure reduces to a sum of two boundary surface integrals. While the change of orientation between these two boundary integrals might make the symplectic structure vanish,  the surface charges might not individually vanish. This is familiar already for the simple example of the phase space of all Schwarzschild black holes with varying mass $M$. The symplectic structure vanishes on-shell but the surface integrals at the two spatial boundaries of the maximal analytic extension of Schwarzschild are $+M$ and $-M$. We therefore expect here that the physical information of the phase space is only partially contained in the symplectic structure but fully in the surface charges. At any rate, all dynamical information is fully contained in the presymplectic form to which we now turn our attention.

According to \eqref{tangent space basis}, the presymplectic structure is completely determined when its action on $\de_{\vec{m}}\Phi,\de_{\vec{n}}\Phi$ for any $\vec{m},\vec{n}$ is known.  
As a consequence of $\xi_-,\xi_0$ invariance, there is no time dependence in the presymplectic structure and the radial dependence is fixed as
\begin{align}
\omega^t \propto \frac{1}{r}, \qquad \omega^r \propto r, \qquad \omega^\theta \propto r^0, \qquad \omega^{\varphi^i} \propto r^0.
\end{align}
Also, since constant $v_b = t + \frac{b}{r}$ surfaces are preserved in the phase space (we keep $b$ arbitrary for book-keeping purposes but we will impose $b=1$ at the end), one has
\bea
\omega^t &= \frac{b}{r^2} \omega^r. \label{omt}
\eea
One usually requires that  $\omega^r$ is zero at the boundary in order to avoid a leaking symplectic flux at spatial infinity. This implies that $\omega^t$ will automatically vanish as well, and the presymplectic structure will be trivial. 
However, it is important to note that we can impose these conditions only on-shell. If the presymplectic structure is zero on-shell but non-zero off-shell, it still allows  to define non-trivial surface charges. 
We conclude that the phase space exists and is non-trivial if and only if the presymplectic form at constant $t$ or constant $r$ is zero on-shell but not off-shell.

\subsection{Lee-Wald symplectic structure}
\label{sec-cov-phase-space}

The standard presymplectic structure as defined by Lee-Wald is given by
\be\label{LW2}
{\boldsymbol \omega}_{(LW)}[\delta_1 \Phi ,\delta_2 \Phi,\Phi ] = \delta_1 {\boldsymbol  \Theta_{(LW)}}[ \delta_2 \Phi,\Phi ] - \delta_2 {\boldsymbol \Theta_{(LW)}}[\delta_1 \Phi, \Phi ] ,
\ee
where for Einstein gravity and for perturbations which preserve the $d$ dimensional volume, $h \equiv g^{\mu\nu}\delta g_{\mu\nu} = 0$, we have
\bea
\Theta_{(LW)}^\mu = \frac{1}{16 \pi G}\nabla_\nu h^{\mu\nu}.
\eea
It is straightforward to check that $\omega^r_{(LW)}$ is non-vanishing. Therefore the set of metrics $\mathcal{G}_{\{ p \}}[F] $ equipped with the Lee-Wald symplectic structure does not define a well-defined phase space. 

More precisely, in four spacetime dimensions we find around the NHEG background 
\begin{align}
\begin{split}
\omega^t[\delta_m g,\delta_n g, \bar g] &= \frac{b}{r^2} \omega^r[\delta_m g,\delta_n g, \bar g] ,\\
\sqrt{-g}\omega^r[\delta_m g,\delta_n g, \bar g] &= \frac{ \Gamma (-1+k^2 \gamma)\;r}{8 \pi G \sqrt{\gamma}} e^{i (m+n)\varphi}\;k^2 m n (m-n)  (m+n-i b k \gamma), \\
\sqrt{-g}\omega^\theta[\delta_m g,\delta_n g, \bar g] &= -i\;\frac{  \Gamma \gamma'}{16 \pi G \sqrt{\gamma}}e^{i (m+n)\varphi}\;b\, k^3 m n (m-n),\\
\sqrt{-g}\omega^\varphi[\delta_m g,\delta_n g, \bar g] &= i\;\frac{ \Gamma  (-1+k^2 \gamma)}{8\pi G \sqrt{\gamma}}e^{i (m+n)\varphi}\;k^2 m  n (m-n).
\end{split}
\end{align}
Given our choice of $b \neq 0$ the integral $\int_\Sigma \boldsymbol{\omega}[\de_m g,\de_n g,\bar{g}]$ over a constant $t$ surface $\Sigma$ is divergent for $m=-n \neq 0$. Also, $\omega^r\propto r$ so the boundary flux is not vanishing and in fact divergent. Also note that since $\gamma\rightarrow 0$ at the poles $\theta=\{0,\pi\} $, $\omega^{\varphi}$ is locally divergent at the poles.

\subsection{Regularization of symplectic structure}\label{section: Y-terms}
As reviewed in appendix \ref{appendix:charges}, the presymplectic potential ${\boldsymbol  \Theta}[ \delta \Phi,\Phi ]$ is ambiguous up to the addition of boundary terms. The total presymplectic potential therefore has the form
\bea
\Theta^\mu[\delta \Phi ,\Phi] = \frac{1}{16 \pi G}\nabla_\nu h^{\mu\nu} + \nabla_\mu Y^{\mu\nu}.
\eea
where $Y^{\mu\nu} = Y^{[\mu\nu]}$ defines a $d-2$ form $\mathbf{Y}[\delta \Phi ,\Phi] $ which is linear in the field variations but non-linear in the fields. This leads to the total presymplectic form
\begin{align}\label{ambiguity omega}
{\boldsymbol{\omega}}[\delta_{1}\Phi,\delta_{2}\Phi,\Phi ] &=\boldsymbol{\omega}_{(LW)} [ \delta_{1}\Phi,\delta_{2}\Phi,\Phi]+d\Big(\delta_{1}\mathbf{Y} [\delta_{2}\Phi,\Phi]-\delta_{2}\mathbf{Y}[ \delta_{1}\Phi,\Phi ]\Big).
\end{align}
Next, we will  define $\mathbf{Y}[\delta \Phi ,\Phi] $ in order to ensure that $\omega^t$ and $\omega^r$ vanish on-shell. 

In the derivation of the finite coordinate transformations we noted that the vector field $\eta_b$ defined in \eqref{defetab} commutes with the generator around the background $\overline \chi$. Since the form of the generator $\chi$ around any point in the set of metrics \eqref{g-F} takes the same functional form, one has 
\bea
[\eta_b , \chi] = 0
\eea
for any metric in the class. It can then be checked that for any two variations tangent to the phase space { around the background
$\delta_{ 1}\bar\Phi,\delta_{ 2}\bar\Phi$ we have
\begin{equation}\label{eta wald}
\mathcal{L}_{\eta_b}\boldsymbol{\omega}_{(LW)} [\delta_{ 1}\bar\Phi,\delta_{ 2}\bar\Phi,\bar\Phi]=\boldsymbol{\omega}_{(LW)} [\delta_{ 1}\bar\Phi,\delta_{ 2}\bar\Phi,\bar\Phi ].
\end{equation}
Applying the finite diffeomorphism \eqref{finite ansatz}, and recalling covariance of $\boldsymbol{\omega}$ and $\eta_b$ one deduces that the equation holds around any element of the phase space,} which we can rewrite on-shell as 
\begin{equation}\label{peq1}
\boldsymbol{\omega}_{(LW)} [\delta_{ 1}\Phi,\delta_{ 2}\Phi,\Phi ] \approx d \left( \eta_b \cdot \boldsymbol{\omega}_{(LW)} [\delta_{ 1}\Phi,\delta_{ 2}\Phi,\Phi ]\right) 
\end{equation}
after using Cartan's identity  $\mathcal{L}_\eta X=\eta\cdot dX+d(\eta\cdot X)$ and recalling the fact that the presymplectic structure is closed on-shell, $d \boldsymbol{\omega} \approx 0$.

Therefore, it is natural to define
\bea
\mathbf{Y}[\delta \Phi ,\Phi]  =- \eta_b \cdot \boldsymbol{\Theta}_{(LW)}[\delta \Phi ,\Phi]  + \mathbf{Y}_{comp}[\delta \Phi ,\Phi] \label{defY1}
\eea
and we obtain from \eqref{ambiguity omega} and \eqref{peq1}, 
\begin{align} 
{\boldsymbol{\omega}}[\delta_{1}\Phi,\delta_{2}\Phi,\Phi ] &\approx d \left( \eta_b \cdot \boldsymbol{\omega}_{(LW)} [\delta_{ 1}\Phi,\delta_{ 2}\Phi,\Phi ] -\delta_1 (\eta_b \cdot  \boldsymbol{\Theta}_{(LW)}[\delta_2 \Phi ,\Phi] )
+\delta_2 (\eta_b \cdot  \boldsymbol{\Theta}_{(LW)}[\delta_1 \Phi ,\Phi] )\right)\nonumber \\
&+ d (\delta_1 \mathbf{Y}_{comp}[\delta_2 \Phi ,\Phi] -\delta_2 \mathbf{Y}_{comp}[\delta_1 \Phi ,\Phi] )\cr
&\approx d (\delta_1 \mathbf{Y}_{comp}[\delta_2 \Phi ,\Phi] -\delta_2 \mathbf{Y}_{comp}[\delta_1 \Phi ,\Phi] )
\end{align}
where we used the fact that $\eta_b$ does not depend upon the fields (its components are identical for the entire family of metrics considered). 
We therefore obtained that for any $ \mathbf{Y}_{comp}$ such that 
\bea
d (\delta_1 \mathbf{Y}_{comp}[\delta_2 \Phi ,\Phi] -\delta_2 \mathbf{Y}_{comp}[\delta_1 \Phi ,\Phi] ) \approx 0,\label{propZ}
\eea
the total symplectic structure is vanishing on-shell. A phase space therefore exists for the set of metrics \eqref{g-F} for all symplectic structures defined off-shell by \eqref{ambiguity omega}-\eqref{defY1}-\eqref{propZ}. In particular $\mathbf{Y}_{comp} = 0$ defines a symplectic structure. The fact that $\mathbf{Y}_{comp}$ is not fixed constitutes a remaining dynamical ambiguity that we need to fix through additional considerations. 
Note that we could only require that the $t,r$ components of the symplectic structure be vanishing instead of fixing all components as in \eqref{propZ} so strictly speaking we did not prove that the conditions \eqref{propZ} are necessary. However, we do not expect that the additional components of the symplectic structure play an important role since the physical observables will be surface charges computed at fixed $t,r$. 

\subsection{Fixation of the dynamical ambiguity}

We fixed most of the ambiguities in the definition of the presymplectic structure by requiring finiteness and conservation of the symplectic structure, up to the remaining ambiguity $\mathbf{Y}_{comp}$ constrained by \eqref{propZ}. 
A first natural question is whether or not this ambiguity matters. In fact, it matters since the value of the charges to be defined in section \ref{sec-NHEG-Algebra} will receive contributions from that term, unless it is of the form  
\bea
\mathbf{Y}_{comp}[\delta\Phi , \Phi ] = \delta \mathbf{Z}[\Phi ] + d \tilde{\mathbf{Z}}[\delta\Phi , \Phi ],\label{trivY}
\eea
see appendix \ref{appendix charge algebra} for a proof. Therefore, we have a cohomological problem: can we find representatives for $\mathbf{Y}_{comp}$ which obey \eqref{propZ} but which are not trivial, i.e. of the form \eqref{trivY}?  Part ot the problem is to clearly specify what are the fields: clearly the metric is the only dynamical field, but non-dynamical fields might enter the expression for $\mathbf{Y}_{comp}$ such as the vector $\eta_b$ \eqref{defetab} defined earlier which we already used to define the presymplectic structure.

We did not find a representative of the cohomology class using only the non-dynamical field $\eta_b$. We were however not exhaustive and we do not claim that such an object does not exist. However, if we introduce one further non-dynamical field and if we also use the binormal tensor to $\mathcal H$ surfaces, we found one representative. Let us define $\eta_2 $ as follows. We first define $\overline \eta_2$ on the background NHEG as
\bea
\overline \eta_2 = \frac{1}{\bar r}\p_{\bar t}.\label{eta2}
\eea
We then extend the definition to an arbitrary element of the phase space using the push-forward of the diffeomorphism generated by $\chi$. Since $[\eta_2,\chi] \neq 0$, the components of $\eta_2$ will depend upon the element of the phase space. One ansatz for such a non-trivial cohomology is 
\bea
Y_{comp}^{\mu\nu}[\delta \Phi , \Phi] = f[\delta \Phi ; \Phi ] \eps_\perp^\mn\label{anz1}
\eea
where  $\eps_\perp^\mn$ is the binormal tensor of $\mathcal{H}$-surfaces (the NHEG bifurcation Killing horizons). The scalar function $f[\delta \Phi ; \Phi ]$ should be linear in the variation of the dynamical field which is the metric, so we construct the function $f$ with the help of two vector fields $t_+,t_-$
\begin{align}
f[\delta \Phi ; \Phi ] = \dfrac{1}{16\pi G} \frac{1}{\Gamma(\theta)} \de g_{\mn}\;t_+^{\;\mu} t_-^{\;\nu }\label{anz2}.
\end{align}
It turns out that if we choose $t_\pm$ as linear combinations of $\eta_b$ and $\eta_2$ as
\begin{align}
t_\pm &=c_\pm \eta_b +d_\pm \eta_2,\label{anz3}
\end{align}
we can obtain a representative for the dynamical ambiguity \eqref{propZ}. Indeed, the central charge of the charge algebra, to be defined in section \ref{sec-charges-and-algebra} from \eqref{central extension}, depends upon this representative. Defining $C_{\vec{m},\vec{n}} = \oint_{\mathcal H} \boldsymbol k_{\chi_{\vec{m}}}[\delta_{\chi_{\vec{n}}} \bar \Phi, \bar \Phi]$ we find after a straightforward computation using the formulas given in appendix \ref{appendix charge algebra}
\begin{align}
i C_{\vec{m},\vec{n}}&= (\vec{k}\cdot \vec{m})^3 \left( (1-b(b+\Delta)) \frac{A_{\mathcal H}}{8 \pi G} + 2 b (b+\Delta) \vec{k}\cdot \vec{J} \right) \delta_{\vec{m}+\vec{n},0} \nonumber \\
& +(\vec{k}\cdot \vec{m}) (2 \vec{k} \cdot \vec{J} ) \delta_{\vec{m}+\vec{n},0}\label{Cmn}
\end{align}
where all dependence in the coefficients $c_\pm,d_\pm$ reduces to a dependence in the single combination $\Delta$,
\bea\label{Delta-b}
\Delta =2d_- d_+ +b(c_+ d_-+c_- d_+). 
\eea
Note that in the original Kerr/CFT ansatz, $b=0$, and this dynamical ambiguity does not appear. 

In order to fix this ambiguity in our case $b = \pm 1$, we now require that the central charge of the charge algebra is independent of the choice $b$. This fixes $\Delta\equiv -b$  where $\Delta$ is defined in \eqref{Delta-b}. We do not have a fundamental justification for imposing such a requirement. We are however motivated by the universality of the computation of central charge obtained using the Kerr/CFT ansatz for which $b=0$ (see e.g. \cite{Lu:2008jk,Azeyanagi:2009wf}) and it seems natural to us to impose that the central charge does not depend upon the particular choice of generator ansatz. 

Up to trivialities (vanishing terms), the choice is then unique in the ansatz \eqref{anz1}-\eqref{anz2}-\eqref{anz3} and given by $\Delta = -b$. A representative is given by 
\begin{align}\label{eta-b-eta2}
c_+&=1,\quad c_-=0, \quad d_+=0, \quad d_-= - 1 \implies t_+=\eta_b,\quad t_-=-\eta_2
\end{align}
Therefore the final symplectic structure is constructed using \eqref{ambiguity omega} with
\bea \label{Y-total}
\boxed{
(16\pi G){Y}^{\mn}[\delta\Phi,\Phi ]= \eta_b^{[\mu} \nabla_\rho h^{\nu ] \rho}- \left( \frac{1}{\Gamma} \de g_{\alpha\beta}\;\eta_b^{\;\alpha} \eta_2^{\;\beta }\right)\;{\epsilon}_\perp^\mn \, .
}
\eea

It would be important to prove that either there is a unique representative for this cohomology class and or that the requirement that the central charge is $b$ independent uniquely fixes the charges. We do not have such a proof. Some properties of special vectors in the phase space are given in appendix \ref{Appendix-eta} for the eager reader who might want to pursue this direction.

\section{Surface Symplectic  Charges and the NHEG Algebra}\label{sec-charges-and-algebra}

In the previous sections we built the NHEG phase space and its symplectic structure. In this section, we show that the set of vector fields which generate the phase space indeed constitutes the set of symplectic symmetries and  analyze their conserved charges and their algebra. To this end, we first observe that any symplectic symmetry is integrable, namely it leads to well-defined charges over the phase space. We then  construct the algebra of charges and  provide an explicit representation of the charges in terms of a Liouville-type stress-tensor on the phase space.

\subsection{Symplectic symmetries and integrability}
\label{secint}

The fundamental theorem of the covariant phase space, see \eqref{eq:f},  states that the symplectic structure contracted with a perturbation generated by the vector field $\chi$ is a boundary term on-shell, 
\begin{align}\label{omega vs k}
\boldsymbol{\omega}[ \de \Phi,\de_\chi \Phi , \Phi]& =d {\boldsymbol k}_\chi [\de\Phi,\Phi ]+ \text{terms that vanish on-shell}. 
\end{align}
{ In the previous section we constructed $\boldsymbol{\omega}$ such that $\boldsymbol{\omega}[ \de_1 \Phi,\de_2 \Phi ,\Phi] \approx 0$  for any two perturbations around an arbitrary element of the phase space $\Phi$.} Therefore, for each generator $\chi$, one has a conserved infinitesimal surface charge
\begin{align}\label{def charge variation}
\de H_\chi&= \oint_{\mathcal{H}} {\boldsymbol k}_\chi [\de\Phi,\Phi ]. 
\end{align}
The charge is conserved upon any smooth deformation of $\mathcal H$ and it is in particular independent of $t$ and $r$. For the Hamiltonian to exist, the integrability condition $\delta \delta H_\chi = 0$ needs to be obeyed. The integrability condition can be written as
\bea
\int_{\mathcal H} \chi \cdot \boldsymbol{\omega}[ \de_1\Phi,\de_2\Phi , \Phi] =  0,
\eea
for any perturbations $\delta_{1,2}\Phi$ and any $\chi$, appendix \ref{appendix:charges}. The integrand is proportional to $\chi^t \omega^r - \chi^r \omega^t$ which is zero off-shell upon using \eqref{omt} with $b= 1$ and $\chi^t = \frac{1}{r^2} \chi^r$. The integrability condition is therefore obeyed off-shell. 

Therefore, to any vector $\chi$ in the class \eqref{Ln expansion} there is a surface charge defined off-shell as 
\begin{align}\label{def charge 1}
H_\chi[\Phi]&= \int_\gamma \oint_{\mathcal{H}} {\boldsymbol k}_\chi [\de\Phi,\Phi] + N_\chi[\bar \Phi],
\end{align}
where $\gamma$ is any path in the phase space between the NHEG background and the solution $\Phi$ and $N_\chi[\bar \Phi]$ is a choice of normalization at the reference solution. The surface charge is conserved on-shell. 

\subsection{Algebra of charges}\label{sec-NHEG-Algebra}

Let us use the Fourier decomposition \eqref{Ln expansion}. We denote the surface charge associated with $\chi_{\vec{n}}$ as $H_{\vec{n}}$. As discussed in section \ref{sec-NHEG-phase-space}, we also have the charges associated with the Killing vectors $\mathrm{m}_i$,  $J_i$, $i=1,\dots d-3$, and charges associated with  $SL(2,\mathbb R)$ Killing vectors $H_{\xi_{\pm,0}}$. $J_i$ are constant over the phase space and $H_{\xi_{\pm,0}}$ are vanishing.
The bracket between charges $H_{\vec{n}}$ is defined as 
\be\label{H-commutator}
\{H_{\vec{m}}, H_{\vec{n}}\}=\delta_{\vec{n}}H_{\vec{m}}=\oint_\mathcal{H}{\boldsymbol k}_{\chi_{\vec{m}}} [\delta_{\vec{n}}\Phi,{\Phi}],
\ee
for an arbitrary point in the phase space $\Phi$ and field variations $\delta_{\vec{n}}\Phi$. The right-hand side  is indeed anti-symmetric as a consequence of the integrability conditions.

Using the representation theorem proven in \cite{Barnich:2007bf} (reviewed in appendix \ref{appendix:charges}), the charges obey the same algebra as the symmetry generators \eqref{chi-algebra} up to a possible central term, i.e.
\be\begin{split}\label{calc central extension}
\{H_{\vec{m}}, H_{\vec{n}}\} &= -i\vec{k} \cdot (\vec{m}- \vec{n}) H_{\vec{m}+\vec{n}} + C_{\vec{m},\vec{n}}\,\cr
\{H_{\vec{p}}, C_{\vec{m},\vec{n}} \}& =\{H_{\vec{m}}, J_{i} \} =\{H_{\vec{m}}, H_{\xi_{\pm,0}} \} =0, \qquad \forall \vec{p},\vec{n},\vec{m}.
\end{split}
\ee
Note that the vanishing bracket between $H_{\vec{m}}$ and the angular momenta follows from either the fact that the angular momenta are constant, or from the fact that the vector fields $\mathrm{m}_i$ are Killing symmetries so that $\oint k_{\chi}[\mathcal L_{\mathrm{m}_i} g,g] = 0$. Even though the Lie bracket $[\chi,\mathrm{m}_i]_{L.B.} \neq 0$, the vanishing charge bracket is also consistent with the representation theorem since the total bracket $[\chi,\mathrm{m}_i] = [\chi,\mathrm{m}_i]_{L.B.} - \delta_\chi^g \mathrm{m}_i = 0$. The same reasoning holds for $H_{\xi_+}$. 

As mentioned in the end of section \ref{sec-NHEG-phase-space}, the angular momenta $J_i$ and the $SL(2,\mathbb R)$ charges are constants over the phase space (the latter are in fact vanishing). To see this, we note that
\begin{equation}\label{Komar-integral}
\delta J_i =- \int_{\mathcal H} \boldsymbol k_{{\mathrm{m}}_i}[\delta_\chi \Phi, \Phi]=- \int_{\mathcal H} \boldsymbol k_{\bar{\mathrm{m}}_i}[\delta_\chi \bar \Phi, \bar \Phi]=0.
\end{equation}	
The second equality follows from general covariance of all expressions and the $\xi_-,\xi_0$ invariance which allows to freely move the surface $\mathcal H$, and the last equality is a result of the fact that $\bar \Phi$ is axisymmetric, and the only $\varphi^i$ dependence coming from $\chi$ makes the integral vanishing. This argument can also be repeated  for $SL(2,\mathbb R)$ charges.

The central extension $C_{\vec{n},\vec{m}}$ is defined in \eqref{central extension} as a constant over the phase space which is computed on the background. The second term on the right-hand side of \eqref{central extension} can be fixed to cancel terms proportional to $(\vec{m}-\vec{n})$ by fixing the reference point for the charges. In this case, it amounts to fixing $N_{\chi_{\vec{n}}} = 0$, $\forall \vec{n} \neq 0$ and $N_{\chi_{\vec{0}}}= -\vec{k} \cdot \vec{J}$ { as we can see from the expression \eqref{Cmn}}.\footnote{This is very similar to the shift of the generators of the Virasoro algebra  $L_0, \bar L_0$ when we move from the cylinder to the plane.}  The central extension is then found to be proportional to the entropy $S$, 
\begin{align}\label{CC}
 C_{\vec{m},\vec{n}}&= -i (\vec{k}\cdot \vec{m})^3 \frac{\hbar S}{2\pi}\,\delta_{\vec{m}+\vec{n},0},
\end{align}
{after multiplying and dividing by one power of $\hbar$, \cf section \ref{section-quantized-NHEG-algebra}.} 
The fact that entropy appears as the central element of the algebra dovetails with the arguments in the end of section \ref{sec-NHEG-phase-space} and especially \eqref{sameA},  ensuring that the area and therefore the entropy does not vary over the phase space.

Therefore we find the classical NHEG algebra
\begin{align}\label{NHEG-algebra}
 i\{H_{\vec{m}}, H_{\vec{n}}\}&= \vec{k} \cdot (\vec{m}- \vec{n}) H_{\vec{m}+\vec{n}} + (\vec{k}\cdot \vec{m})^3 \frac{\hbar  S}{2\pi}\,\delta_{\vec{m}+\vec{n},0} \,,\\
\label{Hn-J-Hxi} \{H_{\vec{m}}, J_{i} \} &=\{H_{\vec{m}}, H_{\xi_{\pm,0}} \}  = \{H_{\vec{m}}, S \} = 0.
\end{align}

\subsection{Charges on the phase space}\label{sec-charge-T}

As discussed earlier, the phase space  $\mathcal{G}_{\{ p \}}[F]$ consists of the one-function family of metrics $g[F]$ given in \eqref{g-F} which is specified by the wiggle function $F(\vec{\varphi})$. This wiggle function defines an auxiliary quantity $\Psi$ defined in \eqref{Psi-def} which we will interpret in the following.

We have proven so far that the charges $H_{\vec{n}}$ are well-defined over phase space and that they obey the algebra \eqref{NHEG-algebra}. We now provide an explicit expression for the charges $H_{\vec{n}}$ as a functional of $\Psi$.
We can plug in the phase space metric and the symplectic symmetries $\chi_{\vec{n}}$ into the explicit formula for the charges in Einstein gravity in order to obtain the explicit expression for the charges $H_{\vec{n}}$. This computation  is explicitly performed in appendix \ref{appendix  charge computation} with the result
\begin{align}\label{charge}
H_{\vec{n}}&= \oint_\mathcal{H} \boldsymbol{\epsilon}_\mathcal{H}\  T[\Psi] e^{-i \vec{n}\cdot \vec{\varphi} },
\end{align}
where $\boldsymbol{\epsilon}_\mathcal{H}$ is the volume form on $\mathcal H$ and
\begin{align}\label{resd4}
T[\Psi]&= \frac{ 1}{16 \pi G}  \Big( (\Psi' )^2 -2 \Psi''+ 2 e^{2 \Psi  } \Big)
\end{align}
where primes are directional derivatives along the vector $\vec{k}$, \ie $\Psi' = \vec{k}\cdot \vec{\partial}_\varphi \Psi$. 
The charges $H_{\vec{n}}$ are therefore the Fourier modes of $T[\Psi]$.

In order to understand this result, it is interesting to first note how the wiggle function $F$ transforms under a symplectic symmetry transformation generated by $\chi[\eps]$. To this end, we recall that by construction 
\be
\mathcal L_{\chi[\eps]}(g_{\mu\nu}[F]) = g_{\mu\nu}[F+\delta_\eps F]-g_{\mu\nu}[F].
\ee
We find 
\bea
\delta_{\eps} F=  (1 +\vec{k}\cdot \vec{\partial}_\varphi F )\eps=e^{{\Psi}}\eps.
\eea
The field $\Psi$ then transforms as
\bea\label{Psi-transform}
\delta_\eps {\Psi} =\eps\,  {\Psi}'+   \eps'.
\eea
where prime denotes again the directional derivative $\vec{k}\cdot \vec{\partial}_\varphi $. Therefore, $\Psi$ transforms like a Liouville field. In particular note that $\delta_\eps e^{{\Psi}}=(e^{{\Psi}}\eps)'$ and hence $e^{{\Psi}}$ resembles a ``weight one operator'' in the terminology of conformal field theory. It is then natural to define the Liouville stress-tensor
\begin{align}\label{T-tensor}
T[\Psi]&= \frac{ 1}{16 \pi G}  \Big( (\Psi' )^2 -2 \Psi''+ \Lambda e^{2 \Psi  } \Big)
\end{align}
with ``cosmological constant'' $\Lambda$ which transforms as 
\bea\label{T-transform}
\delta_\eps T = \eps  T' + 2  \eps' T - \frac{1}{8\pi G}  \eps'''.
\eea
Expanding in Fourier modes as in \eqref{charge}, it is straightforward to check from the transformations law \eqref{T-transform} that the algebra \eqref{NHEG-algebra} is recovered. Using the explicit computation for the surface charges \eqref{charge} we identify the cosmological constant to be $\Lambda = 2$. 

The above resembles the transformation of the energy momentum tensor, a ``quasi-primary operator of weight two''. However, we would like to note that ${\Psi}$ and hence $T[\Psi]$ are not function of time but are functions of all coordinates $\varphi^i$, in contrast with the standard Liouville theory.

Given \eqref{charge} and \eqref{resd4}, one can immediately make the following interesting observation: The charge associated with the zero mode $\vec{n}=0$, $H_{\vec{0}}$, is positive definite over the whole phase space. This is due to the fact that the $\p^2 \Psi$ term does not contribute to $H_{\vec{0}}$ and the other two terms in \eqref{T-tensor} give positive contributions.

\subsection{Quantization of algebra of charges: The \textit{NHEG algebra} }
\label{section-quantized-NHEG-algebra}

Since the symplectic structure is nontrivial off-shell and the resulting surface charges are integrable, we were able to define physical surface charges $H_{{\vec{n}}}$ associated with the symplectic symmetries $\chi[\eps_{\vec{n}}]$, where $\eps_{\vec{n}}=e^{-i \vec{n} \cdot \vec{\varphi}}$, $n_i\in \mathbb{Z}$. The generators of these charges satisfy the same algebra as $\chi$ themselves, but with the entropy as the central extension in \eqref{NHEG-algebra}. One can use the Dirac quantization rules
\be\label{quantization}
\{\quad\}\to \frac{1}{i\hbar}[\quad]\,,\qquad\mathrm{and}\qquad  H_{\vec{n}}\to \hbar \,L_{\vec{n}},
\ee 
to  promote the symmetry algebra to an operator algebra, the \emph{NHEG algebra} $\widehat{\mathcal{V}_{\vec{k},S}}$
\bea\label{NHEG-algebra quantized}
[L_{\vec{m}}, L_{\vec{n}}] = \vec{k} \cdot (\vec{m}- \vec{n}) L_{\vec{m}+\vec{n}} +  \frac{S}{2\pi}(\vec{k}\cdot \vec{m})^3 \delta_{\vec{m}+\vec{n},0}\,.
\eea
The angular momenta  $J_i$ and the entropy $S$ obeying \eqref{entropy law} commute with $L_{\vec{n}}$, in accordance with \eqref{Hn-J-Hxi}, and are therefore central elements of the NHEG algebra  $\widehat{\mathcal{V}_{\vec{k},S}}$. Explicitly, the \emph{full symmetry of the phase space} is
\be\label{full-algebra}
\mathrm{Phase\ Space\ Symmetry\ Algebra}=SL(2,\mathbb{R})\times U(1)^{d-3}\times \nhegalgebra. 
\ee
We reiterate that all geometries in the phase space have vanishing $SL(2,\mathbb{R})$ charges and $U(1)$ charges equal to $J_i$.

\paragraph{The case $d=4$.}
For the four dimensional {Kerr case},  $k=1$ and one obtains the familiar Virasoro algebra
\bea\label{Virasoro}
[L_m, L_n  ]= (m-n) L_{m+n} + \frac{c}{12}m^3\delta_{m+n,0}
\eea
with central charge $c= 12 \frac{S}{2 \pi}=\frac{12J}{\hbar}$, as in  \cite{Guica:2008mu}. We indeed fixed the dynamical ambiguity in the definition of the symplectic structure in order that the resulting central charge be independent of the choice of constant $b$ in the definition of the generator. Since $b=0$ corresponds to the Kerr/CFT generator, we reproduce their central charge.

\paragraph{The cases $d>4$.} In higher dimensions, the NHEG algebra  $\widehat{\mathcal{V}_{\vec{k},S}}$ \eqref{NHEG-algebra} is a more general infinite-dimensional algebra in which the entropy appears as the central extension.
For $d>4$ the NHEG algebra contains infinitely many Virasoro subalgebras. To see the latter, first we note that vectors $\vec{n}$ construct a $d-3$ dimensional lattice. $\vec{k}$ may or may not be on the lattice.  Let $\vec{e}$ be any given vector on this lattice  such that $\vec{e}\cdot \vec{k}\neq 0$. Consider the set of generators $L_{\vec{n}}$ such that $\vec{n}=n \vec{e}$. Then one may readily observe that these generators form a Virasoro algebra of the form \eqref{Virasoro}. If we define
\begin{equation}\label{Virasoro-subalgebra-generators}
\ell_n\equiv \frac{1}{\vec{k}\cdot \vec{e}} L_{\vec{n}}\,,
\end{equation}
then
\begin{align}
[\ell_m,\ell_n]&=[\frac{L_{\vec{m}}}{\vec{k}\cdot \vec{e}} ,\frac{L_{\vec{n}}}{\vec{k}\cdot \vec{e}} ]
=\frac{\vec{k}\cdot(\vec{m}-\vec{n})}{\vec{k}\cdot \vec{e}}\frac{L_{\vec{m}+\vec{n}}}{\vec{k}\cdot \vec{e}}+\frac{(\vec{k}\cdot \vec{m})^3}{(\vec{k}\cdot \vec{e})^2} \frac{S}{2\pi} \,\delta_{\vec{m}+\vec{n},0}\cr
&=(m-n)\ell_{m+n}+\frac{c_{\vec{e}}}{12}m^3 \,\delta_{m+n,0}\,.
\end{align}
As a result, the central charge for the selected subalgebra would be:
\begin{equation}\label{virasoro-subalgenra-central-charge}
c_{\vec{e}}=12 (\vec{k}\cdot \vec{e}) \frac{S}{2\pi}\,.
\end{equation}
The entropy might then be written in the suggestive form $S= \frac{\pi^2}{3} c_{\vec{e}} \,T_{F.T.}$ where
\begin{equation}
T_{F.T.}=\frac{1}{2\pi(\vec{k}\cdot \vec{e})}
\end{equation}
is the { extremal Frolov-Thorne chemical potential associated with $\vec{e}$}, as reviewed in \cite{Compere:2012jk}.

We also comment that  $\widehat{\mathcal{V}_{\vec{k},S}}$ contains many Abelian subalgebras spanned by generators of the form $L_{\vec{n}}$ where $\vec{n} = n \vec{v} $ and $\vec{v} \cdot \vec{k} = 0$, if $\vec{v}$ is on the lattice.

\section{Discussion and Outlook}\label{sec-discussion}

In this work  we elaborated on the main results reported in \cite{Compere:2015mza}. We introduced a consistent phase space for near-horizon { spinning} extremal geometries in four and higher dimensions which we dubbed the NHEG phase space. We identified its symmetries as a direct product of the  $SL(2,\mathbb{R})\times U(1)^{d-3}$ isometries and a class of symmetries that we called symplectic symmetries. The symplectic symmetries form a novel generalized Virasoro algebra which we dubbed the NHEG algebra and denoted as $\nhegalgebra$. The phase space is  generated by diffeomorphisms corresponding to the symplectic symmetries. All elements of the phase space have the same angular momenta and entropy. We will comment below on various aspects of our construction, on the comparison with existing literature and on possible future directions. 

\paragraph{Comments on the NHEG algebra.}
One of our results is the representation of the infinite dimensional NHEG algebra $\nhegalgebra$ \eqref{NHEG-algebra quantized} in the phase space of near-horizon geometries. Its structure constants are specified by the vector $\vec{k}$ obtained from the near-extremal expansion of the black hole angular velocity $\vec{\Omega} = \vec{\Omega}_{ext} + \frac{2 \pi}{\hbar} \vec{k} \, T_H + O(T_H^2)$. The central charge is given by the black hole entropy $S$. 
As discussed, the generators of the isometries $SL(2,\mathbb{R})\times U(1)^{d-3}$ commute with the generators $L_{\vec{n}}$. The total symmetry algebra is therefore a direct product  \eqref{FULL-NHEG-algebra}. Generalized or higher rank Virasoro algebras have been considered in the mathematics literature \cite{Patera91,Mazorchuk98,2006math......7614G} but to our knowledge none of these algebras depends upon a real vector $\vec{k}$. It is desirable to explore further various interesting mathematical aspects of this algebra, including its unitary representations, the corresponding group manifold and its coadjoint orbits. Obtaining a stringy realization of this algebra would also be interesting.

\paragraph{NHEG phase space vs Kerr/CFT.} Our construction shares several features with the original Kerr/CFT proposal \cite{Guica:2008mu}. We both use covariant phase space methods to describe the microscopics of extremal rotating black holes and (at least) a Virasoro algebra appears as a symmetry algebra. However, we would like to emphasize that our results are both conceptually and technically distinct from the Kerr/CFT proposal. 

\begin{enumerate}
\item In four dimensions, we obtained that the symmetry algebra is a direct product $SL(2,\mathbb{R})\times U(1) \times \text{Virasoro}$ while the Kerr/CFT conjectured algebra is $SL(2,\mathbb{R}) \times \text{Virasoro}$.\footnote{In this proposal, there is an obvious tension between requiring the angular momentum to be the Virasoro zero mode (which \emph{does not} commute with the other Virasoro generators) and at the same time the central term in the symmetry algebra (which \emph{does} commute with the other Virasoro generators). We resolve this tension here by identifying an additional $U(1)$ factor.} We obtained that consistency requires the angular momentum $J$ associated with the $U(1)$ isometry to be constant over the phase space. Instead, the Virasoro zero mode $L_0$, associated with the symplectic symmetry $\p_\varphi$, varies over the phase space.

\item As a consequence of invariance under two out of the three generators of $SL(2,\mathbb R)$, the NHEG phase space admits a transitive action which maps any codimension two surface at fixed $t_{\mathcal H},r_{\mathcal H}$ to another such surface at fixed $t,r$. Therefore, surface charges are not only defined at infinity but rather on any sphere $t,r$ in the bulk of spacetime, which leads to the feature that symmetries are symplectic instead of only asymptotic.

\item We explicitly construct the phase space, with a consistent symplectic structure,  and specify the set of smooth metrics. Specifying the phase space in the Kerr/CFT setup has faced various issues, including non-smoothness of the candidate metrics at the poles \cite{Amsel:2009ev,Amsel:2009pu}. We resolve these issues here thanks to the change of symmetry ansatz. While we described the largest symmetric phase space in the main text, we also found that it is consistent to define a phase space which admits only one Virasoro algebra as symmetry algebra in any dimension. We describe the details of this alternative ``Kerr/CFT type'' phase space in appendix \ref{append-Kerr-CFT}. 

\item Our construction in higher dimensions than four is invariant under permutation of the $d-3$ $\varphi^i$ directions. We have provided a democratic treatment of all $U(1)$ directions.

\end{enumerate}

\paragraph{Dynamical ambiguity and central charge.} As our construction shows, the symplectic structure is determined upon the addition of a specific class of boundary terms which might contribute to the central charge. We formulated the existence of such boundary terms as a cohomological problem and  identified a cohomology representative by using specific background structures in the phase space. We then fixed the coefficient in front of this boundary term by requiring that the central charge be identical for a one-parameter (the $b$ parameter) family of symmetry generators. It would  of course be interesting to fully classify this cohomology. Also, one possible more solid way to fix these boundary terms would be to study the boundary terms necessary to obtain a well-defined variational principle and use those to fix the remaining ambiguity in the symplectic structure using the prescription of \cite{Compere:2008us}.

\paragraph{Conserved charges from a Liouville-type stress-tensor.} The phase space is labelled by the periodic wiggle function $F(\vec{\varphi})$ over the $d-3$ dimensional torus which allows defining the periodic function $\Psi$. We showed that the charges defined over the phase space can be expressed in terms of the Fourier modes  of the functional $T[\Psi]$ \eqref{T-tensor} over the torus. The functional $T[{\Psi}]$ has a striking resemblance to (a component of) the energy-momentum tensor of a Liouville field theory. However, there are also major differences since there is no time dependence here and instead there are multidimensional circle directions. 
While the relationship between $3d$ Einstein gravity and Liouville theory is well understood using the Chern-Simons formulation \cite{Coussaert:1995zp}, to our knowledge, it is the first occurrence of a connection between four and higher dimensional gravity and Liouville theory. We also remark that the zero mode of the NHEG algebra $H_{\vec{0}}$  is positive definite over the whole phase space. Therefore, one might be tempted to use $H_{\vec{0}}$  as a defining Hamiltonian for such a Liouville-type theory. It is natural to ask where such a ``holographically dual'' theory would be defined. In that regards, we note that a special role in the construction is played by one null shear-free rotation-free and expansion-free geodesic congruence \cite{Durkee:2010ea} which is kept manifest in the phase space and thereby provides a natural class of null ``holographic screens''.

\paragraph{Diffeomorphism covariance of the phase space.} The phase space that we constructed \eqref{g-F} constitutes a zero-measure set of all metrics diffeomorphic to the background near-horizon geometry \eqref{NHEG-metric}. One may  wonder if there is a physical significance to all other metrics related by diffeomorphisms which are not generated by the symplectic symmetries \eqref{ASK}. In the usual construction of asymptotic boundary conditions, many diffeomorphisms are pure gauge in the sense that they are associated with vanishing asymptotic charges while very large diffeomorphisms are not allowed by the boundary conditions and are associated with infinite charges. Pure gauge transformations do not contain any physics while very large diffeomorphisms are by definition not usually considered. Here since the asymptotics plays no role and we do not strictly impose boundary conditions we propose the alternative following answer. Let us consider two NHEG background metrics related by an arbitrary diffeomorphism $\psi$. If the construction of the phase space is covariant, it will be possible to define a phase space depending upon a wiggle function $F$ for each of these background metrics, and therefore the background metric \eqref{NHEG-metric} and its coordinate system $(t,r,\theta,\vec{\varphi})$ will have no preferred role. The diffeomorphism $\psi$  need not be associated with finite or vanishing charges, or even need not admit an infinitesimal version. The diffeomorphism will just be a map, an isomorphism between the two classical phase spaces which will share an identical functional structure. Most of the steps in our construction are covariant but we did not entirely complete that program, since for example $\eta_2$ defined in \eqref{eta2} does not admit a covariant definition. If the program of defining the phase space in a covariant manner can be completed, it would establish that the phase space is diffeomorphism covariant in the sense above. Note that at the quantum level, the choice of time matters in the definition of quantum states and two phase spaces related by diffeomorphisms may not remain equivalent at quantum level. 

\paragraph{$Z_2$-isometries.}  We mentioned the $t$--$\vec{\varphi}$ and $r$--$\vec{\varphi}$ inversion $Z_2$ isometries of the NHEG background. As is explicitly seen from \eqref{ASK} the phase space generator does not respect these $Z_2$ symmetries. Instead, there is a one-to-one map between the two phase spaces built upon \eqref{NHEG-metric} by the action of $\chi$ with $b = \pm 1$. One may hence ``gauge'' this $Z_2$ by identifying the two phase spaces. The explicit bijection between the two phase spaces is provided in appendix \ref{app Z2}.

\paragraph{Comparison with $3d$ Einstein gravity.} Three dimensional gravity is often considered as a toy model for higher dimensional gravity.  It is instructive to quickly emphasize the similarities and differences between the four and higher dimensional setup and these lower dimensional models.  Specifically for AdS$_3$ Einstein gravity, the most general solution with Brown-Henneaux or Dirichlet boundary conditions \cite{Brown:1986nw} is specified by a ``holomorphic'' and a ``anti-holomorphic'' function, $f_\pm(t\pm \phi)$, where $t,\phi$ are parametrizing  the AdS$_3$ boundary cylinder. The boundary conditions lead to the standard Lee-Wald symplectic structure. Therefore, the set of geometries, nicely summarized by Ba\~nados \cite{Banados:1998gg} constitutes the phase space of AdS$_3$ Einstein gravity with Dirichlet boundary conditions (see also \cite{Sheikh-Jabbari:2014nya} for further analysis). 
Similarly to the geometries analyzed here, one can show that the Brown-Henneaux asymptotic symmetry charges \cite{Brown:1986nw} extend to symplectic charges \cite{Compere:2014cna} and may be formulated in the bulk with the same results for the algebra and central charges. This phase space however does not directly compares to the NHEG phase space considered here, e.g. there is no $SL(2,\mathbb R)$ isometry.

Another class of boundary conditions for AdS$_3$ Einstein gravity exists where the solutions have $SL(2,\mathbb R)$ isometry and are specified with a single ``holomorphic'' function, say $f(t+\phi)$ \cite{Li:2013pra,Sheikh-Jabbari:2014nya}. While Ba\~nados geometries may be viewed as ``descendant geometries'' of the $AdS_3$ vacuum, conical defects and generic BTZ black holes \cite{Banados:1992wn,Banados:1992gq}, these solutions may be viewed as ``descendant geometries'' of the AdS$_3$ self-dual orbifold  \cite{Coussaert:1994tu} which appears in the near-horizon limit of the extremal BTZ black hole. It is therefore the best analogue of a $3$ dimensional NHEG geometry. The relationship between this phase space and the full $AdS_3$ has not been worked out in full details but it has been convincingly argued that the near-horizon limit of extremal geometries will freeze out one chiral Virasoro algebra, say the left-movers, leaving one chiral copy free to vary, the right-movers, which extend the $U(1)$ isometry of the self-dual orbifold \cite{Balasubramanian:2009bg}. It is also expected that the asymptotic symmetries are realized in the bulk as symplectic symmetries with the same Brown-Henneaux central charge. 

In AdS$_3$ gravity, the symmetry algebra of near-horizon extremal geometries is therefore $SL(2,\mathbb R) \times \text{Virasoro}$, in distinction with the higher dimensional case where there are additional $U(1)^{d-3}$ factors. In $3d$ the angular momentum is linearly proportional to the Virasoro zero mode and therefore varies over the phase space. This is qualitatively distinct from the fixed angular momenta which parametrize a higher dimensional NHEG solution. Also, the Virasoro central charge depends upon the theory but does not depend upon the physical parameters of the black hole solution, unlike the higher dimensional case where the entropy, an intrinsic property of the NHEG solution, appears as the central charge. These two features are therefore radically different in $3d$ as compared with higher dimensions. The best map between the NHEG phase space and a $3d$ model, if such a map would be useful, would be to identify the $AdS_3$ scale $\ell$ with the higher dimensional NHEG black hole entropy $S$. One would identify the  $SL(2,\mathbb R) \times \text{Virasoro}$ symmetries between higher $d$ and $d=3$ but the $U(1)^{d-3}$ symmetries with fixed angular momenta would not belong to the $3d$ description. 

A natural question is if, like the AdS$_3$ case, there exists  a bigger algebra which contains the physics before taking the near-horizon limit and/or physics beyond extremality. The AdS$_3$ example, then suggests that such a generalization may require a ``non-chiral'' extension of the NHEG algebra; e.g. by doubling it with left-movers, which is frozen out as a result of extremality and the near-horizon limit. (See \cite{Compere:2014bia} for a step in that direction, e.g. in the case of warped AdS$_3$ { but see also \cite{Baggio:2012db,Castro:2013kea} for limitations of the occurrence of conformal symmetry in an asymptotically flat geometry}.)

\paragraph{Extension to other near-horizon extremal geometries.} In this work 
we focused on the specific example of $d$ dimensional Einstein vacuum solutions with $SL(2,\mathbb{R})\times U(1)^{d-3}$ isometry. More general near horizon geometries exist and we expect our construction to be extendible to any such geometries.  In particular, one may consider the near horizon geometry of the extremal Kerr-Newman solution to the four (or higher dimensional)  Einstein-Maxwell theory, where the symmetries of the solution involves two (or more) $U(1)$'s, one associated with the ``internal'' $U(1)$ of electromagnetism and the rest with Killing isometries. It would be interesting to explicitly explore how this other internal $U(1)$ appears in the NHEG algebra and phase space.

\paragraph{Possible relationship with black hole microstates.} Our main motivation for embarking on this study has been understanding the microstates of extremal black holes. The existence of a large symmetry algebra in near-horizon geometries together with the application of Dirac semi-classical quantization rules, if valid in this case, imply that black hole quantum states, whatever they might be, form a representation of the quantized NHEG algebra $\nhegalgebra$ \eqref{NHEG-algebra quantized}. A stronger statement would be that the low energy description of these microstates is entirely captured by a quantization of the phase space (which might be possible thanks to the existence of a symplectic structure). If such a low energy description is available, $H_{\vec{0}}$ would appear as the natural ``Hamiltonian'' governing the dynamics on this Hilbert space. Alternatively, one might seek for an embedding in string theory. If the supergravity low energy approximation captures a large fraction of the microstates, a possible route would be to build primaries corresponding to the (generalized) Virasoro algebra using classical solitons with non-trivial homological cycles and fluxes by exploiting the loopholes in uniqueness theorems \cite{Gibbons:2013tqa}. Progress in that direction can be found in \cite{Bena:2012wc}.

All the above points discussed here cries for a better understanding and further analysis in these directions are very much needed. We will be exploring them in our future studies. { Anybody} is very welcome to join in this, hopefully fruitful, research.

\section*{Acknowledgments}

GC would like to first thank J. Jottar and P. Mao for sharing earlier work on closely related issues and An Huang for pointing at relevant references. We thank Matthew Headrick for the ``diffgeo" package which was used to enhance some of the Mathematica computations in this work. AS would like to thank Victor Lekeu for a useful discussion on section 2. GC would like to thank the hospitality and feedback of the Center for Mathematical Sciences and Applications as well as the string theory group at Harvard where part of this work was conducted. MMSHJ, AS and KH  would like to thank Allameh Tabatabaii Prize Grant of Boniad Melli Nokhbegan of Iran. MMSHJ, KH and AS would like to thank the ICTP network project NET-68.  G.C. is a Research Associate of the Fonds de la Recherche
Scientifique F.R.S.-FNRS (Belgium) and he acknowledges the current support of the ERC Starting Grant 335146 ``HoloBHC".

\appendix

\section{Generalities on the symplectic structure and charge algebra}\label{appendix:charges}

The construction of symplectic structure,  corresponding surface charges, their algebra and the central charge for diffeomorphic and/or gauge invariant theories has an established framework based on the covariant phase space method. In this appendix, to make our article self-contained, we present a quick review of this framework. Instead of providing { the analysis in the most general case, we concentrate on diffeomorphic invariant theories without additional gauge transformations. We will indicate the explicit expressions for the theory we consider here, namely pure Einstein gravity in generic $d$ dimensions.}

{ {\noindent \bf \emph{Notations.}} We use the standard conventions of \cite{Iyer:1994ys} where boldface symbols are used to denote forms and
\be
(d^{d-p}x)_{\mu_1\cdots \mu_{p}}= \frac{1}{(d-p)! p!} \eps_{\mu_1 \cdots \mu_{p}\nu_{p+1} \cdots \nu_d} dx^{\nu_{p+1}}\wedge \cdots \wedge dx^{\nu_d},
\ee
so that a $d-p$ form is given by $\mathbf{X} = X^{\mu_1 \cdots \mu_p} (d^{d-p}x)_{\mu_1\cdots \mu_{p}}$. Here $\eps_{\mu_1 \cdots \mu_d}$ is the volume-form (it contains $\sqrt{-g}$). We use the conventions of \cite{Barnich:2007bf} for the definition of the variations of fields which imply $\delta_\chi \Phi = \mathcal L_\chi \Phi$ and lead to define the bracket of charges as $\{ H_{\chi_1}, H_{\chi_2}\} = \delta_{\chi_2} H_{\chi_1}$ in order to represent the algebra of symmetry generators with the correct signs. The conventions of \cite{Barnich:2001jy} are opposite in that respect ($\delta_\chi \Phi = -\mathcal L_\chi \Phi$, $\{ H_{\chi_1}, H_{\chi_2}\} = \delta_{\chi_1} H_{\chi_2}$). We use the convention for the overall sign of the surface charges such that the energy of the Schwarzschild black hole is $+M$ with our convention for the orientation, $\eps_{tr\theta\varphi^1 \dots \varphi^{d-3}}=+1$. 
}

\subsection{Symplectic structure}\label{appendix-LW-review}

Let all fields in the theory (including the metric) be collectively denoted as $\Phi$. We assume that all fields are bosonic.  Let the Lagrangian $d$-form be denoted by $\mathbf{ {L}}[\Phi]$.
We define the $d-1$ form presymplectic potential $\boldsymbol \Theta [\d \Phi ,\Phi ]$ via  variation of the Lagrangian
\be
\d \mathbf{{L}} [\Phi]= \mathbf{E}_{\Phi}[\Phi] \d \Phi +  d \boldsymbol \Theta [\d \Phi, \Phi ]\label{deltaL}
\ee
where $\mathbf{E}_{\Phi}[\Phi] = \frac{\delta \cL}{\delta \Phi}$ { are the Euler-Lagrange equations for the fields $\Phi$ and summation on all fields is understood in the first term on the right-hand side.} Here $\delta \Phi$ are Grassmann-even field variations which obey $\delta_1\delta_2\Phi -\delta_2 \delta_1 \Phi = 0$. $\delta$ may be viewed as an exterior derivative operator on the field space while $d$ is the exterior derivative operator on the spacetime. The operator $\de$ commutes with the total derivative operator $ d$. The presymplectic potential $\boldsymbol\Theta$ is hence a $d-1$-form over the spacetime and a one form over the field space. 

The general solution of $\bTheta$ in \eqref{deltaL} has the following form:
\bea \label{omega LW}
\boldsymbol \Theta[\delta \Phi,\Phi]  = \boldsymbol \Theta^{ref}[\delta \Phi,\Phi]  + d \mathbf{Y}[\delta \Phi,\Phi] \label{pres}
\eea
where 
 $\boldsymbol \Theta^{ref}$ is defined by the standard algorithm, which consists in integrating  by parts the variation of the Lagrangian or, more formally, by acting on the Lagrangian with Anderson's homotopy operator $\boldsymbol I^d_{\delta \Phi}$  \cite{Barnich:2001jy,Barnich:2000zw,Barnich:2007bf}, defined for second order theories as
\bea
\boldsymbol \Theta^{ref}=\boldsymbol I^d_{\delta \Phi}  \boldsymbol{\mathcal{L}} \;,\;\; \;\;\;\;\; \boldsymbol I^d_{\delta \Phi}  \equiv \left( \delta \Phi \frac{\p}{\Phi_{\, ,\mu}} - \delta \Phi \p_\nu \frac{\p}{\Phi_{\, ,\nu\mu}} \right) \frac{\p}{\p (d x^\mu)} .\label{homot}
\eea
No universal method exists (so far) to determine $\mathbf{Y}[\delta \Phi,\Phi] $. Instead, a case by case analysis is necessary to fix this ambiguity depending upon the physical problem.

The Lee-Wald presymplectic current $d-1$ form $ \boldsymbol \om [\d_1 \Phi, \d_2 \Phi ,\Phi  ]$ is defined as the antisymmetrized variation  of the presymplectic potential  \cite{Lee:1990nz}
\be\label{LW}
{\boldsymbol \omega}_{(LW)}[\delta_1 \Phi ,\delta_2 \Phi,\Phi ] = \delta_1 {\boldsymbol  \Theta}[ \delta_2 \Phi,\Phi ] - \delta_2 {\boldsymbol \Theta}[\delta_1 \Phi, \Phi ] .
\ee
Under \eqref{omega LW} we find
	\bea
	{\boldsymbol \omega}[ \delta_1 \Phi ,\delta_2 \Phi, \Phi ] = {\boldsymbol \omega}^{ref} [ \delta_1 \Phi ,\delta_2 \Phi,\Phi ] +  d \left( \delta_1 {\mathbf{Y}}[ \delta_2 \Phi,\Phi] - \delta_2 {\mathbf{Y}}[\delta_1 \Phi,\Phi ]\right).
	\eea

The symplectic form contracted with two  vectors $\de_1 \Phi,\de_2 \Phi$ on the tangent space of the phase space is defined as
\begin{align}\label{symplectic form def}
\Omega_{AB}(\Phi)(\de_1\Phi)^A (\de_2 \Phi)^B=\int_\Sigma \boldsymbol{\omega} [\de_1 \Phi,\de_2 \Phi,\Phi]
\end{align}
where the integral is defined over a spacelike surface $\Sigma$. Since the presymplectic form is conserved on-shell $d\boldsymbol{\omega} \approx 0$ the symplectic form does not depend upon continuous deformations of the surface $\Sigma$ when its boundaries are fixed.
	
Physically, we require that the symplectic structure be finite and conserved upon deforming $\Sigma$ including at the boundary. This implies that there is no symplectic flux at the spatial boundary of the spacetime.	
	
	\subsection{Gauge transformations and associated surface charges}
	
	Let $\delta_\eps \Phi$ denote an infinitesimal gauge transformation of the fields. For gravitational theories, $\eps$ is a vector field $\chi$ which generates an infinitesimal diffeomorphism. For all generally covariant fields $\Phi$ we have therefore $\delta_\chi \Phi=\mathcal{L}_\chi \Phi$, the Lie derivative of fields with respect to $\chi$.
	
	The Noether-Wald current for a diffeomorphism $\chi$ is defined as \cite{Iyer:1994ys}
	\begin{align}
	\mathbf{J}_\chi&=\bTheta[\de_\chi\Phi,\Phi ]-\chi\cdot \mathbf{L}
	\end{align}
	One can show that $d \mathbf{J}_\chi$ vanishes on-shell and therefore 
	\begin{align}
	\mathbf{J}_\chi&=d\mathbf{Q}_\chi
	\end{align}
	where the $d-2$ form $\mathbf{Q}_\chi$ is the Noether charge density associated to $\chi$. We define $ \mathbf{Q}^{ref}_\chi[\Phi ]$ up to a total derivative from $d  \mathbf{Q}^{ref}_\chi[\Phi ] = \bTheta^{ref}[\de_\chi\Phi]-\chi\cdot \mathbf{L}$.
	
	The fundamental identity of the covariant phase space formalism is the following. The proof can be found in several references; e.g. \cite{Wald:1993nt,Barnich:2001jy,Wald:1999wa}.
	\begin{theorem}

	If the presymplectic form is contracted with a gauge transformation $\de_\chi\Phi$, there is a unique (up to a total derivative) $d-2$ form $ {\boldsymbol k}_\chi [\delta\Phi,\Phi]$ satisfying the following identity 
	\bea
	\bomega [\delta \Phi , \delta_\chi \Phi,\Phi ] = d \,{\boldsymbol k}_\chi [\delta\Phi,\Phi]\label{eq:f}
	\eea
	provided that the fields $\Phi$ satisfy the equations of motion and the field variations $\delta\Phi$ satisfy the linearized equations of motion around $\Phi$. The form of ${\boldsymbol k}_\chi [\delta\Phi,\Phi]$ is given by 
	\begin{align}\label{charge variation identity}
		{\boldsymbol k}_\chi [ \delta\Phi,\Phi ] = \delta \mathbf{Q}_\chi[\Phi ] - \chi \cdot \bm{\Theta} [\delta\Phi,\Phi ]+ d(\cdot) .
	\end{align}
	\end{theorem}	
{ Here the notation $d(\cdot)$ refers to possible boundary terms which cancel upon integration over a closed surface.} 
	 The surface charge is explicitly given by
	\bea
	{\boldsymbol k}_\chi [ \delta\Phi,\Phi ] &=&\delta \mathbf{Q}^{ref}_\chi[\Phi ] - \chi \cdot \bm{\Theta}^{ref} [\delta\Phi,\Phi ]  +\delta \mathbf{Y}[\delta_\chi \Phi,\Phi ] - \delta_\chi \mathbf{Y} [\delta \Phi,\Phi ] . \label{sc}
	\eea
One can define the associated infinitesimal surface charge  on a closed surface $\mathcal{H}$ as
	\bea
	\slash\hspace{-6pt}\delta H_\chi &=& \oint_\mathcal{H} {\boldsymbol k}_\chi [\delta \Phi,\Phi]  .\label{chargeexact}
	\eea
	
	There are 3 standard physical requirements on this surface charge: (1) It should be finite; (2) it should be conserved upon shifting $\mathcal{H}$ along time; (3) it should also be integrable in the sense that $\delta (\slash\hspace{-6pt} \delta H_\chi)  =0 $ as we detail below.
	
	 For Einstein theory which is the context of this paper, 
	\be
		\mathbf{L}_{Einstein} =\frac{1}{16 \pi G} R \boldsymbol{\eps} , 
	 \ee
	 and
\begin{align}
	(16\pi G) \Theta_{ref}^{\,\mu} &= \nabla_\nu h^{\nu\mu} - \nabla^\mu h ,\qquad
	(16\pi G)Q^{\mu\nu}_\chi   = -\nabla^\mu \chi^\nu + \nabla^\nu \chi^\nu, 
\end{align}	
{ where we denoted $h_{\mu\nu} \equiv \delta g_{\mu\nu}$, $h^{\mu\nu}=g^{\mu\alpha}h_{\alpha\beta}g^{\beta \nu}$, $h=g^{\mu\nu}h_{\mu\nu}$. } Therefore
	\begin{align}\label{kgrav}
 \hspace*{-8mm}{\boldsymbol k}_\chi^{Einstein}&\equiv\delta \mathbf{Q}_\eps[\Phi ] - \chi \cdot \bm{\Theta}^{ref} [\delta\Phi,\Phi ]\cr
&=\dfrac{1}{8 \pi G} (d^{d-2}x)_{\mu \nu}\left(\chi^\nu\nabla^\mu h
-\chi^\nu\nabla_\sigma h^{\mu\sigma}
+\chi_\sigma\nabla^{\nu}h^{\mu\sigma}
+\frac{1}{2}h\nabla^{\nu} \chi^{\mu}
-h^{\rho\nu}\nabla_\rho\chi^{\mu}\right). 
	\end{align}

	\subsection{Integrability condition}\label{app integrability}

	The integrability condition is 
	\bea
	\mathcal I [\delta_1\Phi,\delta_2\Phi,\Phi] \equiv \delta_1 \oint  {\boldsymbol k}_{\chi}[\delta_2 \Phi ; \Phi ] - (1 \leftrightarrow 2) = 0\label{integrability Wald}
	\eea
	for all variations $\delta\Phi$ on the phase space and for an arbitrary point in the phase space $\Phi$. If the integrability condition holds at any point $\Phi$ of the phase space, then $\oint  {\boldsymbol k}_{\chi}[\delta \Phi ; \Phi ]$ is an exact variation. In other words, there exist a function $H_\chi$ on phase space satisfying 
	\begin{align}
	\de H_\chi&=\oint  {\boldsymbol k}_{\chi}[\delta \Phi ; \Phi ].
	\end{align} 
	The function $H_\chi$ is the canonical charge corresponding to the gauge transformation along $\chi$ which is the generator of this transformation through the Poisson bracket
	\begin{align}
	\{H_\chi,f\}&=\de_\chi f.
	\end{align}
	
	To compute $H_\chi$, one can choose any path $\gamma$ in the phase space between a reference configuration $\bar\Phi$ (which can be the background field configuration) and the field of interest $\Phi$ and define the canonical  charge associated with any transformation of the phase space as
	\bea
	H_\chi [\Phi,\bar\Phi]= \int_\gamma \oint {\boldsymbol k}_\chi[d \Phi,\Phi ] + N_\chi[\bar\Phi]
	\eea
	where $d\Phi$ is a phase space variation one-form which is integrated along the path $\gamma$. Here, $N_\chi[\bar\Phi]$ is the freely chosen charge of the reference configuration $\bar\Phi$.\footnote{In the covariant phase space formalism, this reference charge is arbitrary. If a holographic renormalization scheme exists, one would be able to define this reference charge from the first principles, as it is done e.g. in asymptotically AdS spacetimes.} 
	Using \eqref{charge variation identity} and the fact that $\de \mathbf{Q}_\chi$ is an exact variation, we find the simple integrability condition,
	\begin{align}\label{LW-integrability}
		\mathcal I [\delta_1\Phi,\delta_2\Phi,\Phi]&\equiv  - \oint  \chi \cdot \bomega[\delta_1 \Phi, \delta_2 \Phi,\Phi] = 0,
	\end{align}
	for arbitrary variations $\delta_1\Phi, \delta_2\Phi$ and for the $\chi$ of interest.

\subsection{Algebra of gauge transformations}

Given two diffeomorphism generators $\chi_1$, $\chi_2$ one can define the Lie bracket $[\chi_1 ,\chi_2]_{L.B.}$ which define a natural algebra among the gauge parameters $\chi_1,\chi_2$. 
For field-independent diffeomorphism generators, the algebra of field variations is isomorphic to the Lie bracket algebra, up to an overall sign,
	\bea
	[ \delta_{\chi_1}, \delta_{\chi_2}]  = - [\mathcal L_{\chi_1},\mathcal L_{\chi_2}]= -\mathcal L_{[\chi_1 , \chi_2]_{L.B.}}= -\delta_{[\chi_1 , \chi_2]_{L.B.}}. \label{alg}
	\eea
	The first minus sign comes from $$\delta_{\chi_1}\delta_{\chi_2}g_{\mu\nu}\equiv (\mathcal L_{\chi_1}g_{\alpha\beta}\frac{\p}{\p g_{\alpha\beta}}+\p_\gamma \mathcal L_{\chi_1}g_{\alpha\beta} \frac{\p}{\p \p_\gamma g_{\alpha\beta}})\mathcal L_{\chi_2}g_{\mu\nu} = \mathcal L_{\chi_2}\mathcal L_{\chi_1} g_{\mu\nu},$$
	and similarly for other fields $\Phi$. 
	
	Now, for generators $\chi_1[\Phi],\, \chi_2[\Phi]$ whose components depend upon the fields $\Phi$, the field variations $\delta$ also act on the field dependence of the generators themselves. Therefore we instead have 
	\bea
	[ \delta_{\chi_1}, \delta_{\chi_2}] = -[\mathcal L_{\chi_1},\mathcal L_{\chi_2}] + \delta_{\delta^\Phi_{\chi_1} \chi_2 -\delta^\Phi_{\chi_2} \chi_1} =-  \delta_{[\chi_1 , \chi_2]}\label{totalbracket0}
	\eea
	where we emphasize that $\delta$ acts on the fields with a superscript $\delta^\Phi$ and the total bracket is
	\bea
	[\chi_1 , \chi_2] = [\chi_1 , \chi_2]_{L.B.}- \delta^\Phi_{\chi_1} \chi_2 +\delta^\Phi_{\chi_2} \chi_1.\label{totalbracket}
	\eea
	The total bracket is the one that appears in the representation theorem for the charges, see the next section. It appeared previously e.g. in \cite{Barnich:2010eb,Barnich:2010xq,Compere:2014cna}.

\subsection{Charge algebra}
	\label{appendix charge algebra}
	
	We define the bracket between two charges as
	\bea
	\{ H_\chi, H_\xi \} \equiv  \delta_\xi H_\chi =  \oint {\boldsymbol k}_\chi[\delta_\xi \Phi ,\Phi ] .\label{PB}
	\eea

	One can obtain the charge algebra as follows: add and subtract two background terms to obtain,
	\bea
	\{ H_\chi, H_\xi \} &=&  \oint {\boldsymbol k}_\chi[\delta_\xi \Phi ,\Phi ]  -  \oint {\boldsymbol k}_\chi[\delta_\xi \bar \Phi ,\bar\Phi ]  +  \oint {\boldsymbol k}_\chi[\delta_\xi \bar\Phi ,\bar\Phi ] \\
	&=& \int_\gamma  \oint d {\boldsymbol k}_\chi[\delta_\xi \Phi , \Phi ]  +N_{[\chi,\xi]}[\bar\Phi] + \mathcal K_{\chi,\xi}[\bar\Phi]
	\eea
	where
	\bea\label{central extension}
	\mathcal K_{\chi,\xi}[\bar\Phi] &=&  \oint {\boldsymbol k}_\chi[\delta_\xi \bar\Phi ,\bar\Phi ] -N_{[\chi,\xi]}[\bar\Phi]
	\eea
	is the central term. The second part of the central extension \eqref{central extension} is trivial in the sense that it can be absorbed by a shift of the charges of the reference solution {(which is usually fixed using additional physical considerations)}. It was proven in \cite{Barnich:2007bf} that integrability of charges \eqref{integrability Wald} implies that
	\bea
	\oint d {\boldsymbol k}_\chi[\delta_\xi \Phi ,\Phi ]  = \oint \delta_\xi {\boldsymbol k}_\chi[ d  \Phi ,\Phi ]    =   \oint {\boldsymbol k}_{[\chi, \xi]} [d  \Phi ,\Phi ]   
	\eea
	for solutions $\Phi$ and linearized solutions $d\Phi$ and where the bracket is defined in \eqref{totalbracket0}-\eqref{totalbracket}. 
	Therefore, one gets the algebra
	\bea
	\{ H_\chi, H_\xi \} &=& H_{[\chi,\xi]}[\Phi,\bar\Phi]    + \mathcal K_{\chi,\xi}[\bar\Phi].\label{alg2a}
	\eea
	One can also prove that $ \mathcal K_{\chi,\xi}[\bar\Phi] = - \mathcal K_{\xi,\chi}[\bar\Phi]$ and
	\bea
	\mathcal K_{[\chi_1,\chi_2],\xi}[\bar\Phi] +  \mathcal K_{[\xi,\chi_1],\chi_2}[\bar\Phi]+ \mathcal K_{[\chi_2,\xi],\chi_1}[\bar\Phi]=0.\label{2co}
	\eea
Therefore the Jacobi identity is satisfied by the centrally extended charge algebra, which is a central extension of the algebra of corresponding diffeomorphisms \eqref{totalbracket0}.

\paragraph{On trivial $\mathbf{Y}$ terms.}

The contribution of the $\mathbf{Y}$ terms to the surface charge $\boldsymbol k_\chi[\delta_\xi \Phi,\Phi]$ is given by
\bea
\boldsymbol k^Y_\chi[\delta_\xi \Phi,\Phi] = \delta_\xi \mathbf{Y}[\delta_\chi \Phi,\Phi] - \delta_\chi \mathbf{Y}[\delta_\xi \Phi,\Phi] - \mathbf{Y}[\delta_{[\chi,\xi]}\Phi,\Phi]
\eea
after carefully commuting $\delta$ with the operator which contracts $\delta \Phi$ with $\delta_\xi \Phi$. Therefore, for $\mathbf{Y}$ of the form $\mathbf{Y} = \delta \mathbf{Z}[\Phi]$ this contribution is zero as a consequence of the algebra \eqref{totalbracket0}. This implies that such $\mathbf{Z}$ terms do not contribute to the central extension and to the bracket of charges. Therefore, from the charge algebra it does not contribute to $H_{[\chi,\xi]}$ and if the Lie bracket of vector fields is surjective in the space of vectors fields associated with non-trivial charges, as it is the case in this paper, it will not contribute to any charges. Terms of the form  $\mathbf{Y} = d \tilde{\mathbf{Z}}$ {with $\tilde{\mathbf{Z}}$ regular will also not contribute since the integral of such terms on a closed surface are zero}.

\section{Details of Calculations and Proofs}\label{Appendix-proof-details}

Some of the computational details in the construction of our generators $\chi$ and the corresponding charges $H_{\vec{n}}$ are given in this appendix. 

\subsection{Consequences of $\bar\xi_{-,0}$ symmetry of field perturbations}\label{theorem xi12}

Let  $\bar\Phi$ denote the NHEG background \eqref{NHEG-metric} and $\mathcal{A}$ the algebra of background isometries  $sl(2,\mathbb{R})\times u(1)^{d-3}$. 
For notational convenience, we will drop all bars on vector fields in this appendix but it is understood that we are considering generators of diffeomorphisms around the background. 
First, we note 
\be
\mathcal{L}_{\xi_{-,0}}\delta_{\chi} \bar{\Phi}=\mathcal{L}_{\xi_{-,0}}{\mathcal L}_{\chi} \bar{\Phi} = \mathcal{L}_{[\xi_{-,0},\chi]}\bar{\Phi},
\ee
since $\xi_{-,0}$ are Killing vectors of the background. Requiring $\mathcal{L}_{\xi_{-,0}}\delta_{\chi} \bar{\Phi} = 0$ is therefore equivalent to requiring that $[\chi,\xi_{-,0}] \in \mathcal A$.

\paragraph{Proposition.} \emph{ The only vectors $\chi$ for which $[\chi,\xi_{-,0}]\in \mathcal{A}$ are linear combination of members of the sl$(2,\mathbb{R})$ algebra and the ones for which $[\chi,\xi_{-}]=0, [\chi,\xi_{0}]=\beta^i {{\mathrm m}_i}$ with $\beta^i$ fixed constants. 
}
\begin{proof}
$[\chi,\xi_{-,0}]\in \mathcal{A}$ means that
\begin{equation}\label{Sym of pert 2}
\begin{split}
[\chi,\xi_{-}]&=\alpha^1\xi_-+\alpha^2\xi_0+\alpha^3\xi_++\alpha^i \mathrm{m}_i,\\
[\chi,\xi_{0}]&=\beta^1\xi_-+\beta^2\xi_0+\beta^3\xi_++\beta^i \mathrm{m}_i,
\end{split}\end{equation}
for some constants $\alpha$ and $\beta$'s. By the Jacobi identity we have:
\begin{equation}
[[\chi,\xi_-],\xi_0]+[[\xi_0,\chi],\xi_-]+[[\xi_-,\xi_0],\chi]=0.
\end{equation}
Inserting \eqref{Sym of pert 2}  in the above equation, and using the algebra of Killings of NHEG, we get
\begin{equation}
(\alpha^1\xi_--\alpha^3\xi_+)+(\beta^2\xi_-+\beta^3\xi_0)-(\alpha^1\xi_-+\alpha^2\xi_0+\alpha^3\xi_++\alpha^i \mathrm{m}_i)=0.
\end{equation}
Noting that the above should identically vanish, coefficients of $\xi_a$ and $\mathrm{m}_i$ all should be set to zero: 
\begin{equation}
\alpha^3=\alpha^i=\beta^2=0, \qquad \qquad \alpha^2=\beta^3,
\end{equation}
and hence 
\begin{equation}\label{Sym of pert 4}
[\chi,\xi_{-}]=\alpha^1\xi_-+\alpha^2\xi_0\,,\qquad  [\chi,\xi_{0}]=\beta^1\xi_-+\alpha^2\xi_++\beta^i \mathrm{m}_i.
\end{equation}

Using the redefinition
\begin{equation}\label{Sym of pert 9}
\chi'\equiv\chi+\alpha^1\xi_0+\alpha^2\xi_+-\beta^1\xi_-
\end{equation}
then
\begin{align}
[\chi',\xi_{-}]&=0,\label{Sym of pert 6}\\
[\chi',\xi_{0}]&=\beta^i \mathrm{m}_i, \label{Sym of pert 7}
\end{align}
{Therefore,  recalling \eqref{Sym of pert 9}, we have proved that $\chi$ is a linear combination of $\chi'$ with properties \eqref{Sym of pert 6}-\eqref{Sym of pert 7}, and a member of sl$(2,\mathbb{R})$, namely $-\alpha^1\xi_0-\alpha^2\xi_++\beta^1\xi_-$.

It is useful for clarifying the requirement \textbf{(1)} in section \ref{Construction of chi} to find the generic components of $\chi'$ explicitly. }
\eqref{Sym of pert 6} is just $\partial_t \chi'^{\mu}=0$. It means that $\chi'=\chi^\mu\partial_\mu$ where $\chi'^\mu=\chi'^\mu(r,\theta,\varphi^i)$. Inserting it in \eqref{Sym of pert 7}, leads to the following equations:
\begin{equation}\label{Sym of pert 8}
\begin{cases}
(r\partial_r+1)\chi'^t=0\\
(r\partial_r-1)\chi'^r=0\\
r\partial_r\chi'^\theta=0\\
r\partial_r\chi'^{\varphi^i}=\beta^i
\end{cases}
\end{equation}
The above equations fix the $r$ dependence of the $\chi'^\mu$ as follows
\begin{equation}\label{chi-1}
\chi'=\frac{\epsilon^{t}}{r}\partial_t+ r\epsilon^{ r} \partial _r +  \epsilon^\theta \partial _ {\theta}+ (\beta^i\ln r+\epsilon^i)  \partial_{\varphi^i}\, ,
\end{equation}
where $\epsilon^\mu=\epsilon^\mu(\theta,\varphi^i)$. 
\end{proof}

\subsection{$Z_2$ transformations as bijections between $b=\pm 1$ phase spaces}\label{app Z2}
The NHEG background \eqref{NHEG-metric} is manifestly invariant under the two $Z_2$ transformations: $(r\to -r, \vec{\varphi} \to -\vec{\varphi})$ or $(t\to -t, \vec{\varphi} \to -\vec{\varphi})$.
{In section \ref{Construction of chi}, two families of vector fields were distinguished as generators for the NHEG phase space:}
\begin{equation}
\chi_\pm[\epsilon(\vec{\varphi})]=-\vec{k}\cdot \vec{\partial}_\varphi \epsilon (\frac{b }{r}\partial_t+r\partial_r)+\epsilon \vec{k}\cdot\vec{\partial}_\varphi, \qquad b=\pm 1.
\end{equation}
Let us denote the phase spaces generated by $\chi_\pm$ as $\mathcal{G}_\pm [F]$. Here we show that
\begin{center}
\emph{{The two $Z_2$ transformations} maps $\mathcal{G}_+ [F]$ and $\mathcal{G}_- [F]$ onto each other.}
\end{center}
\begin{proof}
The background is mapped to itself under { any of the two} $Z_2$ transformations. The $\chi_+[\epsilon]$ is mapped to the $\chi_-[\tilde{\epsilon}]$ in which:
\begin{equation}
\tilde{\epsilon}(\vec{\varphi})=-\epsilon(-\vec{\varphi})
\end{equation}
This map provides the bijection relation:
\begin{equation}
\mathcal{G}_+[F(\vec{\varphi})]\leftrightarrow \mathcal{G}_-[-F(-\vec{\varphi})]
\end{equation}
\end{proof}

\subsection{Proof of \eqref{Psi-def}}
\label{proof1}

The transformations \eqref{finite ansatz} imply that
\begin{align}\label{basis-change}
\pd_{t}&=\pd_{\,\bar{t}},\cr
\pd_{r}&=e^{-{\Psi}}\pd_{\bar{r}}+\dfrac{b}{r^2}(e^{\Psi}-1)\pd_{\bar{t}},\\
\pd_{\varphi^i}&=-\pd_{\varphi^i}(e^{\Psi})\;(\,\dfrac{b}{r}\, \pd_{\,\bar{t}} +e^{-2\Psi}r\pd_{\,\bar{r}})+(\delta^j_{\;i}+k^j\pd_{\varphi^i} F)\pd_{\bar{\varphi}^j}.\nonumber
\end{align}
Therefore, we have
\be\begin{split}\label{x-barx-deriv-trans}
\vec{k}\cdot\vec{\pd}_\varphi&=-\vec{k}\cdot\vec{\pd}_\varphi {\Psi}(\dfrac{b}{\bar{r}}\pd_{\,\bar{t}}+\bar{r}\pd_{\,\bar{r}})+(1+X)\vec{k}\cdot\vec{\pd}_{\bar{\varphi}},\\
r\pd_{r}&=\bar{r}\pd_{\,\bar{r}}+\dfrac{b}{r}(e^{\Psi}-1)\pd_{\,\bar{t}},
\end{split}
\ee
where $X({x})\equiv \vec{k}\cdot \vec{\pd}_\varphi F(\vec{\varphi})$. We now start from the LHS of \eqref{chi-vs-chi'}:
\begin{align}
\chi[{\epsilon}(\vec{\varphi})]&={\epsilon}(\vec{\varphi})\vec{k}\cdot\vec{\pd}_\varphi-\vec{k}\cdot\vec{\pd}_\varphi{\epsilon}(\dfrac{b}{r}\pd_{t}+r\pd_{r})\cr
&={\epsilon}(1+X)\vec{k}\cdot\vec{\pd}_{\bar{\varphi}} -(\dfrac{b}{\bar{r}}\pd_{\,\bar{t}}+\bar{r}\pd_{\,\bar{r}})\Big(\;{\epsilon}\ \vec{k}\cdot\vec{\pd}_\varphi\nonumber {\Psi}+\vec{k}\cdot\vec{\pd}_\varphi{\epsilon}\Big).
\end{align}
Defining
\begin{align}\label{epsilon}
\bar\epsilon(\vec{\bar\varphi})\equiv (1+X)\ {\epsilon},
\end{align}
if we can find ${\Psi}$ such that
\begin{align}\label{Z-equation}
\vec{k}\cdot\vec{\pd}_{\bar{\varphi}} \bar\epsilon&={\epsilon}\ \vec{k}\cdot\vec{\pd}_\varphi {\Psi}+\vec{k}\cdot\vec{\pd}_\varphi\ {\epsilon},
\end{align}
we would obtain the desired result
\begin{align}
\bar\chi[\bar{\epsilon}(\vec{\varphi})]&=\bar{\epsilon}(\vec{\bar{\varphi}})\vec{k}\cdot\vec{\pd}_{\bar{\varphi}}-\vec{k}\cdot\vec{\pd}_{\bar{\varphi}}{\bar\epsilon}\;(\dfrac{b}{\bar{r}}\pd_{\,\bar{t}}+\bar{r}\pd_{\,\bar{r}}).
\end{align}

To solve \eqref{Z-equation}, we use the fact that, when dealing with functions of  $\vec{\varphi}$ only,
\begin{align}
\vec{k}\cdot\vec{\pd}_{\bar{\varphi}}&=\dfrac{\vec{k}\cdot\vec{\pd}_\varphi}{1+X}\,.
\end{align}
which is a result of \eqref{x-barx-deriv-trans}. Therefore,
\begin{align}
\vec{k}\cdot\vec{\pd}_{\bar{\varphi}} \bar\epsilon&=\dfrac{\vec{k}\cdot\vec{\pd}_\varphi}{1+X}({\epsilon}(1+X))\cr
&=\dfrac{{\epsilon}\vec{k}\cdot\vec{\pd}_\varphi X}{1+X}+\vec{k}\cdot\vec{\pd}_\varphi{\epsilon}.
\end{align}
Comparison with the RHS of \eqref{Z-equation} then implies ${\Psi}=\ln (1+X)$ or \eqref{Psi-def}.
So we have established our ansatz \eqref{finite ansatz} which defines a one-function family of finite coordinate transformations, specified by the function $F(\vec{\varphi})$.

\subsection{Special vector fields $\eta_\pm$}\label{Appendix-eta}

Two special vectors fields are singled out in our construction: 
\begin{equation}
\overline \eta_+=\frac{1}{\bar r}\partial_{\bar t}+\bar r\partial_{\bar r},\qquad \overline \eta_-=\frac{1}{\bar r}\partial_{\bar t}-\bar r\partial_{\bar r}. 
\end{equation}
They obey the commutation relation 
\begin{equation}
[\overline \eta_+,\overline \eta_-]=-(\overline \eta_++\overline \eta_-).
\end{equation}
and therefore they form a closed algebra under the Lie bracket. Here are some of their properties:
\begin{enumerate}
\item Although not Killing vectors, $\overline  \eta_\pm$ commute with $\overline \xi_-,\overline \xi_0$ and the $U(1)^{d-3}$ generators $\overline{ \mathrm{m}}_i$ \eqref{U(1)-generators}.
\item They commute with the respective symmetry generator $\overline \chi_\pm$; i.e.
\begin{equation}\label{commute eta}
\delta_{\overline \eta_+}\overline \chi_+=[\overline \eta_+,\overline \chi_+]=0\,, \qquad \delta_{\overline \eta_-}\overline \chi_-=[\overline \eta_-,\overline \chi_-]=0,
\end{equation}
where $\overline \chi_\pm$ correspond to the choice of $\overline \chi_b$ with $b=\pm1$.
\item As $\overline \eta_+$ commutes with the phase space generating diffeomorphism $\overline \chi_+$, it is invariant in the phase space generated by $\overline \chi_+$ { which in turn implies
\be
\bar\eta_+ =\eta_+ =\frac{1}{r}\p_t+ r\p_r.
\ee
The above may be explicitly checked using \eqref{basis-change}. The same property holds with minuses in the respective phase space. } 

\item  $\eta_\pm$, similarly to $\chi_{\pm}$, are mapped to each other by the $t$--$\vec{\varphi}$ or $r$--$\vec{\varphi}$ $Z_2$-transformations discussed in section \ref{sec-NHEG-review}.  
\end{enumerate} 
One can in fact show that $\eta_+$ (or $\eta_-$) are the only vectors with properties 1. and 2. in the above list. Properties 3. and 4. then follow from the first two.

\subsection{Explicit computation of the surface charges}\label{appendix  charge computation}

{ Here we give the explicit computation of charges $H_{\vec{n}}$ over the phase space as a function of ${\Psi}$. We derived all expressions in dimensions $d=4$ and $d=5$ which allowed us to infer the general expressions for any $d$. As outlined in the appendix \ref{appendix:charges} the charges are defined through an integration of the infinitesimal surface charge over the phase space which we can compute in principle. Since we know that the charges are integrable, we are allowed to use the symmetry algebra to simplify the derivation of the charges. We present such a simpler derivation below.} Explicitly, we start from \eqref{NHEG-algebra} which implies 
\begin{align}
\{H_{\vec{n}}, H_{\vec{0}}\} =-i (\vec{k} \cdot \vec{n}) H_{\vec{n}}\,.
\end{align}
However, recalling \eqref{H-commutator}, we have
\begin{align}
\{H_{\vec{n}}, H_{\vec{0}}\}&=\de_{\vec{0}} H_{\vec{n}}= \oint {\boldsymbol k}_{\chi_{\vec{n}}}[\de_{\chi_{\vec{0}}}\Phi,\Phi]\,.
\end{align}
Therefore,
\begin{align}\label{H_n}
H_{\vec{n}}&= \dfrac{i}{\vec{k} \cdot \vec{n}}\oint {\boldsymbol k}_{\chi_{\vec{n}}}[\de_{\chi_{\vec{0}}}\Phi,\Phi], \hspace{1cm}\vec{n}\neq \vec{0}.
\end{align}
Using \eqref{NHEG-algebra} we can also obtain $H_{\vec{0}}$ from
\begin{align}
H_{\vec{0}}&=\dfrac{1}{2\vec{k} \cdot \vec{n}}\left(i\{H_{\vec{n}}, H_{-\vec{n}}\}-\frac{A_{\mathcal H}}{8 \pi G}(\vec{k}\cdot \vec{n})^3 \right).
\end{align}
In order to determine $H_{\vec{n}}$ in \eqref{H_n}, we need to calculate ${\boldsymbol k}_{\chi_{\vec{n}}}[\de_{\chi_{\vec{0}}}\Phi,\Phi]$. The result is
{\small	\begin{align}\label{Hn charge k term}
 \Big({\boldsymbol k}^{Einstein}_{\chi_{\vec{n}}}[\delta_{\chi_{\vec{0}}}\Phi,\Phi ]\Big)_{\theta\varphi^1\dots\varphi^n}=&\frac{-\sqrt{-g}\,   e^{-i \vec{n}\cdot\vec{\varphi} }}{16 \pi G \Gamma } \Bigg[2k^ik^j\gamma_{ij}\Big(e^\Psi(i\vec{k}\cdot\vec{n}\Psi'-\Psi'^2-\Psi'')\Big)+\Big(i\vec{k}\cdot\vec{n}\Psi''-\Psi'''\Big)\nonumber\\
  &+\Big(e^\Psi(\Psi'^2+\Psi''-i\vec{k}\cdot\vec{n}\Psi')\Big)+2k^ik^j\gamma_{ij}\Big(\Psi''\Psi'-
  i\vec{k}\cdot\vec{n}\Psi''+e^{2\Psi}\Psi'\Big)\Bigg],\nonumber
	\end{align}}
{\small \begin{align}
\Big({\boldsymbol k}^{\mathbf{Y}}_{\chi_{\vec{n}}}[\delta_{\chi_{\vec{0}}}\Phi,\Phi ]\Big)_{\theta\varphi^1\dots\varphi^n}=&\frac{\sqrt{-g}\,i\vec{k}\cdot\vec{n} (k^ik^j\gamma_{ij}-1) e^{-i\vec{n}\cdot\vec{\varphi}}}{16 \pi G \Gamma} \Bigg[\Psi'^2-2\Psi''+(\Psi''-i\vec{k}\cdot\vec{n}\Psi')\Bigg],
	\end{align}}
where prime denotes the directional derivative $\vec{k}\cdot \vec{\partial}$. The first three parenthesis in ${\boldsymbol k}^{Einstein}$ and the last one in ${\boldsymbol k}^{\mathbf{Y}}$ are total derivatives in $\vec{\varphi}$. They are explicitly proportional to  $(\Psi' e^{\Psi-i \vec{n}\cdot\vec{\varphi} })'$, $(\Psi''e^{-i \vec{n}\cdot\vec{\varphi} })'$, $(\Psi' e^{\Psi-i \vec{n}\cdot\vec{\varphi} })'$ and $(\Psi'e^{-i \vec{n}\cdot\vec{\varphi} })'$. Therefore their integration vanishes. Now considering the identity $\int d\theta \sqrt{-g}\frac{k^ik^j\gamma_{ij}}{\Gamma}=2 \int d\theta \frac{\sqrt{-g}}{\Gamma}$, we  have
\begin{align}
H_{\vec{n}}&= \dfrac{i}{\vec{k}\cdot\vec{n}}\oint {\boldsymbol k}^{Einstein}_{\chi_{\vec{n}}}[\delta_{\chi_{\vec{0}}}\Phi,\Phi ]+\dfrac{i}{\vec{k}\cdot\vec{n}}\oint{\boldsymbol k}^{\mathbf{Y}}_{\chi_{\vec{n}}}[\delta_{\chi_{\vec{0}}}\Phi,\Phi ]\\
&=\dfrac{i}{\vec{k}\cdot\vec{n}}\oint\boldsymbol{\epsilon}_{\mathcal{H}} \frac{-4e^{-i \vec{n}\cdot\vec{\varphi}}}{16 \pi G } \Big(\Psi''\Psi'- i\vec{k}\cdot\vec{n}\Psi''+e^{2\Psi}\Psi'\Big)-\oint\boldsymbol \epsilon_{\mathcal H}\ \frac{e^{-i \vec{n}\cdot\vec{\varphi}}}{16 \pi G} \left({\Psi'}^2-2\Psi '' \right)\\
&=\oint\boldsymbol{\epsilon}_{\mathcal{H}} \frac{e^{-i \vec{n}\cdot\vec{\varphi}}}{16 \pi G } \Big(2\Psi'^2- 4\Psi''+2e^{2\Psi}\Big)-\oint\boldsymbol \epsilon_{\mathcal H}\ \frac{e^{-i \vec{n}\cdot\vec{\varphi}}}{16 \pi G} \left({\Psi'}^2-2\Psi '' \right)\,,
\end{align}
where in the last equation we used integration by parts, and dropped some total derivatives of $\vec{\varphi}$. Finally,
\begin{equation}\label{Hn charge}
H_{\vec{n}}=\oint \boldsymbol \epsilon_{\mathcal H}\ \frac{1}{16 \pi G} \left({\Psi'}^2-2 \Psi '' +2e^{2 \Psi }\right) e^{-i \vec{n}\cdot\vec{\varphi}}.
\end{equation}

\section{The Kerr/CFT type Phase Space}\label{append-Kerr-CFT}

In this section we derive the \emph{Kerr/CFT phase space}, defined as the regular phase space resulting from defining symmetry generators which depend on a function of a single angle along an arbitrary direction of the $d-3$ dimensional torus. We show that a symplectic structure exists such that the Kerr/CFT infinitesimal diffeomorphisms (defined however with a different ansatz than in the original proposal \cite{Guica:2008mu}) are symplectic symmetries and we build the set of regular metrics which represent the symplectic symmetries. We also show that there is no larger phase space which contains both the Kerr/CFT phase space and the NHEG phase space defined in the main text. 

The Kerr/CFT ansatz prescribes choosing a particular direction along the $d-3$ dimensional torus spanned by the $\varphi^i$ coordinates and defining an arbitrary diffeomorphism along that direction (see e.g. \cite{Loran:2009cr} for the $5d$ case.). Namely, one fixes a vector $K^i$ and  defines the $\varphi$ angle such that $K^i \p_{\varphi^i}\equiv \p_\varphi$. In order $\varphi$ to be periodic (with period $2\pi$) the direction $K$ should be a vector on the $d-3$ lattice associated with the torus. In other words, one should be able to map $\varphi$ to one of the directions $\varphi^i$'s using $SL(d-3,\mathbb{Z})$ transformations.   The arbitrary function of $\varphi$ is denoted as $\eps(\varphi)$. According to our discussions on the choice of symmetry generator in section \ref{sec-NHEG-phase-space} (\emph{cf.} discussions in the paragraph above \eqref{ASK}), we define the infinitesimal diffeomorphism
\begin{align}
\chi_b[{\epsilon}(\varphi)]&= {\epsilon}(\varphi)\pd_{\varphi} -  \pd_{\varphi}{\epsilon(\varphi) }\;(\dfrac{b}{r}\pd_{t}+r\pd_{r}).\label{KCFT}
\end{align}
In the original Kerr/CFT proposal, one set $b=0$. However, we saw that requiring regularity of the phase space obtained by exponentiating this generator instead fixes $b=\pm 1$. 

Quite nontrivially, a symplectic structure exists such that \eqref{KCFT} are symplectic symmetries. The symplectic structure can be chosen to be exactly the same as for the NHEG phase space, namely, 
\bea
\boldsymbol \omega[\delta_1 \Phi ,\delta_2 \Phi ; \Phi ] = \boldsymbol \omega_{(LW)}[\delta_1 \Phi ,\delta_2 \Phi ; \Phi ]  +  d (\delta_1 \mathbf{Y}[\delta_2 \Phi ; \Phi] - \delta_2 \mathbf{Y}[\delta_1 \Phi ; \Phi] )\label{cs}
\eea
where the boundary term $\mathbf{Y}[\delta \Phi ; \Phi]$ is defined in \eqref{Y-total}. One can readily check that the infinitesimal diffeomorphism \eqref{KCFT} obeys
\bea
\boldsymbol \omega[\delta_{\chi} \Phi ,\delta_{\chi'} \Phi ; \Phi ]  \approx 0.
\eea
Therefore, the charges are conserved and are integrable off-shell using the same reasoning as the one in section \ref{secint}. The charges represent a Virasoro algebra. 

For the simple example $K^i=\delta^i_1$, $\varphi = \varphi^1$ one can check that the Virasoro central charge $c$ is given by $c=12 k^1 \frac{S}{2\pi}$ where $S$ is the black hole entropy. In order to define the central charge for a general choice of cycle, let us define $e_i$ such that $\varphi = e_i \varphi^i$. It then follows that $e_i K^i = 1$ and then $\varphi^i = K^i \varphi + \varphi^i_\perp$ with $e_i \varphi^i_\perp = 0$. We can recycle the computation of the central charge that we performed for the NHEG ansatz to the Kerr/CFT case as follows. In section \ref{section-quantized-NHEG-algebra} we discussed that the NHEG algebra $\nhegalgebra$ has infinitely many Virasoro subalgebras obtained through considering only generators $L_\vn$ where $\vn=n \vec{e}$, for a given vector $\vec{e}$, $\vec{k}\cdot \vec{e} \neq 0$ and $n$ integer. These generators which we denoted by $\ell_n$ \eqref{Virasoro-subalgebra-generators} may be viewed as the Fourier modes of $\ell(\varphi)$, where $n_i\varphi^i=n\varphi$. If along with the $\ell(\varphi)$ we also restrict ourselves to part of the phase space specified by functions of $\varphi=e_i\varphi^i$ (not depending on other combinations of $\varphi^i$), then for this sector the $\ell_n$ NHEG generators reduce to 
$$
\ell_n=  -e^{-i n \varphi } \p_\varphi+\p_\varphi  e^{-i n \varphi } (\dfrac{b}{r}\pd_{t}+r\pd_{r})
$$
and therefore coincide with the Kerr/CFT generators  discussed in this appendix. The central charge is therefore equal to the one given in \eqref{virasoro-subalgenra-central-charge},
\bea
c_{\vec{e}} = 12 (\vec{k} \cdot \vec{e}) \frac{S}{2\pi}. 
\eea
One can also check that for $e_i = \delta_i^1$ we reproduce the explicit result mentioned earlier.  The central charge for general choices of cycles on the torus were also discussed in \cite{Loran:2009cr}.

The regular phase space is obtained by exponentiating the diffeomorphism with the choice $b= 1$ and applying this finite coordinate transformation on the background. The phase space is labelled by an arbitrary function $F(\varphi)=F(\varphi+2\pi)$ from which one defines the Liouville field $e^\Psi = 1+ \p_\varphi F(\varphi)$. Using a similar reasoning as in the main text, one obtains
\begin{align}\label{g-F2}
ds^2=\Gamma(\theta)&\Big[-\left( \boldsymbol\sigma - b d\Psi \right)^2+\Big(\dfrac{dr}{r}-d{\Psi}\Big)^2
+d\theta^2+\gamma_{ij}(d\tilde{\varphi}^i+{k^i}\boldsymbol{\sigma})(d\tilde{\varphi}^j+{k^j}\boldsymbol{\sigma})\Big],
\end{align}
where
\be\label{tilde-varphi2}
\boldsymbol{\sigma}=e^{-{\Psi}}rdv +\dfrac{dr}{r},\qquad \tilde{\varphi}^i=\varphi^i+K^i  F(\varphi) -k^i  {\Psi}(\varphi),
\ee
with $v=t+\frac{1}{r}$ and { $\vec{K}$ is the direction in the $d-3$ torus defining the $\varphi$ direction defined earlier.}

The computation of the charges follows the same route as in the main text. Here the charges are labelled by the single function $\eps[\varphi]$. Explicitly, in any dimension
\bea
H_{\chi[\eps]} =\frac{ 1}{16 \pi G}  \int_{\mathcal H} \boldsymbol \eps_{\mathcal H}  \Big( (\p_\varphi\Psi)^2-2 \p_\varphi^2 \Psi+ \Lambda e^{2 \Psi  } \Big) \eps[\varphi]
\eea
The Virasoro charges are therefore expressed as the modes of a Liouville-type stress-tensor which depends upon a single angle $\varphi$ and is time-independent.

An obvious question is whether or not a larger phase space exists that contains both the NHEG symplectic symmetries and the Kerr/CFT symplectic symmetries. Here, we  show that these phase spaces are mutually incompatible. Let us consider the $5d$ case and consider the vector field
\begin{align}
\chi_{{\vec{\epsilon}_i}}&={\vec{\epsilon}_i}  \cdot \vec{\pd}_{\varphi} - \vec{\pd}_{\varphi}\cdot {\vec{\epsilon}_i}\;(\dfrac{b}{r}\pd_{t}+r\pd_{r}).
\end{align}
with $i=1,2$ and
\bea
\vec{\eps}_1 = (\eps^1(\varphi^1),0), \qquad \vec{\eps}_2 = (k^1,k^2) \eps^2(\varphi^2),
\eea

The first vector $\vec\chi_{\vec\eps_1}$ is part of the Kerr/CFT ansatz and the second one  $\vec\chi_{\vec\eps_2}$ is part of the NHEG ansatz. If a phase space exists where both of these vectors define symmetry generators, then the commutator of these generators should also be a symmetry generator. We have
\bea
[\chi_{\vec{\eps}_1},\chi_{\vec{\eps}_2}] = \chi_{[\vec{\eps}_1,\vec{\eps}_2]}
\eea
Now, $\vec{\eps}_3 \equiv [\vec{\eps}_1,\vec{\eps}_2] = (k^1 \chi^2(\varphi^2)\p_1 \epsilon^1(\varphi^1),0)$. Expanding in Fourier modes, $\vec{\eps}_3 \sim (e^{i m_1 \varphi^1 + i m_2 \varphi^2},0)$. Let us now compute the symplectic structure for two such $\vec\chi_{\vec{\eps}_3}$ vectors with modes $(m_1,m_2)$ and $(n_1,n_2)$. Since we require that the symplectic structure be independent on $b$, we set $b=0$ without loss of generality. Evaluating the symplectic structure \eqref{cs} around the background in $(v,r,\theta,\varphi^1,\varphi^2)$ coordinates we find 
\bea
\omega^v [\delta g[m_1,m_2] , \delta g[n_1,n_2], \bar g] \propto \frac{\sqrt{-g}}{16 \pi G} \frac{1}{r}  e^{-i(m_1+n_1)\varphi^1-i(m_2+n_2)\varphi^2} (m_1+n_1)(n_1 m_2 + m_1 n_2) \nonumber
\eea
In the NHEG phase space, $\omega^v =0$ exactly so this divergence is specific to the extension of the phase space. After an extensive search we didn't find any possible boundary term which we could add to the symplectic structure to cancel the divergence. We conclude that the Kerr/CFT ansatz and NHEG ansatz define mutually incompatible phase spaces in five and higher dimensions. In four dimensions the NHEG phase space and the Kerr/CFT phase space that we constructed in this appendix simply coincide.


\begin{thebibliography}{10}

\bibitem{Strominger:1996sh}
A.~Strominger and C.~Vafa, ``{Microscopic origin of the Bekenstein-Hawking
  entropy},'' \href{http://dx.doi.org/10.1016/0370-2693(96)00345-0}{{\em
  Phys.Lett.} {\bfseries B379} (1996) 99--104},
\href{http://arxiv.org/abs/hep-th/9601029}{{\ttfamily arXiv:hep-th/9601029
  [hep-th]}}.

\bibitem{Dabholkar:2014ema}
A.~Dabholkar, J.~Gomes, and S.~Murthy, ``{Nonperturbative black hole entropy
  and Kloosterman sums},''
  \href{http://dx.doi.org/10.1007/JHEP03(2015)074}{{\em JHEP} {\bfseries 1503}
  (2015) 074},
\href{http://arxiv.org/abs/1404.0033}{{\ttfamily arXiv:1404.0033 [hep-th]}}.

\bibitem{McClintock:2006xd}
J.~E. McClintock, R.~Shafee, R.~Narayan, R.~A. Remillard, S.~W. Davis, {\em et
  al.}, ``{The Spin of the Near-Extreme Kerr Black Hole GRS 1915+105},''
  \href{http://dx.doi.org/10.1086/508457}{{\em Astrophys.J.} {\bfseries 652}
  (2006) 518--539},
\href{http://arxiv.org/abs/astro-ph/0606076}{{\ttfamily arXiv:astro-ph/0606076
  [astro-ph]}}.

\bibitem{2011ApJ...742...85G}
L.~{Gou}, J.~E. {McClintock}, M.~J. {Reid}, J.~A. {Orosz}, J.~F. {Steiner},
  R.~{Narayan}, J.~{Xiang}, R.~A. {Remillard}, K.~A. {Arnaud}, and S.~W.
  {Davis}, ``{The Extreme Spin of the Black Hole in Cygnus X-1},''
  \href{http://dx.doi.org/10.1088/0004-637X/742/2/85}{{\em The Astrophysical
  Journal} {\bfseries 742} (Dec., 2011) 85},
  \href{http://arxiv.org/abs/1106.3690}{{\ttfamily arXiv:1106.3690
  [astro-ph.HE]}}.

\bibitem{Brenneman:2006hw}
L.~W. Brenneman and C.~S. Reynolds, ``{Constraining Black Hole Spin Via X-ray
  Spectroscopy},'' \href{http://dx.doi.org/10.1086/508146}{{\em Astrophys.J.}
  {\bfseries 652} (2006) 1028--1043},
\href{http://arxiv.org/abs/astro-ph/0608502}{{\ttfamily arXiv:astro-ph/0608502
  [astro-ph]}}.

\bibitem{Bardeen:1973gs}
J.~M. Bardeen, B.~Carter, and S.~Hawking, ``{The Four laws of black hole
  mechanics},''
\href{http://dx.doi.org/10.1007/BF01645742}{{\em Commun.Math.Phys.} {\bfseries
  31} (1973) 161--170}.

\bibitem{Bekenstein:1973ur}
J.~D. Bekenstein, ``{Black holes and entropy},''
\href{http://dx.doi.org/10.1103/PhysRevD.7.2333}{{\em Phys.Rev.} {\bfseries D7}
  (1973) 2333--2346}.

\bibitem{Compere:2015mza}
G.~Comp\`ere, K.~Hajian, A.~Seraj, and M.~Sheikh-Jabbari, ``{Extremal Rotating
  Black Holes in the Near-Horizon Limit: Phase Space and Symmetry Algebra},''
\href{http://arxiv.org/abs/1503.07861}{{\ttfamily arXiv:1503.07861 [hep-th]}}.

\bibitem{Wald:1993nt}
R.~M. Wald, ``{Black hole entropy is the Noether charge},''
  \href{http://dx.doi.org/10.1103/PhysRevD.48.R3427}{{\em Phys.Rev.} {\bfseries
  D48} (1993) 3427--3431},
\href{http://arxiv.org/abs/gr-qc/9307038}{{\ttfamily arXiv:gr-qc/9307038
  [gr-qc]}}.

\bibitem{Iyer:1994ys}
V.~Iyer and R.~M. Wald, ``{Some properties of Noether charge and a proposal for
  dynamical black hole entropy},''
  \href{http://dx.doi.org/10.1103/PhysRevD.50.846}{{\em Phys.Rev.} {\bfseries
  D50} (1994) 846--864},
\href{http://arxiv.org/abs/gr-qc/9403028}{{\ttfamily arXiv:gr-qc/9403028
  [gr-qc]}}.

\bibitem{Amsel:2009et}
A.~J. Amsel, G.~T. Horowitz, D.~Marolf, and M.~M. Roberts, ``{Uniqueness of
  Extremal Kerr and Kerr-Newman Black Holes},''
  \href{http://dx.doi.org/10.1103/PhysRevD.81.024033}{{\em Phys.Rev.}
  {\bfseries D81} (2010) 024033},
\href{http://arxiv.org/abs/0906.2367}{{\ttfamily arXiv:0906.2367 [gr-qc]}}.

\bibitem{Bardeen:1999px}
J.~M. Bardeen and G.~T. Horowitz, ``{The Extreme Kerr throat geometry: A Vacuum
  analog of AdS(2) x $S^2$},''
  \href{http://dx.doi.org/10.1103/PhysRevD.60.104030}{{\em Phys.Rev.}
  {\bfseries D60} (1999) 104030},
\href{http://arxiv.org/abs/hep-th/9905099}{{\ttfamily arXiv:hep-th/9905099
  [hep-th]}}.

\bibitem{Kunduri:2007vf}
H.~K. Kunduri, J.~Lucietti, and H.~S. Reall, ``{Near-horizon symmetries of
  extremal black holes},''
  \href{http://dx.doi.org/10.1088/0264-9381/24/16/012}{{\em Class.Quant.Grav.}
  {\bfseries 24} (2007) 4169--4190},
\href{http://arxiv.org/abs/0705.4214}{{\ttfamily arXiv:0705.4214 [hep-th]}}.

\bibitem{Kunduri:2013gce}
H.~K. Kunduri and J.~Lucietti, ``{Classification of near-horizon geometries of
  extremal black holes},'' \href{http://dx.doi.org/10.12942/lrr-2013-8}{{\em
  Living Rev.Rel.} {\bfseries 16} (2013) 8},
\href{http://arxiv.org/abs/1306.2517}{{\ttfamily arXiv:1306.2517 [hep-th]}}.

\bibitem{Hajian:2013lna}
K.~Hajian, A.~Seraj, and M.~Sheikh-Jabbari, ``{NHEG Mechanics: Laws of Near
  Horizon Extremal Geometry (Thermo)Dynamics},''
  \href{http://dx.doi.org/10.1007/JHEP03(2014)014}{{\em JHEP} {\bfseries 1403}
  (2014) 014},
\href{http://arxiv.org/abs/1310.3727}{{\ttfamily arXiv:1310.3727 [hep-th]}}.

\bibitem{Hajian:2014twa}
K.~Hajian, A.~Seraj, and M.~Sheikh-Jabbari, ``{Near Horizon Extremal Geometry
  Perturbations: Dynamical Field Perturbations vs. Parametric Variations},''
  \href{http://dx.doi.org/10.1007/JHEP10(2014)111}{{\em JHEP} {\bfseries 1410}
  (2014) 111},
\href{http://arxiv.org/abs/1407.1992}{{\ttfamily arXiv:1407.1992 [hep-th]}}.

\bibitem{Aharony:1999ti}
O.~Aharony, S.~S. Gubser, J.~M. Maldacena, H.~Ooguri, and Y.~Oz, ``{Large N
  field theories, string theory and gravity},''
  \href{http://dx.doi.org/10.1016/S0370-1573(99)00083-6}{{\em Phys.Rept.}
  {\bfseries 323} (2000) 183--386},
\href{http://arxiv.org/abs/hep-th/9905111}{{\ttfamily arXiv:hep-th/9905111
  [hep-th]}}.

\bibitem{Sen:2008yk}
A.~Sen, ``{Entropy Function and AdS(2) / CFT(1) Correspondence},''
  \href{http://dx.doi.org/10.1088/1126-6708/2008/11/075}{{\em JHEP} {\bfseries
  0811} (2008) 075},
\href{http://arxiv.org/abs/0805.0095}{{\ttfamily arXiv:0805.0095 [hep-th]}}.

\bibitem{Sen:2008vm}
A.~Sen, ``{Quantum Entropy Function from AdS(2)/CFT(1) Correspondence},''
  \href{http://dx.doi.org/10.1142/S0217751X09045893}{{\em Int.J.Mod.Phys.}
  {\bfseries A24} (2009) 4225--4244},
\href{http://arxiv.org/abs/0809.3304}{{\ttfamily arXiv:0809.3304 [hep-th]}}.

\bibitem{Strominger:1998yg}
A.~Strominger, ``{AdS(2) quantum gravity and string theory},''
  \href{http://dx.doi.org/10.1088/1126-6708/1999/01/007}{{\em JHEP} {\bfseries
  9901} (1999) 007},
\href{http://arxiv.org/abs/hep-th/9809027}{{\ttfamily arXiv:hep-th/9809027
  [hep-th]}}.

\bibitem{Maldacena:1998uz}
J.~M. Maldacena, J.~Michelson, and A.~Strominger, ``{Anti-de Sitter
  fragmentation},'' \href{http://dx.doi.org/10.1088/1126-6708/1999/02/011}{{\em
  JHEP} {\bfseries 9902} (1999) 011},
\href{http://arxiv.org/abs/hep-th/9812073}{{\ttfamily arXiv:hep-th/9812073
  [hep-th]}}.

\bibitem{Hartman:2008dq}
T.~Hartman and A.~Strominger, ``{Central Charge for AdS(2) Quantum Gravity},''
  \href{http://dx.doi.org/10.1088/1126-6708/2009/04/026}{{\em JHEP} {\bfseries
  0904} (2009) 026},
\href{http://arxiv.org/abs/0803.3621}{{\ttfamily arXiv:0803.3621 [hep-th]}}.

\bibitem{Castro:2014ima}
A.~Castro and W.~Song, ``{Comments on $AdS_2$ Gravity},''
\href{http://arxiv.org/abs/1411.1948}{{\ttfamily arXiv:1411.1948 [hep-th]}}.

\bibitem{Guica:2008mu}
M.~Guica, T.~Hartman, W.~Song, and A.~Strominger, ``{The Kerr/CFT
  Correspondence},'' \href{http://dx.doi.org/10.1103/PhysRevD.80.124008}{{\em
  Phys.Rev.} {\bfseries D80} (2009) 124008},
\href{http://arxiv.org/abs/0809.4266}{{\ttfamily arXiv:0809.4266 [hep-th]}}.

\bibitem{Amsel:2009ev}
A.~J. Amsel, G.~T. Horowitz, D.~Marolf, and M.~M. Roberts, ``{No Dynamics in
  the Extremal Kerr Throat},''
  \href{http://dx.doi.org/10.1088/1126-6708/2009/09/044}{{\em JHEP} {\bfseries
  0909} (2009) 044},
\href{http://arxiv.org/abs/0906.2376}{{\ttfamily arXiv:0906.2376 [hep-th]}}.

\bibitem{Dias:2009ex}
O.~J. Dias, H.~S. Reall, and J.~E. Santos, ``{Kerr-CFT and gravitational
  perturbations},'' \href{http://dx.doi.org/10.1088/1126-6708/2009/08/101}{{\em
  JHEP} {\bfseries 0908} (2009) 101},
\href{http://arxiv.org/abs/0906.2380}{{\ttfamily arXiv:0906.2380 [hep-th]}}.

\bibitem{Bredberg:2011hp}
I.~Bredberg, C.~Keeler, V.~Lysov, and A.~Strominger, ``{Cargese Lectures on the
  Kerr/CFT Correspondence},''
  \href{http://dx.doi.org/10.1016/j.nuclphysbps.2011.04.155}{{\em
  Nucl.Phys.Proc.Suppl.} {\bfseries 216} (2011) 194--210},
\href{http://arxiv.org/abs/1103.2355}{{\ttfamily arXiv:1103.2355 [hep-th]}}.

\bibitem{Compere:2012jk}
G.~Comp\`ere, ``{The Kerr/CFT correspondence and its extensions: a
  comprehensive review},'' {\em Living Rev.Rel.} {\bfseries 15} (2012) 11,
\href{http://arxiv.org/abs/1203.3561}{{\ttfamily arXiv:1203.3561 [hep-th]}}.

\bibitem{Brown:1986nw}
J.~D. Brown and M.~Henneaux, ``{Central Charges in the Canonical Realization of
  Asymptotic Symmetries: An Example from Three-Dimensional Gravity},''
\href{http://dx.doi.org/10.1007/BF01211590}{{\em Commun.Math.Phys.} {\bfseries
  104} (1986) 207--226}.

\bibitem{Compere:2014cna}
G.~Comp\`ere, L.~Donnay, P.-H. Lambert, and W.~Schulgin, ``{Liouville theory
  beyond the cosmological horizon},''
  \href{http://dx.doi.org/10.1007/JHEP03(2015)158}{{\em JHEP} {\bfseries 1503}
  (2015) 158},
\href{http://arxiv.org/abs/1411.7873}{{\ttfamily arXiv:1411.7873 [hep-th]}}.

\bibitem{Johnstone:2013ioa}
M.~Johnstone, M.~Sheikh-Jabbari, J.~Simon, and H.~Yavartanoo, ``{Extremal black
  holes and the first law of thermodynamics},''
  \href{http://dx.doi.org/10.1103/PhysRevD.88.101503}{{\em Phys.Rev.}
  {\bfseries D88} no.~10, (2013) 101503},
\href{http://arxiv.org/abs/1305.3157}{{\ttfamily arXiv:1305.3157 [hep-th]}}.

\bibitem{Astefanesei:2006dd}
D.~Astefanesei, K.~Goldstein, R.~P. Jena, A.~Sen, and S.~P. Trivedi,
  ``{Rotating attractors},''
  \href{http://dx.doi.org/10.1088/1126-6708/2006/10/058}{{\em JHEP} {\bfseries
  0610} (2006) 058},
\href{http://arxiv.org/abs/hep-th/0606244}{{\ttfamily arXiv:hep-th/0606244
  [hep-th]}}.

\bibitem{Patera91}
J.~Patera and H.~Zassenhaus, ``The higher rank virasoro algebras,''
  \href{http://dx.doi.org/10.1007/BF02096787}{{\em Communications in
  Mathematical Physics} {\bfseries 136} no.~1, (1991) 1--14}.
  \url{http://dx.doi.org/10.1007/BF02096787}.

\bibitem{Mazorchuk98}
V.~Mazorchuk, ``On unitarizable modules over generalized virasoro algebras,''
  \href{http://dx.doi.org/10.1007/BF02525253}{{\em Ukrainian Mathematical
  Journal} {\bfseries 50} no.~9, (1998) 1461--1463}.
  \url{http://dx.doi.org/10.1007/BF02525253}.

\bibitem{2006math......7614G}
X.~{Guo}, R.~{Lu}, and K.~{Zhao}, ``{Classification of irreducible
  Harish-Chandra modules over generalized Virasoro algebras},'' {\em ArXiv
  Mathematics e-prints} (July, 2006) ,
  \href{http://arxiv.org/abs/math/0607614}{{\ttfamily math/0607614}}.

\bibitem{Hollands:2009ng}
S.~Hollands and A.~Ishibashi, ``{All vacuum near horizon geometries in
  arbitrary dimensions},''
  \href{http://dx.doi.org/10.1007/s00023-010-0022-y}{{\em Annales Henri
  Poincare} {\bfseries 10} (2010) 1537--1557},
\href{http://arxiv.org/abs/0909.3462}{{\ttfamily arXiv:0909.3462 [gr-qc]}}.

\bibitem{Kunduri:2008rs}
H.~K. Kunduri and J.~Lucietti, ``{A Classification of near-horizon geometries
  of extremal vacuum black holes},''
  \href{http://dx.doi.org/10.1063/1.3190480}{{\em J.Math.Phys.} {\bfseries 50}
  (2009) 082502},
\href{http://arxiv.org/abs/0806.2051}{{\ttfamily arXiv:0806.2051 [hep-th]}}.

\bibitem{Schiffrin:2015yua}
J.~S. Schiffrin and R.~M. Wald, ``{Reflection Symmetry in Higher Dimensional
  Black Hole Spacetimes},''
  \href{http://dx.doi.org/10.1088/0264-9381/32/10/105005}{{\em
  Class.Quant.Grav.} {\bfseries 32} no.~10, (2015) 105005},
\href{http://arxiv.org/abs/1501.02752}{{\ttfamily arXiv:1501.02752 [gr-qc]}}.

\bibitem{Coley:2004jv}
A.~Coley, R.~Milson, V.~Pravda, and A.~Pravdova, ``{Classification of the Weyl
  tensor in higher dimensions},''
  \href{http://dx.doi.org/10.1088/0264-9381/21/7/L01}{{\em Class.Quant.Grav.}
  {\bfseries 21} (2004) L35--L42},
\href{http://arxiv.org/abs/gr-qc/0401008}{{\ttfamily arXiv:gr-qc/0401008
  [gr-qc]}}.

\bibitem{Godazgar:2009fi}
M.~Godazgar and H.~S. Reall, ``{Algebraically special axisymmetric solutions of
  the higher-dimensional vacuum Einstein equation},''
  \href{http://dx.doi.org/10.1088/0264-9381/26/16/165009}{{\em
  Class.Quant.Grav.} {\bfseries 26} (2009) 165009},
\href{http://arxiv.org/abs/0904.4368}{{\ttfamily arXiv:0904.4368 [gr-qc]}}.

\bibitem{Durkee:2010ea}
M.~Durkee and H.~S. Reall, ``{Perturbations of near-horizon geometries and
  instabilities of Myers-Perry black holes},''
  \href{http://dx.doi.org/10.1103/PhysRevD.83.104044}{{\em Phys.Rev.}
  {\bfseries D83} (2011) 104044},
\href{http://arxiv.org/abs/1012.4805}{{\ttfamily arXiv:1012.4805 [hep-th]}}.

\bibitem{Bengtsson:2005zj}
I.~Bengtsson and P.~Sandin, ``{Anti de Sitter space, squashed and stretched},''
  \href{http://dx.doi.org/10.1088/0264-9381/23/3/022}{{\em Class.Quant.Grav.}
  {\bfseries 23} (2006) 971--986},
\href{http://arxiv.org/abs/gr-qc/0509076}{{\ttfamily arXiv:gr-qc/0509076
  [gr-qc]}}.

\bibitem{Unruh:1976db}
W.~Unruh, ``{Notes on black hole evaporation},''
\href{http://dx.doi.org/10.1103/PhysRevD.14.870}{{\em Phys.Rev.} {\bfseries
  D14} (1976) 870}.

\bibitem{Arnowitt:1962hi}
R.~L. Arnowitt, S.~Deser, and C.~W. Misner, ``{The Dynamics of general
  relativity},'' \href{http://dx.doi.org/10.1007/s10714-008-0661-1}{{\em
  Gen.Rel.Grav.} {\bfseries 40} (2008) 1997--2027},
\href{http://arxiv.org/abs/gr-qc/0405109}{{\ttfamily arXiv:gr-qc/0405109
  [gr-qc]}}.

\bibitem{Regge:1974zd}
T.~Regge and C.~Teitelboim, ``{Role of Surface Integrals in the Hamiltonian
  Formulation of General Relativity},''
\href{http://dx.doi.org/10.1016/0003-4916(74)90404-7}{{\em Annals Phys.}
  {\bfseries 88} (1974) 286}.

\bibitem{Brown:1986ed}
J.~D. Brown and M.~Henneaux, ``{On the Poisson Brackets of Differentiable
  Generators in Classical Field Theory},''
\href{http://dx.doi.org/10.1063/1.527249}{{\em J.Math.Phys.} {\bfseries 27}
  (1986) 489--491}.

\bibitem{Lee:1990nz}
J.~Lee and R.~M. Wald, ``{Local symmetries and constraints},''
\href{http://dx.doi.org/10.1063/1.528801}{{\em J.Math.Phys.} {\bfseries 31}
  (1990) 725--743}.

\bibitem{Barnich:2001jy}
G.~Barnich and F.~Brandt, ``{Covariant theory of asymptotic symmetries,
  conservation laws and central charges},''
  \href{http://dx.doi.org/10.1016/S0550-3213(02)00251-1}{{\em Nucl.Phys.}
  {\bfseries B633} (2002) 3--82},
\href{http://arxiv.org/abs/hep-th/0111246}{{\ttfamily arXiv:hep-th/0111246
  [hep-th]}}.

\bibitem{Barnich:2007bf}
G.~Barnich and G.~Comp\`ere, ``{Surface charge algebra in gauge theories and
  thermodynamic integrability},''
  \href{http://dx.doi.org/10.1063/1.2889721}{{\em J.Math.Phys.} {\bfseries 49}
  (2008) 042901},
\href{http://arxiv.org/abs/0708.2378}{{\ttfamily arXiv:0708.2378 [gr-qc]}}.

\bibitem{Compere:2008us}
G.~Comp\`ere and D.~Marolf, ``{Setting the boundary free in AdS/CFT},''
  \href{http://dx.doi.org/10.1088/0264-9381/25/19/195014}{{\em
  Class.Quant.Grav.} {\bfseries 25} (2008) 195014},
\href{http://arxiv.org/abs/0805.1902}{{\ttfamily arXiv:0805.1902 [hep-th]}}.

\bibitem{Lu:2008jk}
H.~Lu, J.~Mei, and C.~Pope, ``{Kerr/CFT Correspondence in Diverse
  Dimensions},'' \href{http://dx.doi.org/10.1088/1126-6708/2009/04/054}{{\em
  JHEP} {\bfseries 0904} (2009) 054},
\href{http://arxiv.org/abs/0811.2225}{{\ttfamily arXiv:0811.2225 [hep-th]}}.

\bibitem{Azeyanagi:2009wf}
T.~Azeyanagi, G.~Compere, N.~Ogawa, Y.~Tachikawa, and S.~Terashima,
  ``{Higher-Derivative Corrections to the Asymptotic Virasoro Symmetry of 4d
  Extremal Black Holes},'' \href{http://dx.doi.org/10.1143/PTP.122.355}{{\em
  Prog.Theor.Phys.} {\bfseries 122} (2009) 355--384},
\href{http://arxiv.org/abs/0903.4176}{{\ttfamily arXiv:0903.4176 [hep-th]}}.

\bibitem{Amsel:2009pu}
A.~J. Amsel, D.~Marolf, and M.~M. Roberts, ``{On the Stress Tensor of
  Kerr/CFT},'' \href{http://dx.doi.org/10.1088/1126-6708/2009/10/021}{{\em
  JHEP} {\bfseries 0910} (2009) 021},
\href{http://arxiv.org/abs/0907.5023}{{\ttfamily arXiv:0907.5023 [hep-th]}}.

\bibitem{Coussaert:1995zp}
O.~Coussaert, M.~Henneaux, and P.~van Driel, ``{The Asymptotic dynamics of
  three-dimensional Einstein gravity with a negative cosmological constant},''
  \href{http://dx.doi.org/10.1088/0264-9381/12/12/012}{{\em Class.Quant.Grav.}
  {\bfseries 12} (1995) 2961--2966},
\href{http://arxiv.org/abs/gr-qc/9506019}{{\ttfamily arXiv:gr-qc/9506019
  [gr-qc]}}.

\bibitem{Banados:1998gg}
M.~Banados, ``{Three-dimensional quantum geometry and black holes},''
  \href{http://dx.doi.org/10.1063/1.59661}{{\em AIP Conf.Proc.} {\bfseries 484}
  (1999) 147--169},
\href{http://arxiv.org/abs/hep-th/9901148}{{\ttfamily arXiv:hep-th/9901148
  [hep-th]}}.

\bibitem{Sheikh-Jabbari:2014nya}
M.~Sheikh-Jabbari and H.~Yavartanoo, ``{On quantization of AdS$_{3}$ gravity I:
  semi-classical analysis},''
  \href{http://dx.doi.org/10.1007/JHEP07(2014)104}{{\em JHEP} {\bfseries 1407}
  (2014) 104},
\href{http://arxiv.org/abs/1404.4472}{{\ttfamily arXiv:1404.4472 [hep-th]}}.

\bibitem{Li:2013pra}
C.~Li and J.~Lucietti, ``{Three-dimensional black holes and descendants},''
  \href{http://dx.doi.org/10.1016/j.physletb.2014.09.012}{{\em Phys.Lett.}
  {\bfseries B738} (2014) 48--4},
\href{http://arxiv.org/abs/1312.2626}{{\ttfamily arXiv:1312.2626 [hep-th]}}.

\bibitem{Banados:1992wn}
M.~Banados, C.~Teitelboim, and J.~Zanelli, ``{The Black hole in
  three-dimensional space-time},''
  \href{http://dx.doi.org/10.1103/PhysRevLett.69.1849}{{\em Phys.Rev.Lett.}
  {\bfseries 69} (1992) 1849--1851},
\href{http://arxiv.org/abs/hep-th/9204099}{{\ttfamily arXiv:hep-th/9204099
  [hep-th]}}.

\bibitem{Banados:1992gq}
M.~Banados, M.~Henneaux, C.~Teitelboim, and J.~Zanelli, ``{Geometry of the
  (2+1) black hole},'' \href{http://dx.doi.org/10.1103/PhysRevD.48.1506,
  10.1103/PhysRevD.88.069902}{{\em Phys.Rev.} {\bfseries D48} no.~6, (1993)
  1506--1525},
\href{http://arxiv.org/abs/gr-qc/9302012}{{\ttfamily arXiv:gr-qc/9302012
  [gr-qc]}}.

\bibitem{Coussaert:1994tu}
O.~Coussaert and M.~Henneaux, ``{Selfdual solutions of (2+1) Einstein gravity
  with a negative cosmological constant},'' {\em Teitelboim, C. (ed.): The
  black hole} (1994) 25--39,
\href{http://arxiv.org/abs/hep-th/9407181}{{\ttfamily arXiv:hep-th/9407181
  [hep-th]}}.

\bibitem{Balasubramanian:2009bg}
V.~Balasubramanian, J.~de~Boer, M.~Sheikh-Jabbari, and J.~Simon, ``{What is a
  chiral 2d CFT? And what does it have to do with extremal black holes?},''
  \href{http://dx.doi.org/10.1007/JHEP02(2010)017}{{\em JHEP} {\bfseries 1002}
  (2010) 017},
\href{http://arxiv.org/abs/0906.3272}{{\ttfamily arXiv:0906.3272 [hep-th]}}.

\bibitem{Compere:2014bia}
G.~Comp\`ere, M.~Guica, and M.~J. Rodriguez, ``{Two Virasoro symmetries in
  stringy warped AdS$_{3}$},''
  \href{http://dx.doi.org/10.1007/JHEP12(2014)012}{{\em JHEP} {\bfseries 1412}
  (2014) 012},
\href{http://arxiv.org/abs/1407.7871}{{\ttfamily arXiv:1407.7871 [hep-th]}}.

\bibitem{Baggio:2012db}
M.~Baggio, J.~de~Boer, J.~I. Jottar, and D.~R. Mayerson, ``{Conformal Symmetry
  for Black Holes in Four Dimensions and Irrelevant Deformations},''
  \href{http://dx.doi.org/10.1007/JHEP04(2013)084}{{\em JHEP} {\bfseries 1304}
  (2013) 084},
\href{http://arxiv.org/abs/1210.7695}{{\ttfamily arXiv:1210.7695 [hep-th]}}.

\bibitem{Castro:2013kea}
A.~Castro, J.~M. Lapan, A.~Maloney, and M.~J. Rodriguez, ``{Black Hole
  Monodromy and Conformal Field Theory},''
  \href{http://dx.doi.org/10.1103/PhysRevD.88.044003}{{\em Phys.Rev.}
  {\bfseries D88} (2013) 044003},
\href{http://arxiv.org/abs/1303.0759}{{\ttfamily arXiv:1303.0759 [hep-th]}}.

\bibitem{Gibbons:2013tqa}
G.~Gibbons and N.~Warner, ``{Global structure of five-dimensional fuzzballs},''
  \href{http://dx.doi.org/10.1088/0264-9381/31/2/025016}{{\em
  Class.Quant.Grav.} {\bfseries 31} (2014) 025016},
\href{http://arxiv.org/abs/1305.0957}{{\ttfamily arXiv:1305.0957 [hep-th]}}.

\bibitem{Bena:2012wc}
I.~Bena, M.~Guica, and W.~Song, ``{Un-twisting the NHEK with spectral flows},''
  \href{http://dx.doi.org/10.1007/JHEP03(2013)028}{{\em JHEP} {\bfseries 1303}
  (2013) 028},
\href{http://arxiv.org/abs/1203.4227}{{\ttfamily arXiv:1203.4227 [hep-th]}}.

\bibitem{Barnich:2000zw}
G.~Barnich, F.~Brandt, and M.~Henneaux, ``{Local BRST cohomology in gauge
  theories},'' \href{http://dx.doi.org/10.1016/S0370-1573(00)00049-1}{{\em
  Phys.Rept.} {\bfseries 338} (2000) 439--569},
\href{http://arxiv.org/abs/hep-th/0002245}{{\ttfamily arXiv:hep-th/0002245
  [hep-th]}}.

\bibitem{Wald:1999wa}
R.~M. Wald and A.~Zoupas, ``{A General definition of 'conserved quantities' in
  general relativity and other theories of gravity},''
  \href{http://dx.doi.org/10.1103/PhysRevD.61.084027}{{\em Phys.Rev.}
  {\bfseries D61} (2000) 084027},
\href{http://arxiv.org/abs/gr-qc/9911095}{{\ttfamily arXiv:gr-qc/9911095
  [gr-qc]}}.

\bibitem{Barnich:2010eb}
G.~Barnich and C.~Troessaert, ``{Aspects of the BMS/CFT correspondence},''
  \href{http://dx.doi.org/10.1007/JHEP05(2010)062}{{\em JHEP} {\bfseries 1005}
  (2010) 062},
\href{http://arxiv.org/abs/1001.1541}{{\ttfamily arXiv:1001.1541 [hep-th]}}.

\bibitem{Barnich:2010xq}
G.~Barnich, ``{A Note on gauge systems from the point of view of Lie
  algebroids},'' \href{http://dx.doi.org/10.1063/1.3527427}{{\em AIP
  Conf.Proc.} {\bfseries 1307} (2010) 7--18},
\href{http://arxiv.org/abs/1010.0899}{{\ttfamily arXiv:1010.0899 [math-ph]}}.

\bibitem{Loran:2009cr}
F.~Loran and H.~Soltanpanahi, ``{5D Extremal Rotating Black Holes and CFT
  duals},'' \href{http://dx.doi.org/10.1088/0264-9381/26/15/155019}{{\em
  Class.Quant.Grav.} {\bfseries 26} (2009) 155019},
\href{http://arxiv.org/abs/0901.1595}{{\ttfamily arXiv:0901.1595 [hep-th]}}.

\end{thebibliography}

\providecommand{\href}[2]{#2}\begingroup\raggedright\endgroup

\end{document}